\documentclass[twocolumn,twocolappendix]{aastex63}

\usepackage{comment}
\usepackage{multirow}
\usepackage{longtable,subfigure,float}

\usepackage{lineno}

\usepackage{chngcntr}

\usepackage{amsmath}	% Advanced maths commands
\usepackage{amssymb}	% Extra maths symbols

\newcommand{\f}{$f$}
\newcommand{\faver}{$\langle f \rangle$}
\newcommand {\ergpersec} {\ifmmode {\rm erg~s}^{-1} \else erg~s$^{-1}$ \fi}
\def\arcsec{{\mbox{$^{\prime \prime}$}}}

\newcommand{\kms}{\ifmmode {\rm km\,s}^{-1} \else ${\rm km\,s}^{-1}$\fi}

\received{13 December 2020}%June 1, 2019}
\revised{28 August 2021}%January 10, 2019}
\accepted{31 August 2021}%\today}
\submitjournal{ApJ}

\shorttitle{Near-infrared view of the BLR}
\shortauthors{Ricci et al.}
\begin{document}
	
	\title{BASS XXIX: The near-infrared view of the BLR: the effects of obscuration in BLR characterisation\footnote{This paper includes data gathered with the 6.5 meter Magellan Telescopes located at Las Campanas Observatory, Chile.}}

	\correspondingauthor{Federica Ricci}
	\email{federica.ricci21@unibo.it}
	
	\author[0000-0001-5742-5980]{Federica Ricci}
	\affiliation{Instituto de Astrof\'{\i}sica  and Centro de Astroingenier{\'{\i}}a, Facultad de F\'{i}sica, Pontificia Universidad Cat\'{o}lica de Chile, Casilla 306, Santiago 22, Chile}
	\affiliation{Dipartimento di Fisica e Astronomia, Università di Bologna, via Gobetti 93/2, I-40129 Bologna, Italy}
	\affiliation{INAF - Osservatorio di Astrofisica e Scienza dello Spazio di Bologna, via Gobetti 93/3, I-40129 Bologna, Italy}
	
	\author[0000-0001-7568-6412]{Ezequiel Treister}
	\affiliation{Instituto de Astrof\'{\i}sica  and Centro de Astroingenier{\'{\i}}a, Facultad de F\'{i}sica, Pontificia Universidad Cat\'{o}lica de Chile, Casilla 306, Santiago 22, Chile}
	
	\author[0000-0002-8686-8737]{Franz E. Bauer}
	\affiliation{Instituto de Astrof\'{\i}sica  and Centro de Astroingenier{\'{\i}}a, Facultad de F\'{i}sica, Pontificia Universidad Cat\'{o}lica de Chile, Casilla 306, Santiago 22, Chile}
	\affiliation{Millennium Institute of Astrophysics (MAS), Nuncio Monse{\~{n}}or S{\'{o}}tero Sanz 100, Providencia, Santiago, Chile}
	\affiliation{Space Science Institute, 4750 Walnut Street, Suite 205, Boulder, Colorado 80301}
	
	\author[0000-0001-8450-7463]{Julian E. Mej\'ia-Restrepo}
	\affiliation{European Southern Observatory, Alonso de Cordova 3107, Vitacura,Casilla 19001, Santiago de Chile, Chile}
	
	\author[0000-0002-7998-9581]{Michael J. Koss}
	\affiliation{Eureka Scientific, 2452 Delmer Street Suite 100, Oakland, CA 94602-3017, USA}
	
	\author[0000-0002-8760-6157]{Jakob S. den Brok}
	\affiliation{Institute for Particle Physics and Astrophysics, ETH Z{\"u}rich, Wolfgang-Pauli-Strasse 27, CH-8093 Z{\"u}rich, Switzerland}
	\affiliation{Argelander Institute for Astronomy, Auf dem H{\"u}gel 71, 53231, Bonn, Germany}
	
	\author[0000-0003-0476-6647]{Mislav Balokovi\'{c}}
	\affiliation{Center for Astrophysics | Harvard \& Smithsonian, 60 Garden Street, Cambridge, MA 02138, USA}
	\affiliation{Black Hole Initiative at Harvard University, 20 Garden Street, Cambridge, MA 02138, USA}
	\affiliation{Yale Center for Astronomy \& Astrophysics, Physics Department, New Haven, CT 06520, USA}
	
	\author[0000-0001-5481-8607]{Rudolf B\"{a}r}
	\affiliation{Institute for Particle Physics and Astrophysics, ETH Z{\"u}rich, Wolfgang-Pauli-Strasse 27, CH-8093 Z{\"u}rich, Switzerland}

	\author[0000-0002-0205-5940]{Patricia Bessiere}
	\affiliation{Instituto de Astrof\'isica de Canarias, 38205, C/ Vía L\'actea, s/n, La Laguna, Tenerife, Spain}
	
	\author[0000-0002-9144-2255]{Turgay Caglar}
	\affiliation{Leiden Observatory, PO Box 9513, 2300 RA Leiden, The Netherlands}
	
	\author{Fiona Harrison}
	\affiliation{Cahill Center for Astronomy and Astrophysics, California Institute of Technology, Pasadena, CA 91125, USA}
	
	\author[0000-0002-4377-903X]{Kohei Ichikawa}
	\affiliation{Frontier Research Institute for Interdisciplinary Sciences, Tohoku University, Sendai 980-8578, Japan}
	
	\author[0000-0002-2603-2639]{Darshan Kakkad}
	\affiliation{European Southern Observatory, Alonso de Cordova 3107, Vitacura,Casilla 19001, Santiago de Chile, Chile}

	\author[0000-0003-3336-5498]{Isabella Lamperti}
	\affiliation{Department of Physics and Astronomy, University College London, Gower Street, London WC1E 6BT, UK}
	\affiliation{European Southern Observatory, Karl-Schwarzschild-Str. 2, 85748 Garching bei M{\"u}nchen, Germany}
	
	\author[0000-0002-7962-5446]{Richard Mushotzky}
	\affiliation{Department of Astronomy and Joint Space-Science Institute, University of Maryland, College Park, MD 20742, USA}
	
	\author[0000-0002-5037-951X]{Kyuseok Oh}
	\affiliation{Korea Astronomy \& Space Science Institute, 776, Daedeokdae-ro, Yuseong-gu, Daejeon 34055, Republic of Korea}
	\affiliation{Department of Astronomy, Kyoto University, Kitashirakawa-Oiwake-cho, Sakyo-ku, Kyoto 606-8502, Japan}
	\affiliation{JSPS Fellow}
	
	\author[0000-0003-2284-8603]{Meredith C. Powell}
	\affiliation{Institute of Particle Astrophysics and Cosmology, Stanford University, 452 Lomita Mall, Stanford, CA 94305, USA}
	
	\author[0000-0003-3474-1125]{George C. Privon}
	\affiliation{Department of Astronomy, University of Florida, 211 Bryant Space Science Center, Gainesville, FL 32611, USA}
	
	\author[0000-0001-5231-2645]{Claudio Ricci}
	\affiliation{N\'ucleo de Astronom\'ia de la Facultad de Ingenier\'ia, Universidad Diego Portales, Av. Ej\'ercito Libertador 441, Santiago, Chile}
	\affiliation{Kavli Institute for Astronomy and Astrophysics, Peking University, Beijing 100871, People's Republic of China}
	\affiliation{George Mason University, Department of Physics \& Astronomy, MS 3F3, 4400 University Drive, Fairfax, VA 22030, USA}
	
	\author[0000-0002-1321-1320]{Rogerio Riffel}
	\affiliation{Departamento de Astronomia, Universidade Federal do Rio Grande do Sul Porto Alegre, Brazil}
	
	\author[0000-0003-0006-8681]{Alejandra F. Rojas}
	\affiliation{Centro de Astronomía (CITEVA), Universidad de Antofagasta, Avenida Angamos 601, Antofagasta, Chile}
	
	\author[0000-0002-3140-4070]{Eleonora Sani}
	\affiliation{European Southern Observatory, Alonso de Cordova 3107, Vitacura,Casilla 19001, Santiago de Chile, Chile}
	
	\author[0000-0001-5785-7038]{Krista L. Smith}
	\affiliation{KIPAC at SLAC, Stanford University}
	
	\author[0000-0003-2686-9241]{Daniel Stern}
	\affiliation{Jet Propulsion Laboratory, California Institute of Technology, 4800 Oak Grove Drive, MS 169-224, Pasadena, CA 91109, USA}

	\author[0000-0002-3683-7297]{Benny Trakhtenbrot}
	\affiliation{School of Physics and Astronomy, Tel Aviv University, Tel Aviv 69978, Israel}
	
	\author[0000-0002-0745-9792]{C. Megan Urry}
	\affiliation{Yale Center for Astronomy \& Astrophysics, Physics Department, New Haven, CT 06520, USA}
	
	\author[0000-0002-3158-6820]{Sylvain Veilleux}
	\affiliation{Department of Astronomy and Joint Space-Science Institute, University of Maryland, College Park, MD 20742, USA}

	\begin{abstract}
		Virial black hole mass ($M_{BH}$) determination directly involves knowing the broad line region (BLR) clouds velocity distribution, their distance 
		from the central supermassive black hole ($R_{BLR}$) and the virial factor (\f). 	
		Understanding 
		whether biases arise in $M_{BH}$ estimation with increasing obscuration is possible
		only by studying a large (N$>$100) statistical sample of obscuration unbiased (hard) X-ray selected active galactic nuclei (AGN)
		in the rest-frame near-infrared (0.8--2.5~$\mu$m) since it penetrates deeper
		into the BLR than the optical.
		We present a detailed analysis of 65 local BAT-selected Seyfert galaxies observed with Magellan/FIRE.
		Adding these to the near-infrared BAT AGN spectroscopic survey (BASS) database, we study a total of 314
		unique near-infrared spectra. While the FWHMs of H$\alpha$ and
		near-infrared broad lines (He~\textsc{i}, Pa$\beta$, Pa$\alpha$) remain unbiased to either BLR extinction or
		X-ray obscuration, the H$\alpha$ broad line luminosity 
		is suppressed when $N_H\gtrsim10^{21}$~cm$^{-2}$, systematically underestimating  
		$M_{BH}$ by $0.23-0.46$~dex.
		Near-infrared line luminosities should
		be preferred to H$\alpha$
		until $N_H<10^{22}$~cm$^{-2}$, while at higher obscuration
		a less biased $R_{BLR}$ proxy should be adopted. 
		We estimate 
		\f\ for Seyfert 1 and 2 using two obscuration-unbiased $M_{BH}$ measurements,
		i.e. the stellar velocity dispersion 
		and
		a BH mass prescription based on near-infrared and X-ray,
		and find that the virial factors do not depend on redshift or obscuration, but for some broad lines show a mild anti-correlation with $M_{BH}$. Our results show the critical impact obscuration can have on BLR characterization and the importance of the near-infrared and X-rays for a less biased view of the BLR.
		%250 word max
		
	\end{abstract}
	
	
%	\linenumbers
	\section{Introduction} \label{sec:intro}
	Supermassive black holes (SMBHs, with black hole masses $M_{BH}\sim10^5 - 10^{9}$~M$_\odot$)
	are ubiquitous in the local Universe, lurking in the spheroid of almost all local 
	galaxies \citep{kormendyrichstone95}. During active accretion phases, the SMBH is no longer dormant but shines as
	an active galactic nucleus (AGN), due to a surrounding
	accretion disk of matter, which releases gravitational energy as it infalls 
	toward the central dark attractor. The ultraviolet emission from the inner accretion disk
	photoionizes nearby clouds located in the broad line region (BLR).
	Under the hypothesis of a virialized BLR, whose dynamics are 
	dominated by the central SMBH, the $M_{BH}$ can 
	be simply determined from the velocity $\Delta V_{BLR}$ of the emitting gas clouds 
	located at distance $R_{BLR}$ in the BLR as $M_{BH}=G^{-1} \Delta V_{BLR}^2 R_{BLR}$, with $G$ being the gravitational constant.
	To model the unknown emission-weighted{\footnote{Note that the geometry and dynamics of the BLR in so-called single-epoch mass estimates is emission-weighted, but in full reverberation mapping studies it is more properly responsivity-weighted.}
	geometry and dynamics of the BLR, 
	the observed width $\Delta W_{obs}$ (either the full-width-at-half maximum FWHM or the second moment of the line profile, i.e., the line
	dispersion $\sigma_{line}$) of a doppler-broadened photoionized element at distance $R_{BLR}$ from the SMBH is used as a tracer of the true velocity in the BLR and a correction
	factor \f\ known as the virial factor is introduced \citep{onken04}:
	\begin{equation}
		M_{BH} =  f \,  G^{-1} \Delta W_{obs}^2 R_{BLR}= f M_{vir} \, ,
		\label{eq:Mvir}
	\end{equation}
	where $M_{vir}$ is the so-called virial product or virial mass. 
	Since it is in practice impossible to spatially resolve the BLR for statistically sized samples, 
	time-resolved observations substitute for spatially-resolved information
	to estimate the BLR radius, adopting the so-called reverberation-mapping technique \citep[RM; ][]{blandformcknee82}. 
	This is gradually changing due to campaigns with the VLTI/GRAVITY instrument, which has opened the path to spatially-resolved observations of the BLR around nearby AGN, allowing high angular resolution ($<0.1$ milliarcsec) spectral-spatial interferometric observations of the Pa$\alpha$ \citep[in 3C 273, see,][]{gravity18} and Br$\gamma$ \citep[in IRAS 09149-6206, see,][]{gravity20} line. 
	However, this powerful technique remains limited to a small sample of AGN. 
	
	Extensive RM campaigns have
	found that the radius of the BLR is linked to the AGN luminosity, $R_{BLR}\propto L_{AGN}^\alpha$ \citep{bentz06,bentz09,bentz13}, where the slope $\alpha$ is consistent with expectations from photoionization ($\alpha\simeq0.5$).
	The AGN continuum and the broad-line luminosity \citep[see, e.g., ][]{shen13} have been both used as a proxy for the BLR radius, which has allowed an efficient calibration for single-epoch (SE) BH mass estimation.
	
	The virial factor \f\ has been directly inferred only for a limited subsample ($<$20)
	of RM AGN with sufficient high quality data available 
	\citep{pancoast14,pancoast18,grier17,williams18,li18}.
	Hence, previous studies have often adopted an ensemble virial factor
	\faver\, that is determined using the $M_{BH}-\sigma_\star$ relation observed in local samples of quiescent galaxies with dynamically-based BH massses \citep{grier13,hk14,batiste17,yu19}
	\begin{equation}
		M_{BH,\sigma_\star} =  \langle f \rangle M_{vir} \, .
		\label{eq:msigma}
	\end{equation}
	
	However it is still unclear whether the $M_{BH}-\sigma_\star$ relation is universally followed by all types of galaxies:  
	e.g., barred/unbarred hosts \citep{graham08}, early/late type hosts \citep{mcconnellma13, sahu19}, elliptical and classical-/pseudo- bulges \citep{kormendyho13, saglia16, denicola19}. 
	It is also unsettled whether AGN should follow the scaling relations determined by quiescent galaxies 
	\citep{woo13,ricci17b,shankar19}, since the methods used to measure the BH masses in active galaxies are not expected to suffer from the resolution-dependent 
	bias that instead affect dynamical-based BH mass estimates \citep{bernardi07,shankar16}.
	Moreover, the \f-factor could change on an object-by-object basis if it depends on some AGN properties, such as the bolometric luminosity, Eddington ratio ($\lambda_{Edd}$), $M_{BH}$,  obscuration or line-of-sight inclination angle $\theta$. 
	The only statistically sound correlation found so far is between \f\ and $\theta$, although this is based on a limited number of RM objects with directly inferred \f. This correlation is consistent with expectations of the BLR being a thick disk with clouds moving in a combination of elliptical and inflowing motions \citep{williams18}. 
	This $f-\theta$ relation is corroborated by statistical studies that
	used a variety of \f-independent BH mass measurements to infer the virial factor, such as 
	the bulge luminosity based BH mass \citep{decarli08},
	the stellar velocity dispersion $\sigma_\star$ based BH mass \citep{shenho14}
	and the accretion disk based BH mass \citep{mejia18}. 
	These studies only focused on optical broad-line AGN, for which optical virial-based $M_{BH}$ estimates were available.
	
	While the use of the H$\alpha$ emission is common practice to derive $M_{BH}$ in statistical samples to study the demography and evolution of the AGN population, it is 
	unclear whether the H$\alpha$ is completely reliable, particularly in so-called Sy 1.9 \citep{osterbrock81}, where the level of extinction due to dust is more relevant than what is usually experienced in optical broad line Seyferts. Indeed \citet{reines15}, 
	using a sample of 262 broad-line AGN in the nearby Universe ($z <$ 0.055) with H$\alpha$ broad-lines measured from the Sloan Digital Sky Survey (SDSS) DR8, find that AGN-hosts with SE H$\alpha$-based BH masses define a separate $M_{BH}-M_\star$ relation, with a slope similar to that of early-type galaxies with dynamically detected BHs but with a normalization $\sim1.2$~dex lower. 
	Similarly, \citet{koss17} show that the SE H$\alpha$-based BH masses in Sy 1.9 are 
	undermassive than what is expected from the $M_{BH}-\sigma_\star$ relation of elliptical/classical-bulges, with the BH mass deviation being more extreme in sources with broad H$\alpha$ equivalent width $EW<50$~\AA. 
	Additionally, \citet{caglar20} find an offset of $\sim0.6$~dex between the SE H$\alpha$-based BH masses and those based on the stellar velocity dispersion in a sample of 19 partially obscured local hard X-ray selected Seyferts from the LLAMA sample. This discrepancy is reduced in the LLAMA sample only after accounting for optical extinction, in the H$\alpha$ measurement, and galaxy rotation, in the $\sigma_\star$ estimate \citep{caglar20}.
	These results can be either explained with the fact that BH-host scaling relations should be different in active vs inactive galaxies, or that in some cases the H$\alpha$-based BH masses could be biased low in presence of extinction (or with a combination of these two effects).
	
	To this end, the rest-frame near-infrared (NIR, 0.8--2.5$\mu$m) band allows to deeper probe the physical condition of the BLR gas, being at least a factor of 10 less affected by dust extinction than the rest-frame optical emission \citep{goodrich94, veilleux97, veilleux02}. 
	However, NIR ground-based spectroscopic observations are more complex and time-consuming than optical spectroscopy, due to the lower atmospheric transmission that reduces the observable windows, bright sky background and strong OH sky line emissions. For these reasons, local AGN samples studied so far have remained limited to few objects, usually $\lesssim50$ \citep{riffel06,glikman06,landt08,mason15,riffel15,onori17a}.

	Determining the BLR properties and \f-factors for a less biased AGN sample 
	is of paramount importance to assess the uncertainties and systematics in $M_{BH}$ measurement in active galaxies as a function of obscuration. This is particularly important for Sy 1.9s, where the only optical broad line available is the H$\alpha$, and Sy 2s, where sometimes so-called hidden BLR are found in the NIR \citep[$\sim$30\% of cases, see, e.g.,  ][]{veilleux97,riffel06,cai10,smith14,lamperti17,onori17a}. 
	Such an investigation is only possible by constructing an obscuration unbiased AGN sample, by means of hard X-ray ($>10$~keV) AGN selection that is almost 
	unaffected by interviening obscuring material, at least up to $N_H\sim10^{23.5} - 10^{24}$~cm$^{-2}$ \citep{cricci15,koss16}, due to the decline of the photoelectric cross-section with increasing photon energy. 
	A sensitive all-sky survey in the ultra-hard X-ray band (14–195~keV), such as the one carried out by the 
	the Burst Alert Telescope \citep[BAT,][]{barthelmy00} on the \textit{Neil Gehrels Swift Observatory} (\textit{Swift}/BAT), coupled with NIR spectroscopy and ancillary optical spectroscopic information, is the ideal 
	database to quantify the effects of obscuration on BLR characterisation and thus BH mass estimation.  The BAT AGN Spectroscopic Survey (BASS)\footnote{https://www.bass-survey.com/} has for the first time increased the sample size of AGN surveyed with NIR spectroscopy to more than 100 beginning with the DR1 \citep{lamperti17}.
	\begin{figure}[ht!]
		\hspace{-0.5in}\includegraphics[scale=0.52]{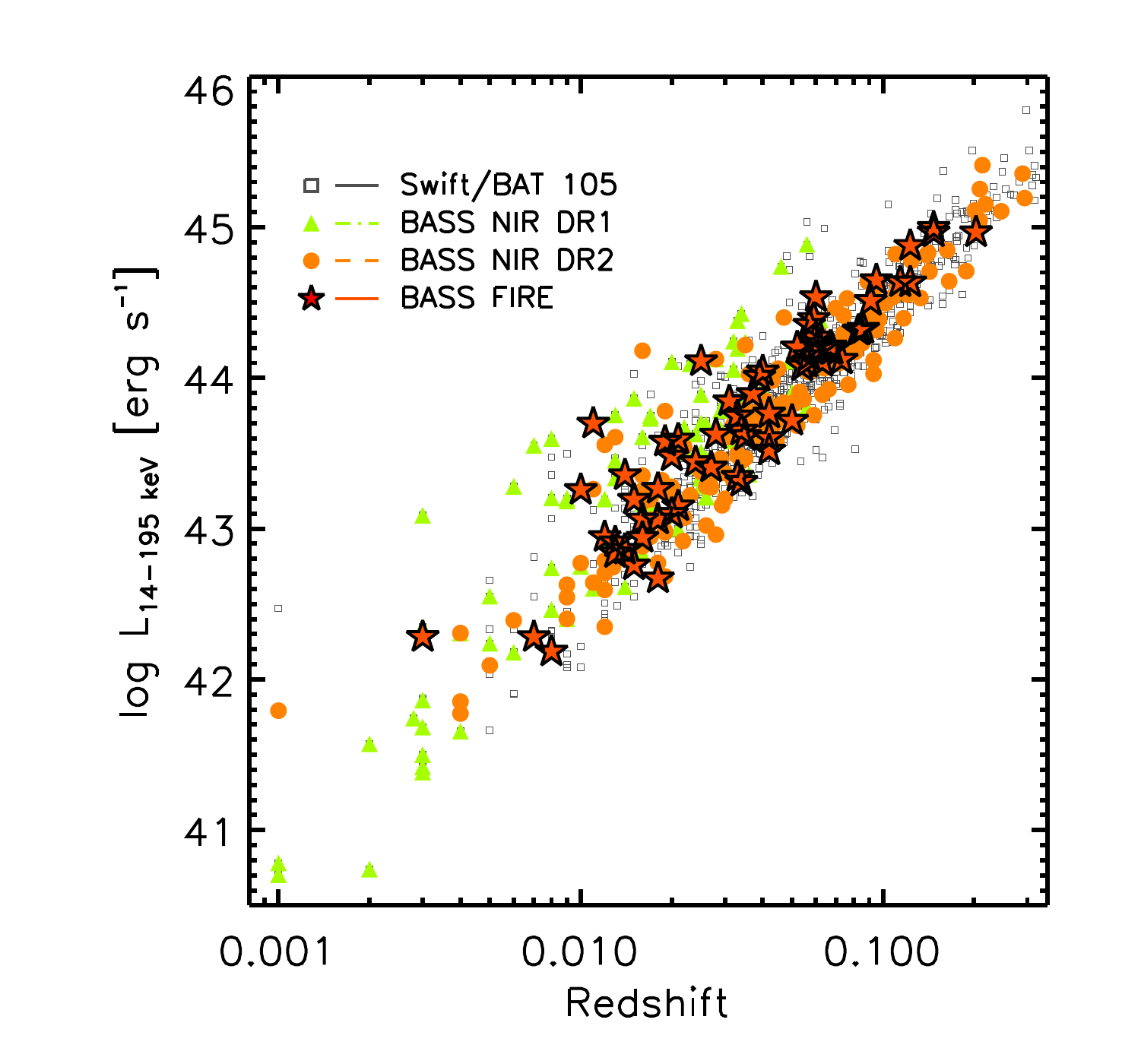}
		
		\vspace{-3.56in}\hspace{1.97in}{\includegraphics[scale=0.54, angle=-90]{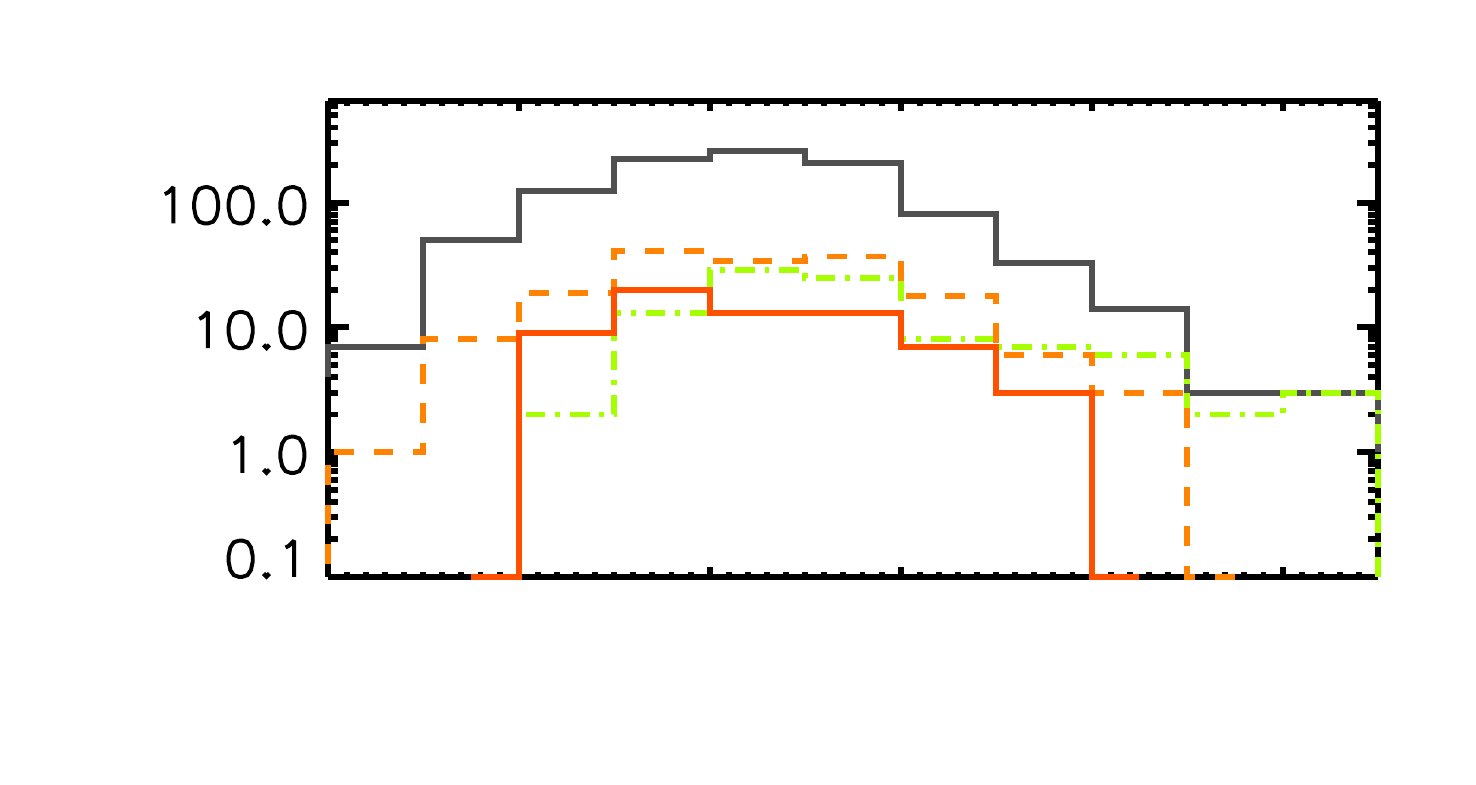}} 
		\vspace{0.1in}
		\caption{\textit{Left:} The $L_{14-195\,\rm{keV} } - z$ distribution of the FIRE sample, BASS NIR DR1 \citep{lamperti17}, BASS NIR DR2 \citep{denBrok_DR2_NIR} and BAT 105-month sample \citep{oh18}. Colors and symbols according to the legend. 
			\textit{Right:} histogram of the hard X-ray luminosities.}
		\label{fig:lxz}
	\end{figure}
	
	In this work, we present NIR Magellan spectra of 65 local Seyferts, selected from the 70-month \textit{Swift}/BAT catalog \citep{baumgartner13}, as part of the BASS survey. We complement our near-infared Magellan sample with the NIR BASS database, NIR DR1 \citep{lamperti17}, DR2 \citep{denBrok_DR2_NIR}\footnote{We note that the NIR BASS DR2 \citep{denBrok_DR2_NIR} does not contain the spectral measurements already published in the NIR BASS DR1 \citep{lamperti17} since the data analysis is very consistent between the two different BASS data releases.}, to construct the largest sample of local Seyferts with NIR and X-ray spectral information available to date, for a total of 314 unique NIR spectra. 
	We further complement the NIR analysis with optical spectral information on the broad component of the H$\alpha$ from the optical BASS DR2 \citep{Mejia_Broadlines}, to quantify and compare the BLR characterisation in both optical and NIR. We finally make use of the optical stellar velocity dispersion measurements available in BASS DR2 \citep{Koss_DR2_sigs,Caglar_DR2_Msigma}, to 
	infer the individual \f\ factors of our sample.
	
	The work is organized as follows: Sect. \ref{sec:data} presents the Magellan data selection and reduction; Sect. \ref{sec:specfit} describes the Magellan spectral fitting of the most important NIR emission lines; Sect. \ref{sec:bhm} is devoted to the descriptions of two independent BH mass measurement methods adopted to derive the virial factors \f. Results are described in Sect. \ref{sec:res}, where we investigate the fraction of hidden BLRs detected in Sy 1.8-1.9-2 (Sect. \ref{ss:bldet}), the effects of X-ray obscuration and BLR extinction on the BLR velocity and radius tracers derived from optical (i.e.,  H$\alpha$) and NIR emission lines (He~\textsc{i}$\lambda$10830~\AA, Pa$\beta$~$\lambda$12821~\AA, Pa$\alpha$~$\lambda$18756~\AA) in Sy 1-1.9 (Sect. \ref{ss:ha$NIR$}) and in Sy 1-2 (Sect. \ref{sec:$NIR$nh}). 
	In Sect. \ref{ss:av} we explore the connection between the material repsonsible for BLR extinction and 
	the one absorbing the X-rays.	
	The virial factors of our sample are derived in Sect. \ref{ss:fvir}, 
	where we test whether \f\ depends on some parameters, e.g., $z$, $N_H$ and BH mass.
	Section \ref{sec:disc} is devoted to discussions and conclusions, while Sect. \ref{sec:sum} briefly summarises our main results. We adopt the concordance cosmological model, $\Omega_M=0.3$, $\Omega_\Lambda=0.7$ and $h=0.7$.

	\section{Data}\label{sec:data} 
	Here we present the NIR spectroscopic data (PI: E. Treister, F. Ricci, M. Balokovi\'{c})\footnote{Proposals number: CN2018A-70, CN2018B-85, CN2019A-10} obtained at Magellan using the Folded-port InfraRed Echellette \citep[FIRE;][]{simcoe08}.
	The Magellan/FIRE sample was selected from the 
	hard X-ray (14-195~keV)
	70-month catalog \citep{baumgartner13} without near-infared coverage in the BASS DR1 \citep{lamperti17},
	as part of an effort within the BASS collaboration to obtain NIR coverage for
	an obscuration unbiased census of local accreting SMBHs. 
	The 70-month catalog \citep{baumgartner13} listed 838 AGN, 102/838 were 
	targeted in the BASS DR1 \citep{lamperti17} and 118/838 were observed as part of the DR2 \citep{denBrok_DR2_NIR} which selected as well additional 50/1016 AGN from the latest 105-month source catalog \citep{oh18}. 
	The combined sample totals 314
	hard X-ray selected AGN with unique NIR spectra.  
	The FIRE sample was chosen to target
	the more obscured sources, to estimate the $M_{BH}$ also in obscured Seyfert class AGN (i.e., Sy 1.8-1.9-2).
	The FIRE sample is composed of 65 targets selected at $z\lesssim0.2$, divided into 52 
	obscured AGN (i.e., Sy 1.8-1.9-2) and 13 optical broad-line AGN (i.e., Sy 1-1.2-1.5), whose Seyfert classification is defined according to the \citet{osterbrock81} standard criteria, using
	optical spectra collected by BASS 
	\citep[e.g., BASS optical DR2,][]{Mejia_Broadlines}. 
	Figure \ref{fig:lxz} shows the $L_{14-195\,\rm{keV} } - z$ distribution of the FIRE targets (red stars), BASS NIR DR1 \citep[green triangles,][]{lamperti17}, BASS NIR DR2 
	\citep[orange filled circles,][]{denBrok_DR2_NIR}
	and the latest 105-month \textit{Swift}/BAT catalog \citep[dark gray open squares,][]{oh18}.
	\startlongtable
	\begin{deluxetable*}{rllcccllccc}  %%%%%%%%%%%%%%%%%%%%%%%%%%%%%%%%%%%%%%%%%%%%%%%%%%%%%%%%%%%%%%%%%%%%%%%
		\tablecaption{ Magellan/FIRE observation log \label{tab:obs}}
		\tablehead{
			\colhead{BAT ID} & 
			\colhead{Counterpart Name} & 
			\colhead{Class} & 
			\colhead{Obs. date} & 
			\colhead{Exposure} & 
			\colhead{Airmass} & 
			\colhead{J} & 
			\colhead{$z$} &
			\multicolumn{2}{c}{slit/aperture} &
			\colhead{$\log L_{2-10\,\rm{intr}}$} \\
			\colhead{} &
			\colhead{} &
			\colhead{} &
			\colhead{dd.mm.yy} & 
			\colhead{[s]} &  
			\colhead{}&  
			\colhead{[mag]}& 
			\colhead{}&
			\colhead{[$\arcsec$]}&
			\colhead{[kpc]}&
			\colhead{[erg~s$^{-1}$]}\\
			\colhead{(1)}&\colhead{(2)}&\colhead{(3)}&\colhead{(4)}&\colhead{(5)}&\colhead{(6)}&\colhead{(7)}&\colhead{(8)}&\colhead{(9)}&\colhead{(10)}&\colhead{(11)}
		}
		\startdata 
		7	 &	SDSSJ000911.57-003654.7	 & Sy2		&30.09.18& $4\times306$ & 1.14	& 14.74	& 0.073  &0.6/0.74  &  0.93/1.15&43.60	\\
		10	 &	LEDA1348				 & Sy1.9	&30.09.18& $4\times370$ & 1.02	& 14.67	& 0.095  &0.6/0.68  &  1.28/1.44&44.40	\\
		80	 &	2MASXJ01290761-6038423	 & Sy2		&30.09.18& $4\times370$ & 1.17	& 15.17	& 0.203  &0.6/0.82  &  4.23/5.80&45.23	\\
		118	 &	3C62					 & Sy2		&30.09.18& $4\times370$ & 1.04	& 15.43	& 0.147  &0.6/0.69  &  2.39/2.74&44.50	\\
		238	 &	LEDA745026				 & Sy2		&05.04.18& $8\times190$ & 1.33	& 15.44	& 0.147  &0.6/0.34  &  2.41/1.36&44.42	\\
		262	 &	ESO553-22				 & Sy2		&30.09.18& $4\times190$ & 1.13	& 13.89	& 0.042  &0.6/0.76  &  0.52/0.66&43.38	\\
		272	 &	IRAS05189-2524			 & Sy2		&30.09.18& $4\times190$ & 1.07	& 13.11	& 0.042  &0.6/0.54  &  0.52/0.46&43.40	\\
		305	 &	LEDA17883				 & Sy2		&30.09.18& $4\times190$ & 1.28	& 14.63	& 0.050	 &0.6/0.86  &  0.62/0.89&43.51	\\
		329	 &	ESO121-28				 & Sy1.9	&05.04.18& $8\times190$ & 1.29	& 13.89	& 0.040  &0.6/1.00  &  0.50/0.83&43.63	\\
		372	 &	1RXSJ072720.8-240629	 & Sy1.9	&30.09.18& $4\times370$ & 1.33	& 15.06	& 0.123	 &0.6/0.52  &  1.83/1.59&44.34	\\
		\enddata 
		\tablenotetext{}{Columns are: (1) 70-month \textit{Swift}/BAT ID (https://swift.gsfc.nasa.gov/results/bs70mon/); (2) associated counterpart name; (3) optical Seyfert classification, as defined by \citet{osterbrock81}; (4) observation date; (5) exposure time; (6) airmass at the midpoint of the observation; (7) J-band Vega mag \citep[from 2MASS extended, e.g., ][]{jarrett00}; (8) redshift from the [O~\textsc{iii}] BASS DR2; 
			(9-10) slit width and aperture of the spectral extraction, in arcsec and kpc;	
			(11) intrinsic 2-10~keV luminosity from \citet{cricci17}. (This table is available in its entirety in a machine-readable form in the online journal. A part is shown as guidance for the reader regarding its content.)}
		\tablenotetext{*}{Readout mode used was Fowler 4 rather than SUTR.} 
	\end{deluxetable*} 
	The FIRE data is complementary to the BASS NIR DR1 and DR2, 
	spanning $\sim$3~dex of overlap in X-ray luminosity (see, right panel in Fig. \ref{fig:lxz}), and aside from some lower $L_X$ and lower-$z$ object of the DR1 and some higher $L_X$ and higher redshift source in the DR2, all the three BASS samples should 	
	be studying AGN with similar properties. The combined NIR dataset is representative of the parent BAT AGN sample.

	\begin{figure*}[ht]
		\vspace{-0.1in}
		\includegraphics[width=\textwidth]{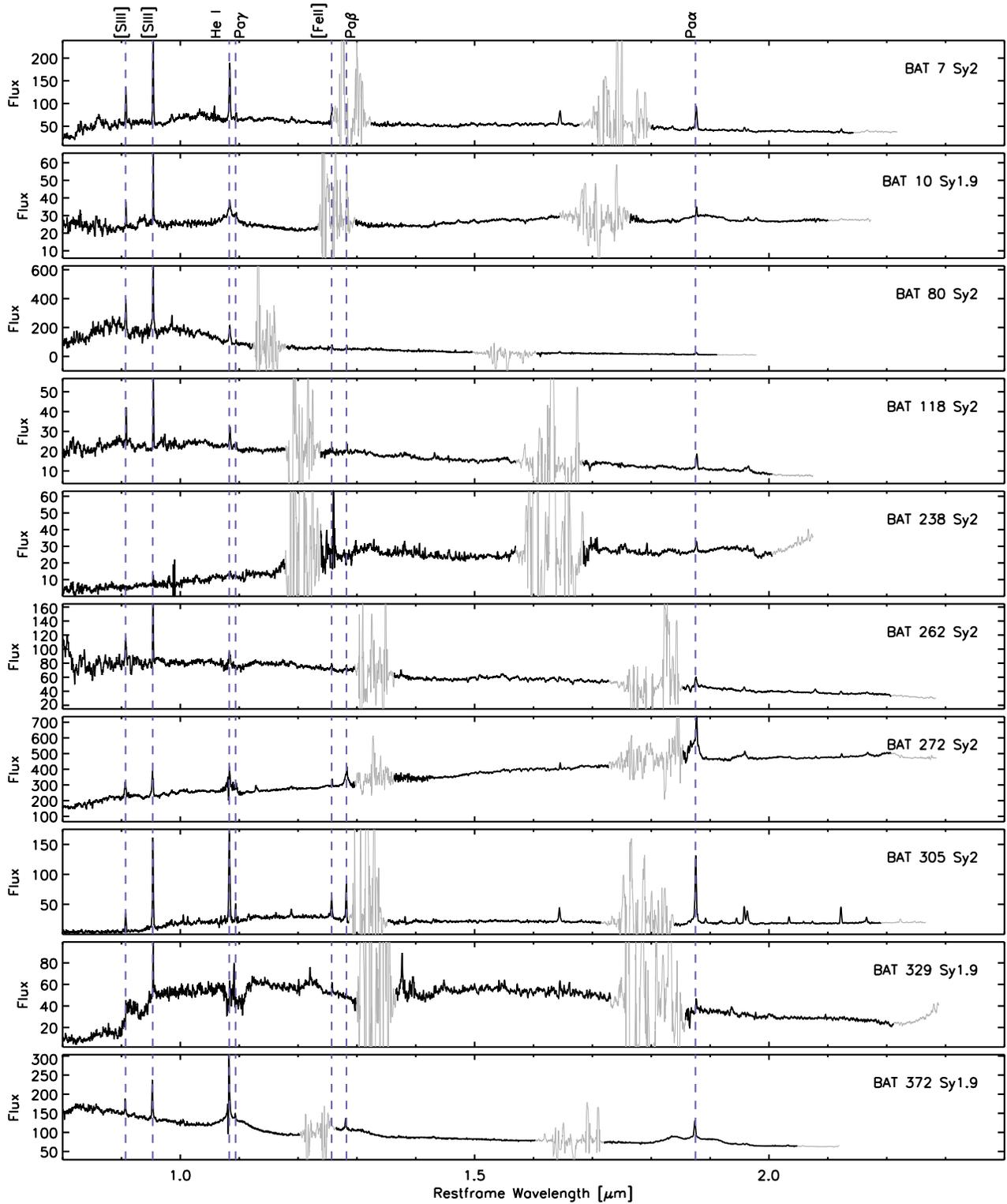}
		\vspace{-1.in}
		\caption{Examples of FIRE/Magellan NIR spectra in units of $\times 10^{-17}$~erg/s/cm$^2$/\AA, redshift-corrected and flux calibrated. 
			The wavelength position of some of the most relevant emission lines are indicated with dashed vertical lines and labeled at the top. Regions of low telluric transmission are shown in gray. The complete figure set for the 65 AGN is available in 7 images in the online journal, ordered by BAT ID.}
		\label{fig:spec}
	\end{figure*}
	The 65 NIR 0.8--2.5 $\mu$m spectra were observed using the FIRE
	instrument in the high-resolution echelle mode in four visiting runs carried out between April 2018 and April 2019.
	FIRE is a dual-mode IR spectrometer mounted at the Magellan Baade telescope at Las Campanas Observatory (LCO), Chile. Its primary mode employs a combination of a diffraction grating and four prisms to deliver cross-dispersed spectra covering the whole NIR bandpass in a single exposure, with nominal wavelength resolution of
	$R=\lambda/\Delta \lambda \approx 6000$ for a 0$\farcs$6-slit width, i.e., $\Delta v \simeq 50$~\kms.
	This slit width was adopted for our program, with the exception of two targets, namely BAT 677 and 1085, which were observed with a 0$\farcs$45-slit width (i.e., $R\approx8000$, $\Delta v \simeq 37$~\kms), since those were expected to have low $M_{BH}$.
	Our observations took place the nights UT 2018 April 5, UT 2018 September 30, UT 2019 March 9 and on UT 2019 April 14-15 during gray time. The observations were executed under clear skies with different airmass conditions (see Tab. \ref{tab:obs}) and visual seeing variying between 1$\farcs$5 to 0$\farcs$4, with an average of 0$\farcs$75.
	
	For each target, the individual spectra were obtained using the nodding technique in a sequence of ABBA acquisitions, with exposure times ranging between 190 to 623~s (multiples of 10.6~s, since sample-up-the-ramp SUTR readout mode was used for all targets but the ones marked with an asterisk in Tab. \ref{tab:obs}, which were observed using Fowler 4 readout), depending on target magnitude and observing conditions. The acquisition sequence involved a short arc frame (ThAr) just after the target observation in order to correct for telescope flexure and obtain the wavelength solution. For long ($>$300~s) exposures, OH airglow was used to improve the wavelength calibration. 
	By default, the wavelengths are calibrated in vacuum. 
	Sky and dome (Qz) flats were acquired to correct for detector illumination and pixel gain variations across the slit, respectively.
	Data was reduced with the Interactive Data Language (IDL) pipeline \texttt{FireHose} v2 package \citep{gagne15}, which performs
	2D sky subtraction and extracts an optimally weighted 1D spectrum. 
	Nearby A0V stars were observed during the night in order to derive relative flux calibrations.
	We corrected the atmospheric absorption features (H$_2$O, CO$_2$, CH$_4$ and O$_2$) using the software tool \texttt{molecfit} \citep{smette15}. 
	\texttt{Molecfit} uses a radiative transfer code to simulate the atmospheric transmission taking into account local weather parameters (temperature, pressure, humidity, etc.), recorded at the LCO/Magellan site. 
	
	Flux-calibrated and redshift-corrected NIR spectra of the 65 AGN are shown in Fig. \ref{fig:spec} (as a figure set)  
	smoothed using a Savitzky-Golay filter, which preserves the average resolving power. 
	The spectra are ordered by their BAT ID in Fig. \ref{fig:spec}. 
	Regions of low telluric transmission are plotted in gray. The locations of some of the most intense NIR emission lines are labeled and indicated with dashed purple lines in Fig. \ref{fig:spec}. 
	The reduced spectra will be available on the BASS survey website\footnote{https://www.bass-survey.com/}.

	\begin{figure*}
		\vspace{-5.88in}
		\hspace{-2.5in}\includegraphics[width=1.7\textwidth]{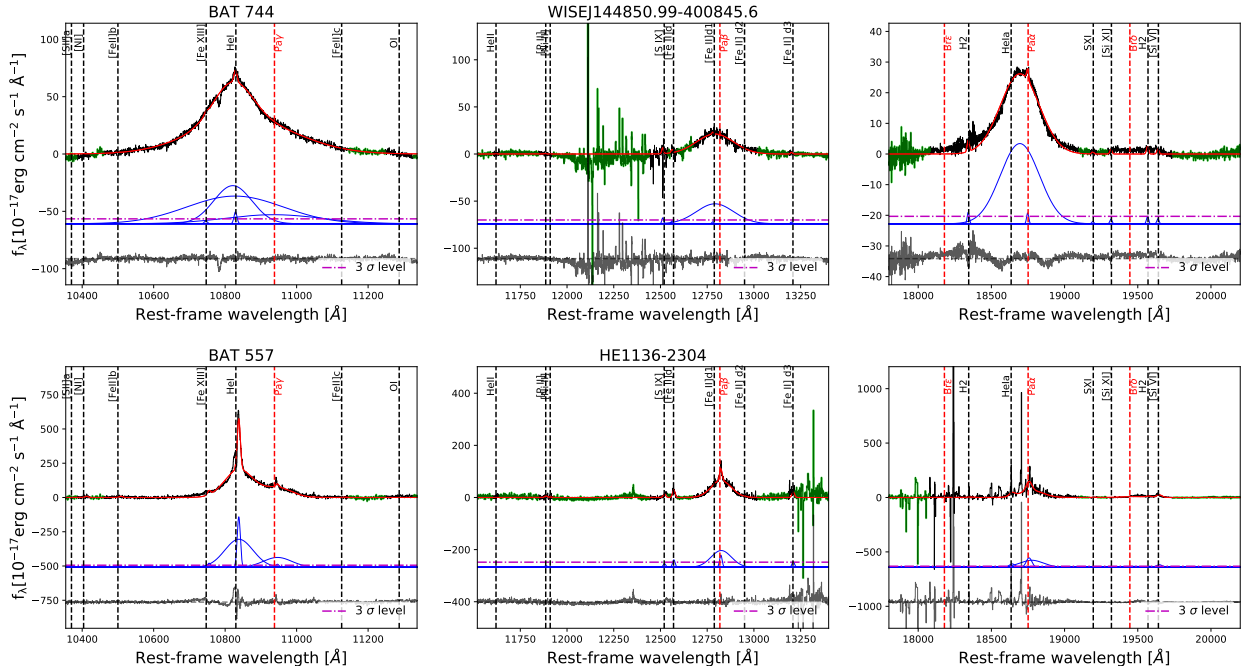}
		
		\vspace{-6.3in}\caption{Examples of spectral emission-line fits of the Pa$\gamma$ (\textit{left}), Pa$\beta$ (\textit{center}) and Pa$\alpha$ (\textit{right}) regions for a Sy 1 (BAT 744, \textit{top}) and a Sy 1.9 (BAT 557, \textit{bottom}). The best-fit Gaussians are plotted in blue. As explained in Sect. \ref{subsec:fit}, a broad Gaussian component is discarded if its intensity is below $3\sigma$ (dashed purple line) with respect to the fitting continuum. The continuum 
			has been evaluated in the regions highlighted in green and subtracted off. In red the resulting total model is shown.
			Residuals are plotted offsetted in dark gray.
			The fit quality flags are: 2, 2 and 1 in the Pa$\gamma$, Pa$\beta$ and Pa$\alpha$ regions of BAT 744, respectively; and  2, 1 and 4 in the Pa$\gamma$, Pa$\beta$ and Pa$\alpha$ regions of BAT 557, respectively.
			The complete figure set with all the 65 FIRE/Magellan spectral emission line fits in the Pa$\gamma$, Pa$\beta$ and Pa$\alpha$ regions is available in the online journal.}
		\label{fig:lines}
	\end{figure*}
	
	\section{Spectral Measurements}\label{sec:specfit}
	Below we describe the NIR emission line fitting analysis of the FIRE spectra.  
	
	\subsection{Near-infrared Spectral fitting procedure}\label{subsec:fit}
	Our emission-line fitting approach is similar to the one adopted for the BASS NIR DR1 \citep[e.g., ][]{lamperti17} and DR2 \citep{denBrok_DR2_NIR}.
	We use \texttt{PySpecKit} (v0.1.21), an extensive spectroscopic analysis toolkit for astronomy, which uses a 
	Levenberg-Marquardt algorithm \citep{ginsburg11}. 
	We fit the 
	Pa$\zeta$ (0.9$-$0.96~$\mu$m), 
	Pa$\gamma$ (1.04$-$1.15~$\mu$m), 
	Pa$\beta$ (1.15$-$1.30~$\mu$m),  
	and Pa$\alpha$ (1.80$-$2.00~$\mu$m) 
	spectral regions separately, to ease the fitting convergence. 
	Each so-defined independent spectral region contains at maximum three lines coming from permitted species in the BLR.
	There are only few differences with the emission lines considered 
	in \citet[][see their Tab. A1]{lamperti17}, which are the following: we fit also
	the [N~\textsc{i}]$\lambda$10404~\AA\ in the Pa$\gamma$ region and the Br$\epsilon$$\lambda$18179.1~\AA, 
	H~\textsc{2}$\lambda$18345~\AA, He~\textsc{i}$\lambda$18635~\AA, and [S~\textsc{xi}]$\lambda$19196~\AA\ in the Pa$\alpha$ region, while we did not include in the fit [Fe~\textsc{ii}]$\lambda$9227~\AA\ in the Pa$\zeta$ spectral region, since it is a faint iron emission and was not detected in our FIRE observations.
	
	We first deredden the spectra using the Galactic extinction value 
	$E(B-V)$ \citep{schlafly11} as listed in the IRSA Dust Extinction Service\footnote{https://irsa.ipac.caltech.edu/applications/DUST/}, and redshift correct the spectra.
	We employ a single first-order power-law fit to model and remove the continuum.
	For each spectral region, we estimate the continuum level 
	using sections of the wavelength range free of emission lines, i.e., excluding 20 \AA\ around the narrow lines and 150 \AA\ where a broad component was expected, e.g., in permitted species.
	
	As the focus of our investigation is to study the BLR properties, the main goal is to derive the FWHM and flux of broad line species.
	Emission lines are modeled using a combination of Gaussian profiles, namely one component for all line species associated with the NLR and an additional Gaussian to account for the BLR component in all permitted transitions of each spectral region. 
	The relative central wavelength of the narrow lines are tied together, but are allowed to shift together by a maximum of 500 \kms\ with respect to the systemic redshift listed in Tab. \ref{tab:obs}. 
	As an initial input value for the width of the narrow lines, we used the best-fit width of the [S~\textsc{iii}]$\lambda$9531 \AA\ line, which is the strongest \citep[and the narrowest, see, e.g., Fig. 3 in][]{riffel13} narrow emission line in the rest-frame NIR wavelength range, with only a minor blending with the Pa$\epsilon$ on its red side. 
	The width of the narrow components in each spectral region are then tied together.
	
	The threshold between narrow and broad components is set at FWHM~$=1200$~\kms, as consistent with the division defined in BASS NIR DR1 \citep{lamperti17}.
	The central wavelengths of the broad components are allowed to shift up to $\approx$1000 \kms\ \citep{shen16}. 
	Since the widths of the broad components in each spectral region are likely to be similar, we 
	tie them together to avoid unnecessary complexity and
	degeneracy\footnote{We checked that the best-fit NIR broad lines in separate regions, that have been fitted independently, thus without imposing any common constraint on their widths, are consistent within the uncertainties, as also previously reported in the literature 
			\citep[see][]{landt08,ricci17a,lamperti17,onori17a}.}. As each spectral region is fitted separately, the resulting best-fitting BLR component can be different in each region. The same applies to the NLR estimate, which might vary in different spectral regions.
	The fit is thus run a first time. Lines are detected if their amplitude is $>3\sigma$, where $\sigma$ is the root-mean-square of a line-free zone in each spectral region.
	After the first minimization, $1)$
	if the strongest narrow transition of each region is not detected, we fixed the width 
	of the narrow component to the [S~\textsc{iii}]$\lambda$9531 \AA, and 
	$2)$ if the broad component of a line is below the detection threshold, we discard the component as an unreliable detection,
	and we run again the fit, using only a single narrow Gaussian component. 
	In a few cases, after visual inspection, additional Gaussian components were added to the model in order to improve the fit. These additional velocity components do not 
	contribute to the broad component used to compute the BH mass (Sect. \ref{sec:bhm}) as they are not considered as BLR tracers\footnote{We note that even considering the additional component as coming from the BLR, our results do not change since those (few) cases do not fulfill one or more of the criteria we adopt in the subsequent analysis.}.
	In particular, in the Pa$\gamma$ region 7/65 spectra required one additional intermediate velocity component in the He~\textsc{i}$\lambda$10830~\AA,
	(i.e., BAT IDs 372, 488, 577, 698, 744, 1064, and 1079), 
	in the Pa$\alpha$ region 1/65 required an additional intermediate velocity component in the Pa$\alpha$ (BAT 1079) and 
	1/65 required one additional intermediate velocity component in the Pa$\alpha$ and two additional lines H~\textsc{2}$\lambda$18345~\AA\ and H~\textsc{2}$\lambda$189205~\AA\ (BAT 372).   
	The model with additional Gaussians was run twice as in the normal case, in order to discard the components in case those were below the detection threshold.
	We then correct the measured FWHMs to account for instrumental resolution, even though it is not a substantial correction for broad lines. 
	
	To estimate the uncertainties related to emission line measurements in each spectral region, we repeated the fit ten times, adding each time an amount of noise $\sigma$ randomly drawn from a normal distribution with the deviation equal to the noise level. We computed the median absolute deviation of the ten measurements and we used this value as an estimate of the uncertainty at the one-sigma confidence level (c.l.).
	We estimate the flux upper limits (at $3\sigma$) on the (undetected) broad line components by assuming a FWHM=4200~\kms, which is the average FWHM of the broad line detections in our FIRE dataset.
	We then visually inspected all the fits and assigned quality flags, following the classification nomenclature of the first BASS paper \citep{koss17}. Quality flag 1 refers to spectra that have small residuals and very good fits. Flag 2 means that the fits are not perfect, but still acceptable. Flag 3 is assigned to not completely satisfactory fits for high S/N sources due to the presence of either absorption lines, additional components in the fit or structure in the residuals, making the fit decomposition more uncertain. 
	Flag 4 refers to spectra with low S/N and/or strongly affected by telluric residuals, the best-fit NLR and BLR estimates are highly uncertain.
	Flag 9 refers to spectra where no emission line is detected.
	The results of the broad emission line fits are presented in Tables \ref{tab:bHei}-\ref{tab:bPaa} for the Pa$\gamma$, Pa$\beta$ and Pa$\alpha$ broad lines. Tables \ref{tab:addHei}-\ref{tab:addPaa} report the few cases that needed additional components in their spectral fit.
	Figure \ref{fig:lines} shows examples of emission-line fits for a Sy 1, BAT 744 (top panels), with fit quality flags 2 (2,1) in the Pa$\gamma$ (Pa$\beta$, Pa$\alpha$) regions, and for a Sy 1.9, BAT 557 (bottom panels), with fit quality flags 
	2 (1,4) in the Pa$\gamma$ (Pa$\beta$, Pa$\alpha$) regions.
	The remainder of the  best-fit models derived for the 
	full FIRE dataset in the spectral regions Pa$\gamma$, Pa$\beta$ and Pa$\alpha$ are shown as figure set in Fig. \ref{fig:lines} (available on the online journal) ordered by increasing BAT number.

	\section{$M_{BH}$ measurements}\label{sec:bhm}
	Below we briefly describe the two independent $M_{BH}$ estimates adopted in order to 
	understand if the NIR view of the BLR gives consistent BH mass estimates 
	compared to the more often adopted
	H$\alpha$-based BH mass estimate, in case of Sy 1 up to Sy 1.9 AGN, and the $\sigma_\star$-based BH mass estimates, which include Sy 2 AGN lacking broad H$\alpha$ line and strong AGN continua but with NIR broad lines. 
	Combining Eqs. \ref{eq:Mvir}-\ref{eq:msigma}, we evaluate the individual virial factors \f\ of our sample as:
	\begin{equation}
		f = \frac{ M_{BH,\sigma_\star} }{M_{vir,line}} \, ,
		\label{eq:f}
	\end{equation}
	where $M_{BH,\sigma_\star}$ is the BH mass estimated from the optical stellar velocity dispersion (Sect. \ref{ss:bhs}), and
	$M_{vir,line}$ is the virial-based BH mass estimated for each near-infared broad line considered in our study (see  Sect. \ref{ss:mvir}).
	\subsection{BH mass estimate from the $M_{BH}-\sigma_\star$ scaling relation}\label{ss:bhs}
	The relation between $M_{BH}$ and the bulge stellar velocity dispersion is probably the most fundamental and most used BH-host scaling relation 
	due to its instrinsic small scatter \citep[$\sim0.3$~dex][]{kormendyho13, vandenbosch16, saglia16, shankar16, denicola19, marsden20}, and lack of strong redshift evolution, at least until $z\sim1$ \citep[e.g.,][]{shen15, sexton19}.
	There are several $M_{BH}-\sigma_\star$ calibrations available in the literature, some of which take into account the host morphology (late/early type \citealt{mcconnellma13, sahu19}; barred/unbarred \citealt{graham08}; elliptical and classical-/pseudo-bulges \citealt{kormendyho13, saglia16, denicola19}). 
	We adopt the $M_{BH}-\sigma_\star$ relation proposed by \citet{kormendyho13}, that is calibrated on dynamical-based $M_{BH}$ measurements of local ellipticals and classical-bulges, and for which an average virial factor \faver\ has already been estimated \citep{hk14,yu19}. More recent estimates of the average virial factor determined by \citet{batiste17} and \citet{yu19} are consistent, within their (large) uncertainties, with the values found by \citet{grier13} and \citet{hk14}.

	The $M_{BH}-\sigma_\star$ relation of elliptical and classical bulges from \citet{kormendyho13} is amongst 
	the highest relations in normalization in the $M_{BH}$ vs $\sigma_\star$ plane up to date \citep[see, e.g., Fig. 2 in][]{ricci17b}. 
	This means that at given velocity dispersion it predicts the largest SMBH masses, at least until predicted 
	$M_{BH}\approx10^9$~M$_\odot$.
	In other words, the \f-factors derived using Eq. \ref{eq:f} are the largest that can be possibly 
	predicted with the currently calibrated $M_{BH}-\sigma_\star$ relations, even considering the most recent updates \citep{saglia16, denicola19}. We will discuss later how this assumption affects the virial factors \f\ and our analysis.

	\subsection{BH mass estimate from the virial method}\label{ss:mvir}
	The virial $M_{BH}$ estimate implicitly assumes that the motion of clouds in the BLR is virialized. Thanks to the $R-L$ relation established by RM campaigns \citep[see, e.g.,][]{bentz13}, Eq. \ref{eq:Mvir} can be rewritten using easily-accessible quantities, like broad emission line or continuum luminosity. 
	We adopted the mixed virial BH mass estimator put forward by 
	\citet[][]{ricci17a}, for a similar mixed $M_{BH}$ virial estimator see also \citet[][]{bongiorno14,LF15}.
	The virial BH mass estimator proposed by \citet{ricci17a} is based on either optical 
	(e.g., H$\beta$ or H$\alpha$)
	or NIR (Pa$\alpha$, Pa$\beta$ or He~\textsc{i}$\lambda$10830~\AA\footnote{From now onward, He~\textsc{i}$\lambda$10830~\AA\ is called just He~\textsc{i}.})
	FWHM and on hard X-ray luminosities, either 2-10 or 14-195 keV.
	Thus, we can compute the virial BH mass $M_{vir,line}$ using the 14-195 keV luminosity $L_X$ and any single broad line reliably detected, i.e., He~\textsc{i}, Pa$\beta$ and Pa$\alpha$. 
	The use of the observed BAT luminosity does not affect our analysis since the majority of our sample has $N_H<10^{23.5}$~cm$^{-2}$, and the observed 14-195 keV luminosity is almost unaffected by X-ray absorbing columns up to at least $N_H\sim10^{23.5} - 10^{24}$~cm$^{-2}$ \citep{cricci15,koss16}.
	
	We can also compute the mass with the FWHM of all the reliable detections and with the weighted\footnote{The weights are the square of the inverse of the measured broad-line FWHM uncertainties, $1/\sigma_i^2$} average, in presence of more than one reliable detection. We call this case $NIR$\footnote{The term $NIR$ in \textit{italic} should not be confused with the general term NIR that indicate the 0.8--2.4~$\mu$m  wavelength range.} in all subsequent analysis. The mixed virial mass is thus $M_{vir,line} \propto FWHM(line)^2 \times L_X^{0.5}$, with $line$ being He~\textsc{i}, Pa$\beta$, Pa$\alpha$, and $NIR$ samples. 
	The statistical uncertainties on the virial BH mass estimate are then the combination of the errors on the broad FWHM and on the X-ray luminosity $L_{14-195\,\rm{keV}}$. For simplicity, we assumed a 5\% uncertainty on the hard X-ray luminosity, while for the FWHM we used the uncertainties determined from the spectral fit. The aforementioned statistical uncertainty does not take into account the intrinsic spread of virial BH mass estimates, which is of the order of $\sim0.5$~dex \citep{mj02, vp06,ricci17a}. 
	
	As for the majority of the virial BH mass estimators, the relations in \citet{ricci17a} were calibrated against a sample of local RM AGN with H$\beta$ measurements, and therefore an average virial factor \faver\ was adopted. 
	This \faver\ changes depending on whether the RM calibrating virial masses are based on the FWHM or the second moment of the line profile $\sigma_{line}$ \citep[see, e.g.,][]{onken04,collin06} such that
	\begin{equation}
		f_\sigma = f_{FWHM} \times \left( \frac{FWHM}{\sigma_{line}} \right)^2  \, .
		\label{eq:fstoffw}
	\end{equation}
	The relations calibrated by \citet{ricci17a} adopt as a calibration sample the RM virial masses based on
	the H$\beta$ line dispersion measured from the rms spectra,
	$\sigma_{H\beta,rms}$ \citep{hk14}, 
	and in particular the value $\langle f_\sigma \rangle = 4.31\pm1.05$, derived in \citet{grier13} by requiring that RM AGN reproduce the $M_{BH}-\sigma_\star$ relation found in quiescent galaxies by \citet{woo13}.
	Alternatively, \citet{ricci17a} also calibrated BH mass estimators considering the bulge morphology, proposing BH mass estimators also for classical-bulges, $\langle f_\sigma \rangle=6.3\pm1.5$, and pseudo-bulges, $\langle f_\sigma \rangle=3.2\pm0.7$. These $\langle f_\sigma \rangle$ values were determined in \citet{hk14} to put RM AGN virial BH masses on the $M_{BH}-\sigma_\star$ relation of classical-bulges, given by \citet{kormendyho13}, and the one followed by pseudo-bulges, determined by \citet{hk14}. For more details on the BH mass estimators, see Table 4 in \citealt{ricci17a}.

	Since the chosen reference BH-$\sigma_\star$ relation is the one from \citet{kormendyho13} (see Sect. \ref{ss:bhs}), we adopt the BH mass relation based on the $\langle f_\sigma \rangle=6.3$, that is 
	relation b3 in Tab. 4 of \citet{ricci17a}, in the classical bulge case
	
	\begin{equation}
		\left(\frac{M_{BH}}{M_\odot} \right)=10^a \cdot \left[  \left(\frac{FWHM}{10^4~\rm km~s^{-1}}\right)^2 +\left(\frac{L_X}{10^{42}~\rm erg~s^{-1}}\right)^{0.5}\right]\, ,
	\end{equation}
	with $a= (7.96 \pm 0.02)$. 
	
	In order to more easily compare with literature works about the \f-factor 
	\citep[see, e.g.,][]{collin06,decarli08,shenho14,mejia18}, 
	we convert the individual \f-factors from $f_\sigma$ to $f_{FWHM}$ 
	using Eq. \ref{eq:fstoffw} and
	assuming the FWHM/$\sigma_{line}$ ratio for a single Gaussian case, i.e., $\sqrt{8 \ln2} \approx 2.355$.
	This is equivalent to rescale the $\langle f_\sigma \rangle$ adopted as $6.3/(2.355)^2$. We denote this rescaled average virial factor as  
	$f_0$. The symbol $f_0$ is hence the equivalent to writing $\langle f_{FWHM} \rangle$, and it is the average value adopted to convert the virial BH mass $M_{vir, line}$ to the BH mass $M(line)$, i.e., $M(line)=f_0 \times M_{vir, line}$.
	With this notation, the \f\ derived in Sect. \ref{ss:fvir} using Eq. \ref{eq:f} will be, strictly speaking, a $f_{FWHM}$. From now on, we will drop the suffix FWHM and simply write \f\ when referring to $f_{FWHM}$.
	
	When optical broad H$\alpha$ mass estimates are available from the BASS DR2 \citep{Mejia_Broadlines},
	we compare them to the NIR-based virial BH mass estimated here (see, Sect. \ref{ss:ha$NIR$}).
	These optical BH masses are computed with the prescription proposed by \citet[][see, e.g., their Eq. 6]{gh05}, based on the FWHM and luminosity of the broad H$\alpha$, with virial factor of unity \citep{Mejia_Broadlines}.
	
	\section{Results}\label{sec:res}
	\begin{figure*}
		\begin{center}
			\hspace{-0.65in}\includegraphics[width=0.37\textwidth]{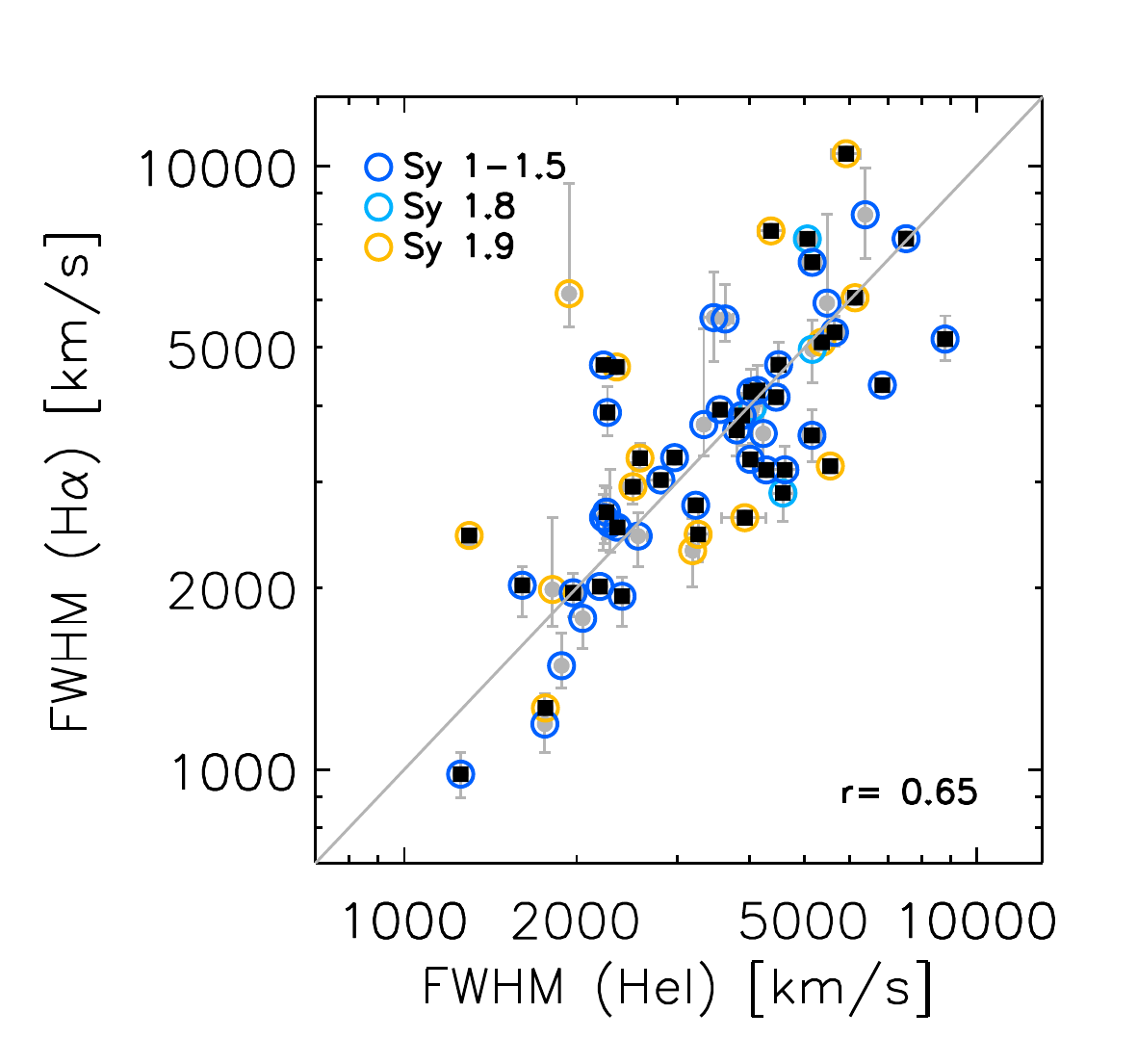}
			\hspace{-.96in}\includegraphics[width=0.37\textwidth]{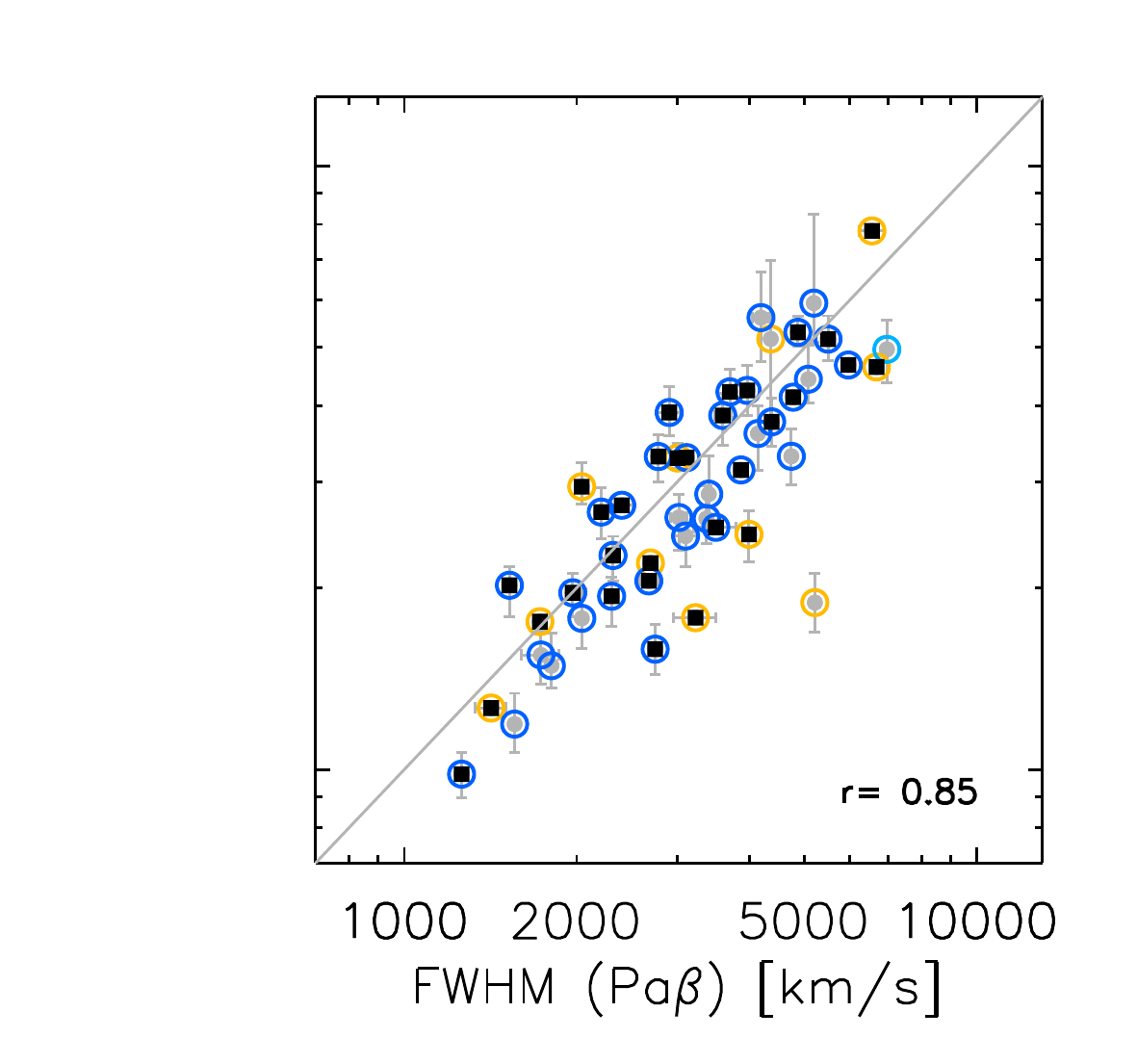}
			\hspace{-.96in}\includegraphics[width=0.37\textwidth]{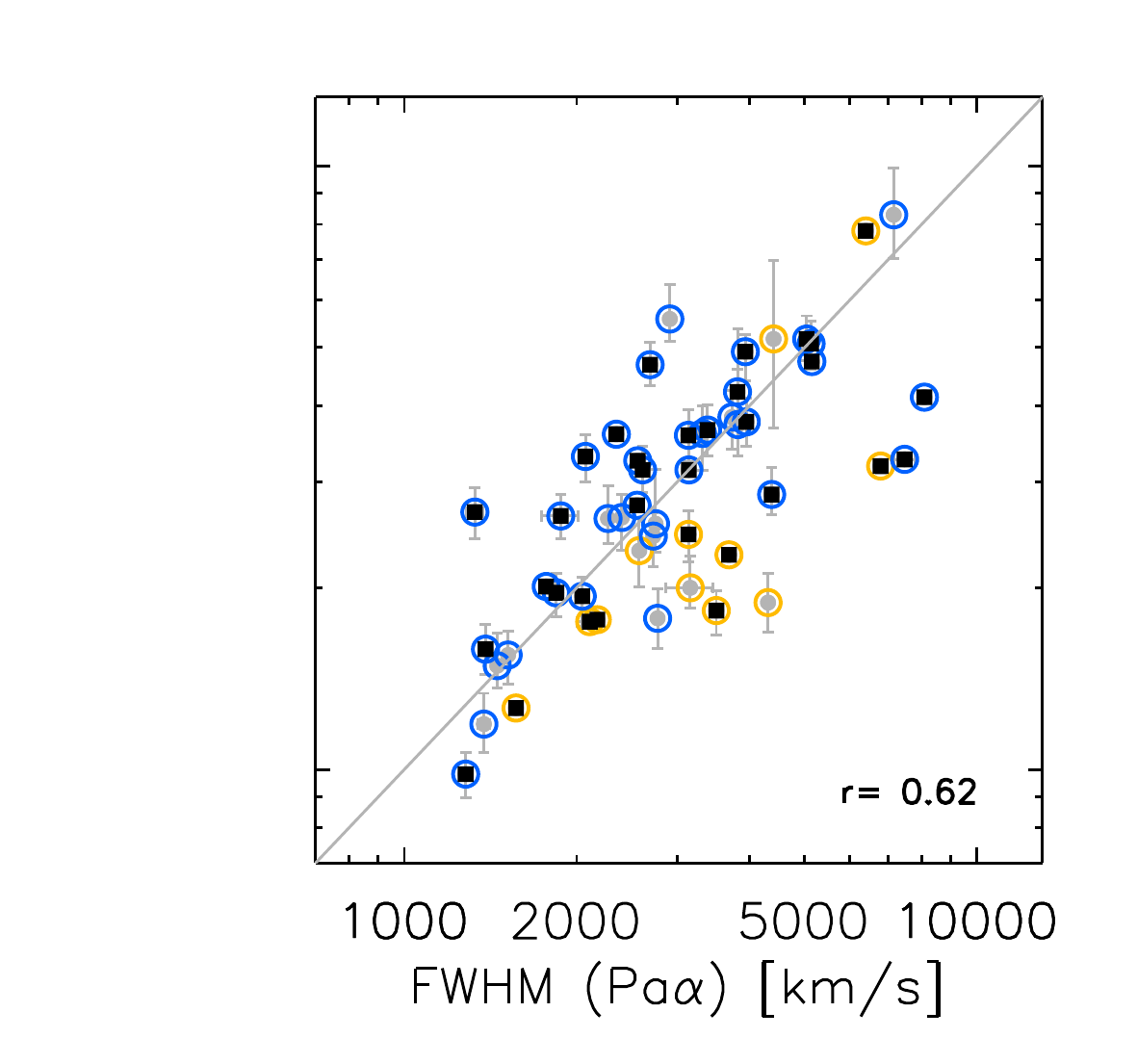}	
			\hspace{-.96in}\includegraphics[width=0.37\textwidth]{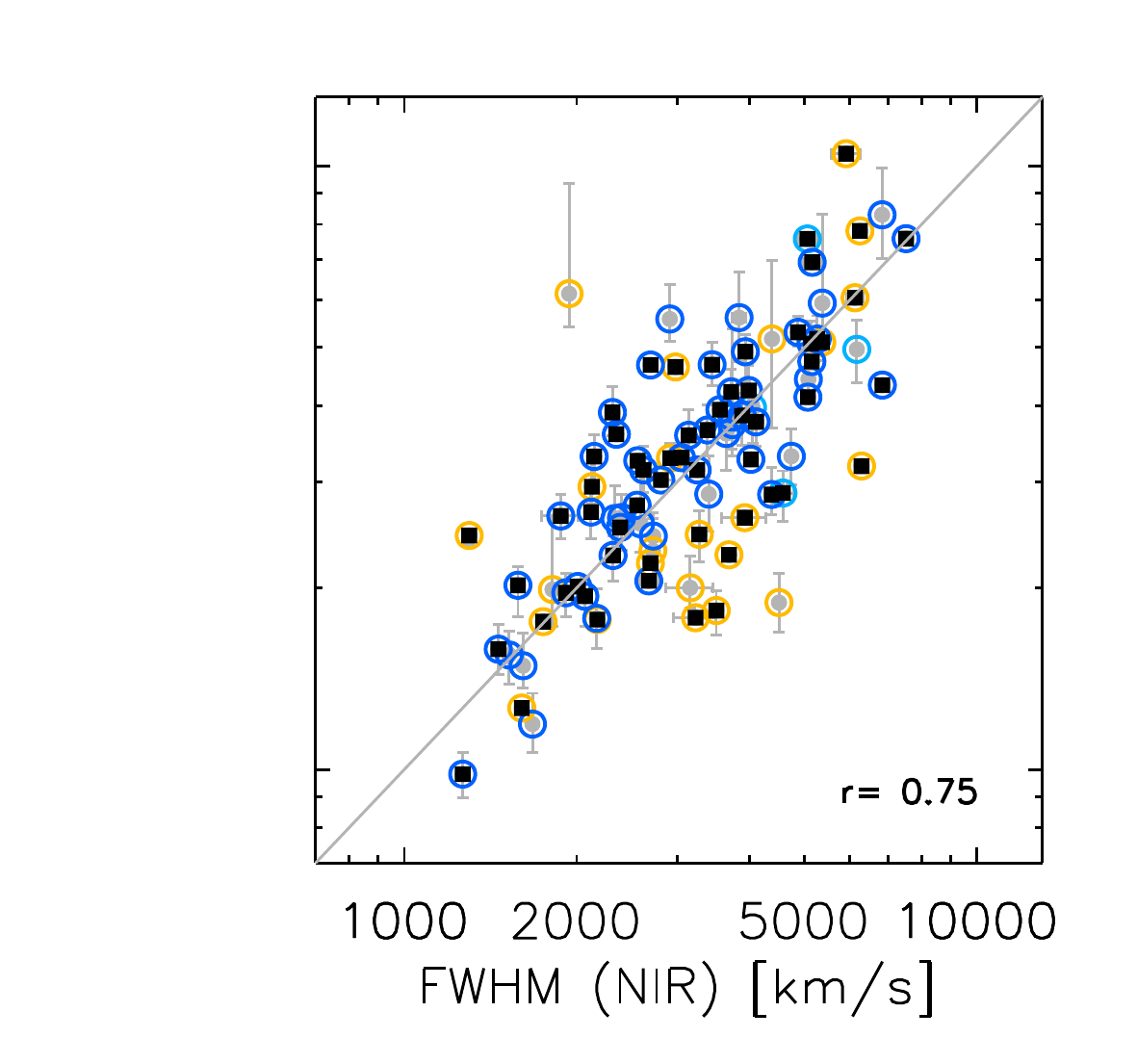}	
		\end{center}	
		
		\vspace{-0.1in} \caption{H$\alpha$ vs NIR BLR velocities for BASS targets, 
			from left to right: He~\textsc{i}, Pa$\beta$, Pa$\alpha$ and $NIR$ broad lines samples, respectively. The $NIR$ sample is the collection of reliably detected NIR broad-line FWHM, being the weighted average FWHM in case there is more than one reliable broad-line detection.
			The colored circles mark the Sy 1-1.5 (blue), Sy 1.8 (cyan), and Sy 1.9 (orange) types.
			Only targets with reliable H$\alpha$ and NIR detection in each line, as defined in Sect. \ref{ss:bldet} and \ref{ss:blr} are shown. 
			The black filled squares are the more robust optical measurements, having $<10\%$ uncertainties in the broad FWHM (H$\alpha$). 
			Pearson correlations coefficients $r$ computed for the more robust subsample (black squares), without separating between Sy subclasses, are reported in each panel. }
		\label{fig:fwha-fw$NIR$}
	\end{figure*}
	To help the reader, we here present a brief summary of the topics in the results section.
	In Sect. \ref{ss:bldet} we introduce the sample of Sy 1.8-1.9-2 types for which we reliably detect NIR broad lines.
	We compare in Sect. \ref{ss:ha$NIR$} the NIR and optical, i.e., H$\alpha$, broad line widths to understand 
	if the ability of measuring the BLR velocity and radius changes with increasing obscuration/extinction.
	We then examine how the medium responsible for obscuring the BLR is related to the material absorbing the 
	X-rays in Sect. \ref{ss:av}.
	By investigating the effect of X-ray obscuration on H$\alpha$ and NIR broad line luminosities we demonstrate that the use of H$\alpha$ (Sect. \ref{sss:habias}) and NIR (Sect. \ref{sec:$NIR$nh})
	broad line luminosities in SE BH mass determinations induce a bias in the BH mass measurements when the column density is above a certain threshold. 
	Finally in Sect. \ref{ss:fvir} we compare two independent obscuration-unbiased BH mass measurements, derive the 
	virial factors and examine whether they depend on some additional parameters.
	
	\subsection{Broad line detection in reddened Seyferts}\label{ss:bldet}
	presented in this work is composed of 33 Sy 2, 18 Sy 1.9 and 1 Sy 1.8, or a total of 52 reddened Sy types.
	As stated in the introduction, in $\sim$30\% of cases
	Sy 1.8 to 2 with narrow emission lines in the rest-frame optical spectra have been
	found to exhibit broad hydrogen and helium recombination lines, i.e., Pa$\alpha$, Pa$\beta$ and He~\textsc{i}. 
	We focus mainly on these three emission lines since all higher order Paschen transitions are
	strongly blended with emission lines from other elements, and He~\textsc{i}, even though blended with Pa$\gamma$, is among the most intense rest-frame NIR lines \citep[see, e.g., Fig. 9 in ][]{riffel06}, and therefore it  
	is possible to detect and deblend faint broad components relatively easily.   
	The rest-frameNIR also contains the hydrogen Brackett series, the strongest emission being the Br$\gamma$~$\lambda$21661\AA. The Br$\gamma$ is isolated, located in a region less affected by low atmospheric transmission, and redder than Pa$\alpha$, thus effects of dust extinction are expected to be even lower. However, its intensity is rather low compared to the Paschen and helium lines \citep[Br$\gamma$/Pa$\beta \approx 0.16$  in case B with T=10$^4$~K and density $n=10^6$~cm$^{-3}$, see e.g., ][]{osterbrock06}, thus we do not investigate its properties in this work. 
	Based on the fact that the He~\textsc{i}, Pa$\beta$ and Pa$\alpha$ lines have been found to have similar FWHMs in Sy 1 and in intermediate types \citep[][see also Sect. \ref{ss:ha$NIR$} and in particular Fig. \ref{fig:fwha-fw$NIR$}]{landt08,ricci17a, lamperti17, onori17a}, we can also examine the collection of near-infrared broad-line measurements, that we call $NIR$.
	When more than one observation was present, we chose the one with the highest S/N evaluated on the continuum, in each spectral region. 
	Combining the DR1, DR2 and FIRE observations the total parent sample is thus composed of 235 obscured BASS AGN, divided into 168 Sy2, 60 Sy 1.9 and 7 Sy 1.8.
	
	Considering only the most reliable measurements, i.e., only the cases where the spectral fit quality is good (quality flag equal to 1 or 2) and the relative uncertainty on the measured broad FWHM is $<10\%$, 
	the final reliably detected broad lines are 
	63/235 ($27^{+4}_{-3} \%$) 
	combining the number of detections of at least one of the above transitions.
	The broad-line detection rate is expected to be consistent in the three BASS samples since 
	the targeted AGN should be similar in their properties (see, Fig. \ref{fig:lxz})\footnote{The FIRE/BASS sample shows a higher detection rate of hidden BLRs compared to the other two BASS works, even though consistent within $\lesssim$2$\sigma$ given the poissonian uncertainties. This might be caused by the higher resolving power of this work (R$\sim$6000) with respect to the BASS DR1 (most observations have R=800, $\Delta v=375$~\kms, see, Tab. 4 in \citealt{lamperti17}) and with the better observing conditions, resulting in slightly higher S/N, of this FIRE dataset in comparison to the DR2 (average seeing 0$\farcs$7 in the BASS/FIRE sample vs 1$\arcsec$ in the DR2, \citealt{denBrok_DR2_NIR}).}.

	\begin{deluxetable*}{rlcccc}
		\tablecaption{Statistical information of the optical and near-infrared BLR velocity measurements.}\label{tab:ha$NIR$-stat}
		\tablehead{
			\colhead{N}&	
			\colhead{sample}&	
			\colhead{$\langle$~FWHM(H$\alpha)~\rangle$}&		\colhead{$\langle$~FWHM(He~\textsc{i})~$\rangle$}&		
			\colhead{r}&	
			\colhead{P(r)}\\
			\colhead{} &
			\colhead{} &
			\colhead{[\kms]} &
			\colhead{[\kms]} & 
			\colhead{} &   
			\colhead{}\\
			\colhead{(1)}&\colhead{(2)}&\colhead{(3)}&\colhead{(4)}&\colhead{(5)}&\colhead{(6)}
		}
		\startdata
		60&	all						&3872$\pm$249	&3669$\pm$212	&0.67	&3.4E-09	\\
		41&	Sy 1-1.5				&3665$\pm$260	&3644$\pm$266	&0.77	&5.1E-09	\\
		19&	Sy 1.8-1.9				&4319$\pm$549 	&3724$\pm$356	&0.58	&9.5E-03	\\
		42&	all more robust			&3942$\pm$298	&3869$\pm$265	&0.65	&2.9E-06	\\
		28&	Sy 1-1.5 more robust	&3674$\pm$275	&3853$\pm$340	&0.75	&4.1E-06	\\
		14&	Sy 1.8-1.9 more robust	&4477$\pm$701	&3900$\pm$430	&0.64	&1.3E-02	\\
		\hline
		N&	sample&	$\langle$~FWHM(H$\alpha)~\rangle$&			$\langle$~FWHM(Pa$\beta)~\rangle$&		r&	P(r)\\					
		& & [\kms] & [\kms] & &\\
		(1) & (2) & (3) & (4) & (5)& (6) \\
		\hline
		47&	all						&3176$\pm$214   &3478$\pm$215	&0.81	&6.4E-12	\\
		35&	Sy 1-1.5				&3117$\pm$220	&3304$\pm$210	&0.87	&1.0E-11	\\
		12&	Sy 1.8-1.9				&3346$\pm$559	&3987$\pm$571	&0.74	&6.0E-03	\\
		31&	all more robust			&3157$\pm$260	&3342$\pm$261	&0.85	&1.2E-09	\\
		22&	Sy 1-1.5 more robust	&3169$\pm$254	&3285$\pm$267	&0.87	&1.3E-07	\\
		9&	Sy 1.8-1.9 more robust	&3125$\pm$673	&3480$\pm$648	&0.84	&4.7E-03	\\
		\hline									
		N&	sample&	$\langle$~FWHM(H$\alpha)~\rangle$&		$\langle$~FWHM(Pa$\alpha)~\rangle$&		r&	P(r)\\					
		& & [\kms] & [\kms] & & \\
		(1) & (2) & (3) & (4) & (5)& (6) \\
		\hline					
		50&	all						&3159$\pm$218	&3299$\pm$235	&0.66	&1.4E-07	\\
		38&	Sy 1-1.5				&3269$\pm$233	&3188$\pm$272	&0.69	&1.4E-06	\\
		12&	Sy 1.8-1.9				&2808$\pm$539	&3650$\pm$468	&0.71	&9.6E-03	\\
		33&	all more robust			&3192$\pm$244	&3403$\pm$313	&0.62	&1.2E-04	\\
		25&	Sy 1-1.5 more robust	&3320$\pm$225	&3318$\pm$357	&0.61	&1.1E-03	\\
		8&	Sy 1.8-1.9 more robust	&2792$\pm$744	&3669$\pm$688	&0.75	&3.1E-02	\\
		\hline									
		N&	sample&	$\langle$~FWHM(H$\alpha)~\rangle$&		$\langle$~FWHM($NIR$)~$\rangle$&		r&	P(r)\\					
		& & [\kms] & [\kms] & &\\
		(1) & (2) & (3) & (4) & (5)& (6)\\
		\hline									
		86	&all					&3593$\pm$192	&3467$\pm$158	&0.73	&1.6E-15	\\
		58	&Sy 1-1.5				&3557$\pm$199	&3358$\pm$187	&0.83	&8.4E-16	\\
		28	&Sy 1.8-1.9				&3669$\pm$427	&3694$\pm$295	&0.62	&4.0E-04	\\
		60	&all more robust		&3620$\pm$232	&3486$\pm$194	&0.75	&5.5E-12	\\
		40	&Sy 1-1.5 more robust	&3573$\pm$216	&3362$\pm$227	&0.82	&6.0E-11	\\
		20	&Sy 1.8-1.9 more robust	&3715$\pm$555	&3734$\pm$365	&0.72	&3.5E-04	\\
		\enddata
		\tablenotetext{}{Columns: (1)-(2) sample size and type, (3) average FWHM of the broad H$\alpha$, (4) average broad near-infrared line, (5)-(6) Pearson correlation coefficient and probability.}
	\end{deluxetable*}
	
	\subsection{Optical and near-infrared view of the BLR in Sy 1-1.9}\label{ss:ha$NIR$}
	\begin{figure*}	
		\begin{center}	
			\hspace{-0.65in}\includegraphics[width=0.37\textwidth]{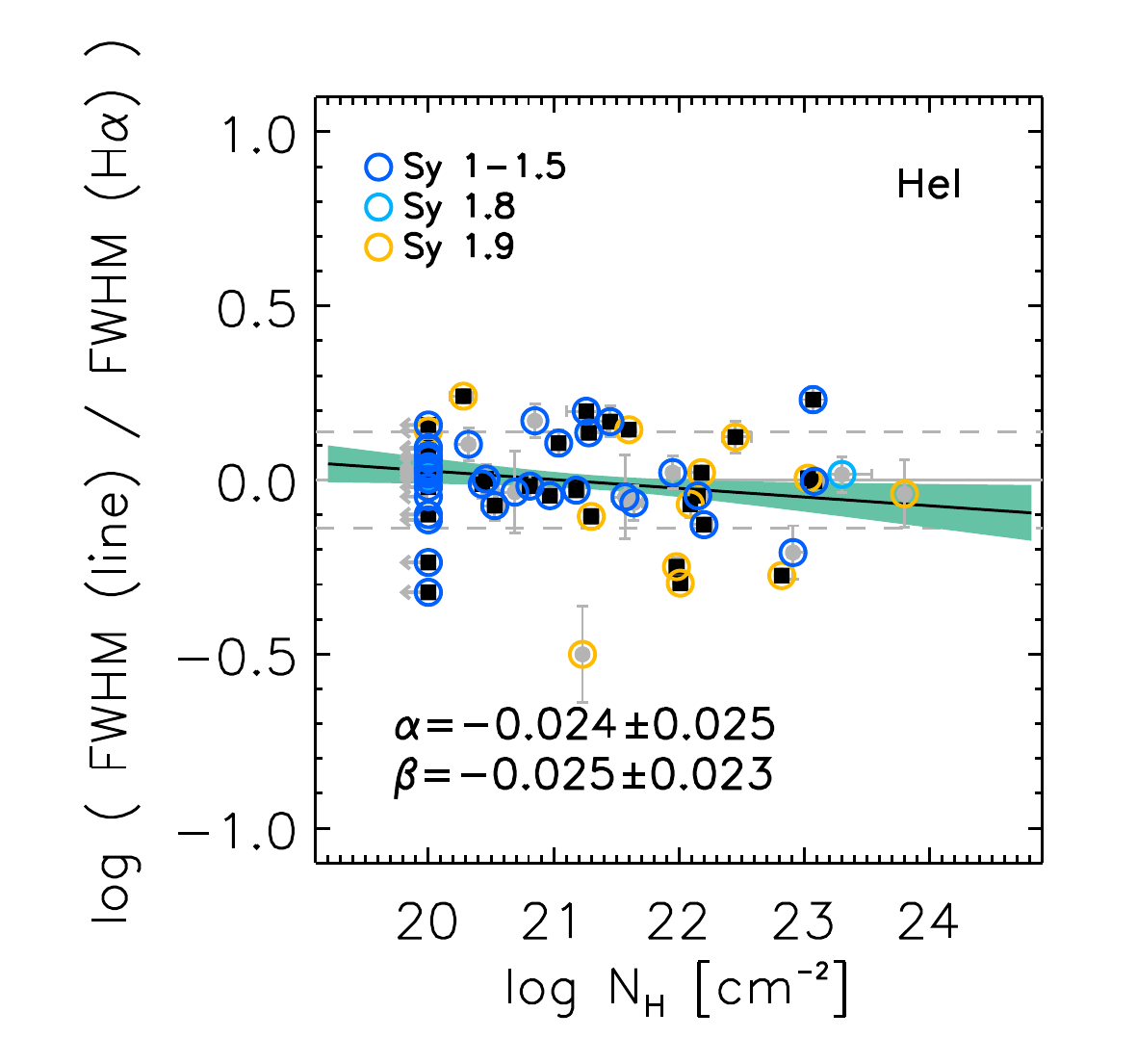}
			\hspace{-.96in}\includegraphics[width=0.37\textwidth]{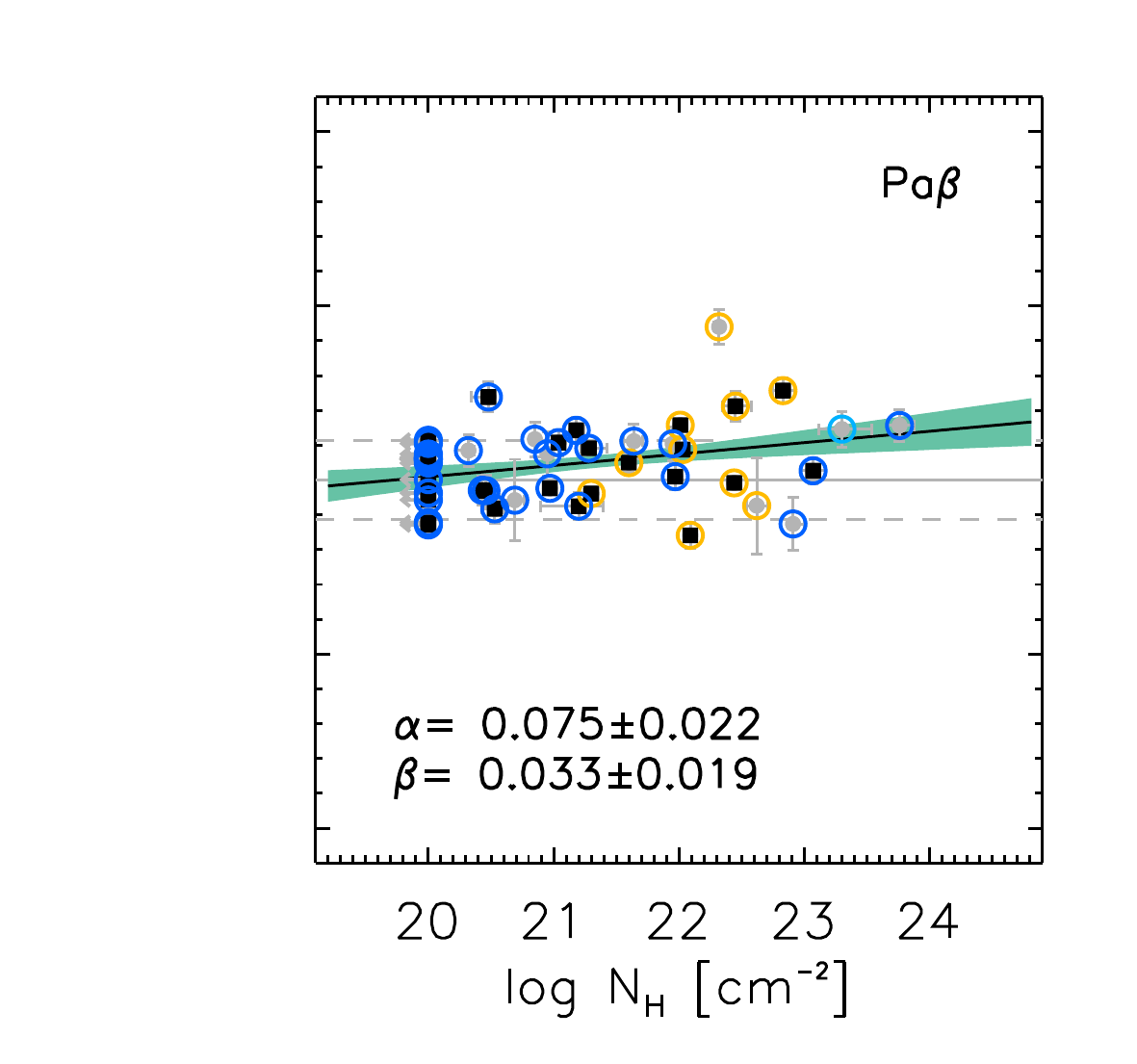}
			\hspace{-.96in}\includegraphics[width=0.37\textwidth]{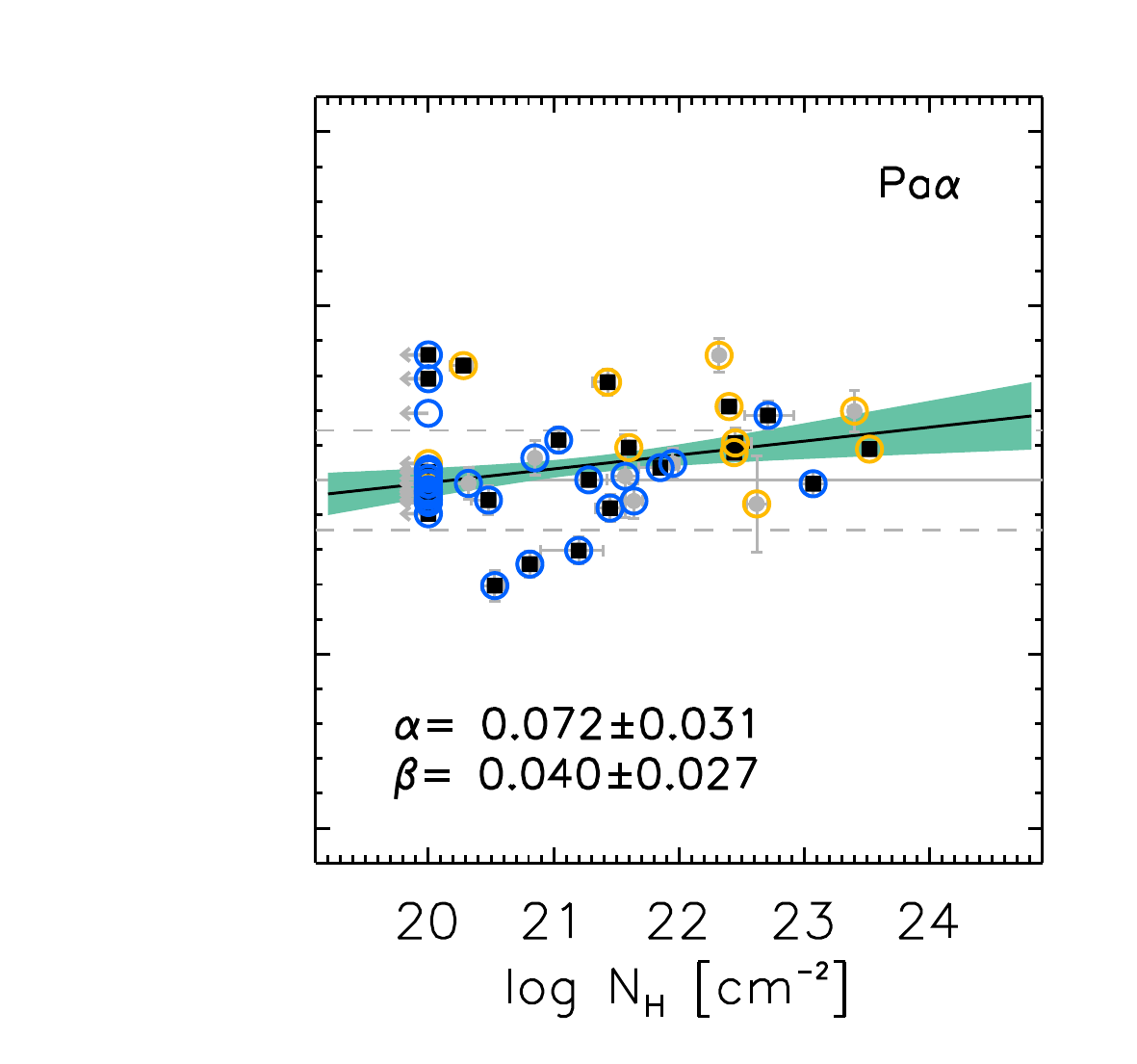}	
			\hspace{-.96in}\includegraphics[width=0.37\textwidth]{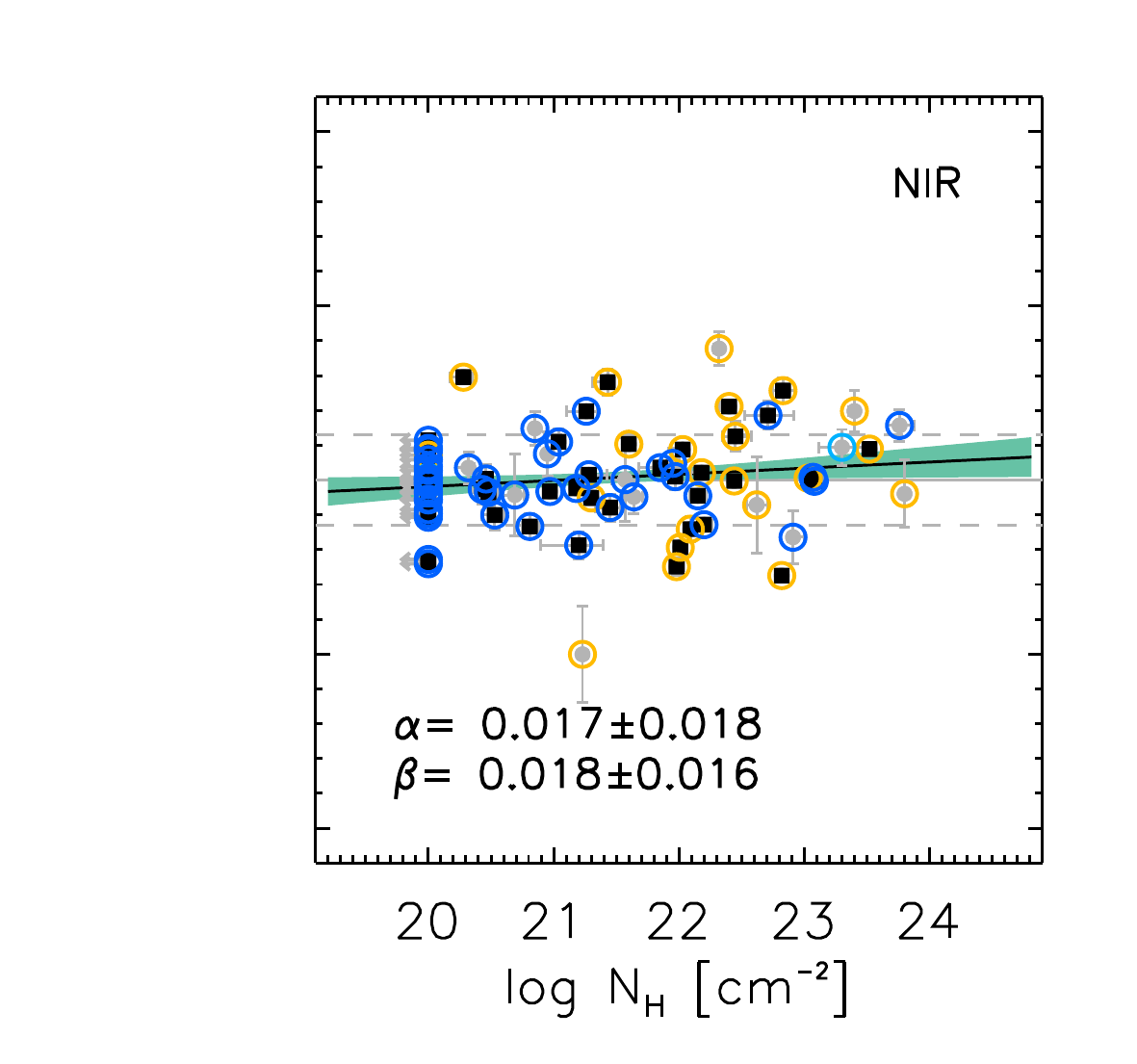}	
		\end{center}
		
		\vspace{-0.1in}	
		\caption{Ratio of the FWHM velocities measured in the NIR to the ones measured from the broad H$\alpha$ as a function of the line-of-sight X-ray column $N_H$.
			The panels from left to right compare He~\textsc{i}, Pa$\beta$, Pa$\alpha$ and $NIR$, respectively.
			Symbols are plotted with the same color scheme as in Fig. \ref{fig:fwha-fw$NIR$}.
			For each panel, the gray solid line shows where the velocities are equal, and the dashed lines
			mark the $\pm$1$\sigma$ intrinsic scatter. 
			The solid black lines are the best-fit Bayesian relations derived using \texttt{linmix\_err} together with the
			68\% c.l. region plotted in green. The best-fit slope $\beta$ and intercept $\alpha$ with their uncertainties are reported as well. The normalizations $y_0$ and $x_0$ in Eq. \ref{eq:offset} are set to 1 and $10^{22}$~cm$^{-2}$, respectively. }
		\label{fig:ratio_nh}
	\end{figure*}
	It is important to understand whether the measured H$\alpha$ broad line width
	and luminosity
	are accurately tracing the underlying BLR motion and radius, particularly for intermediate Sy 1.8 and 1.9
	where the extinction is higher than in normal optical broad-line AGN.
	The NIR offers a view of the BLR complementary to the optical, as it can help penetrate into the innermost and fastest moving BLR clouds. Together with X-ray ancillary data provided by BASS \citep{cricci17}, NIR spectroscopy can help to understand biases or systematics in BLR measurements, e.g., the BLR velocity and radius, obtained from the 
	H$\alpha$ 
	\citep[BASS DR2][]{Mejia_Broadlines}.
	
	Indeed, a large community effort is being spent to find intermediate mass black holes 
	\citep[IMBHs, $M_{BH}<10^5$~M$_\odot$,][]{gh04,gh07, reines13, moran14, baldassare17, mezcua17,mezcua18, chillingarian18, martinez20}, 
	as this population has a huge impact on several aspects concerning BH seed formation and BH growth at high redshift 
	\citep[see, e.g.,][]{volonteri08, treister13,pacucci18,inayoshi20}.
	One of the most commonly used lines in the local Universe to this end is H$\alpha$ \citep{reines15,baldassare16}, 
	since it is about three times more intense than H$\beta$, and it is in practice the 
	only line available in the rest-frame optical in case of Sy 1.9.
	\begin{figure*}
		\begin{center}	
			\hspace{-0.55in}\includegraphics[width=0.37\textwidth]{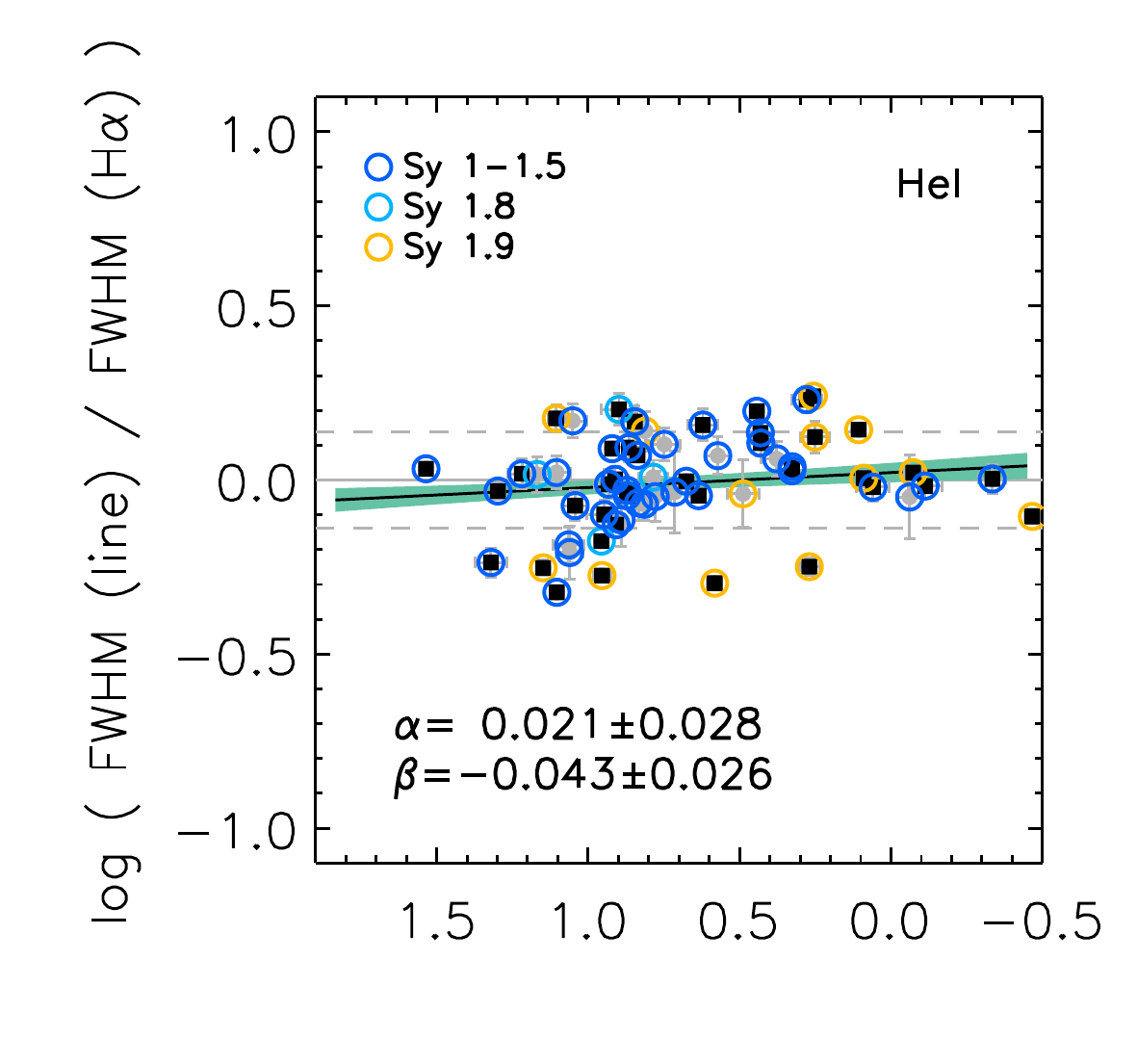}
			\hspace{-.95in}\includegraphics[width=0.37\textwidth]{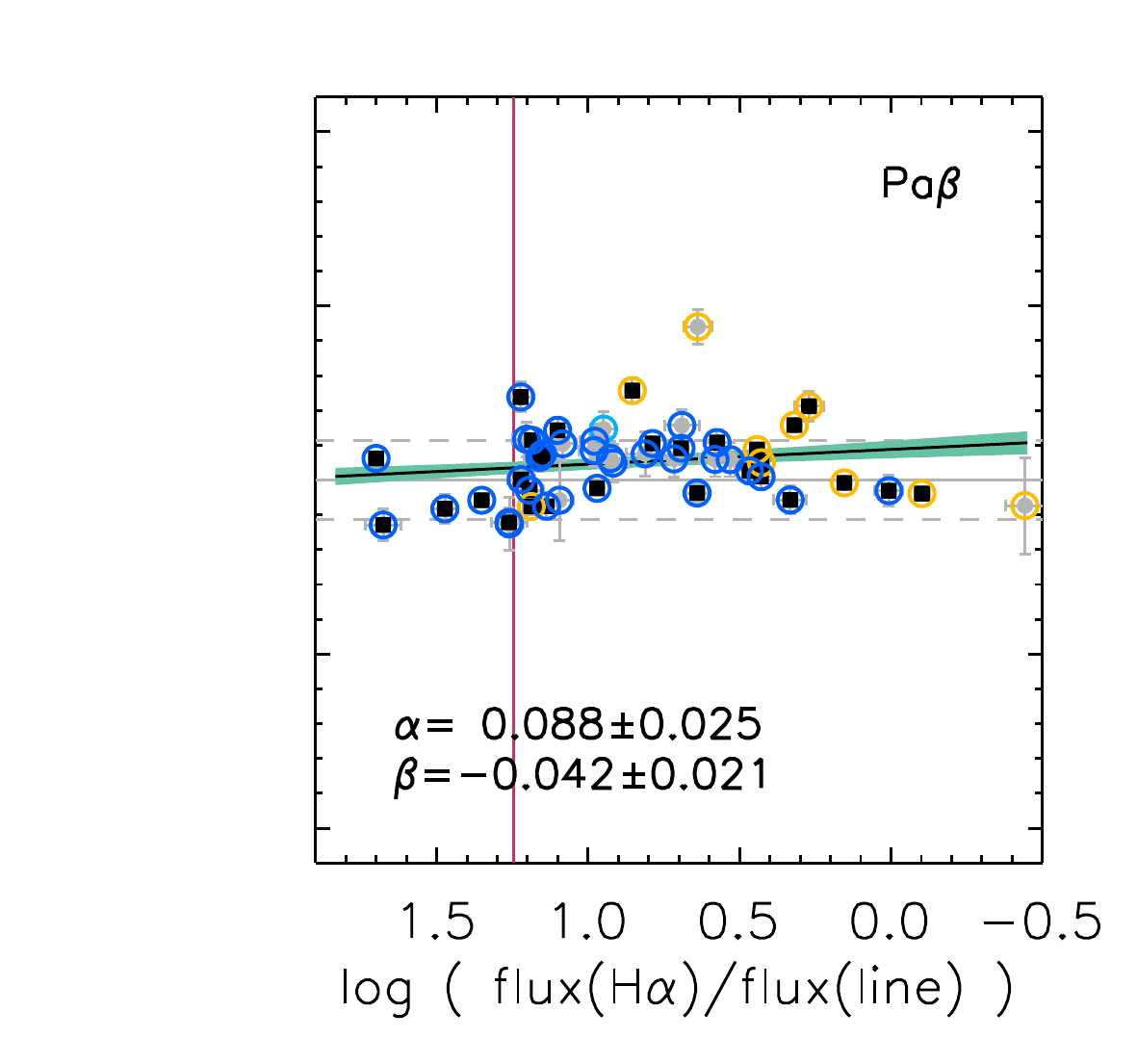}
			\hspace{-.95in}\includegraphics[width=0.37\textwidth]{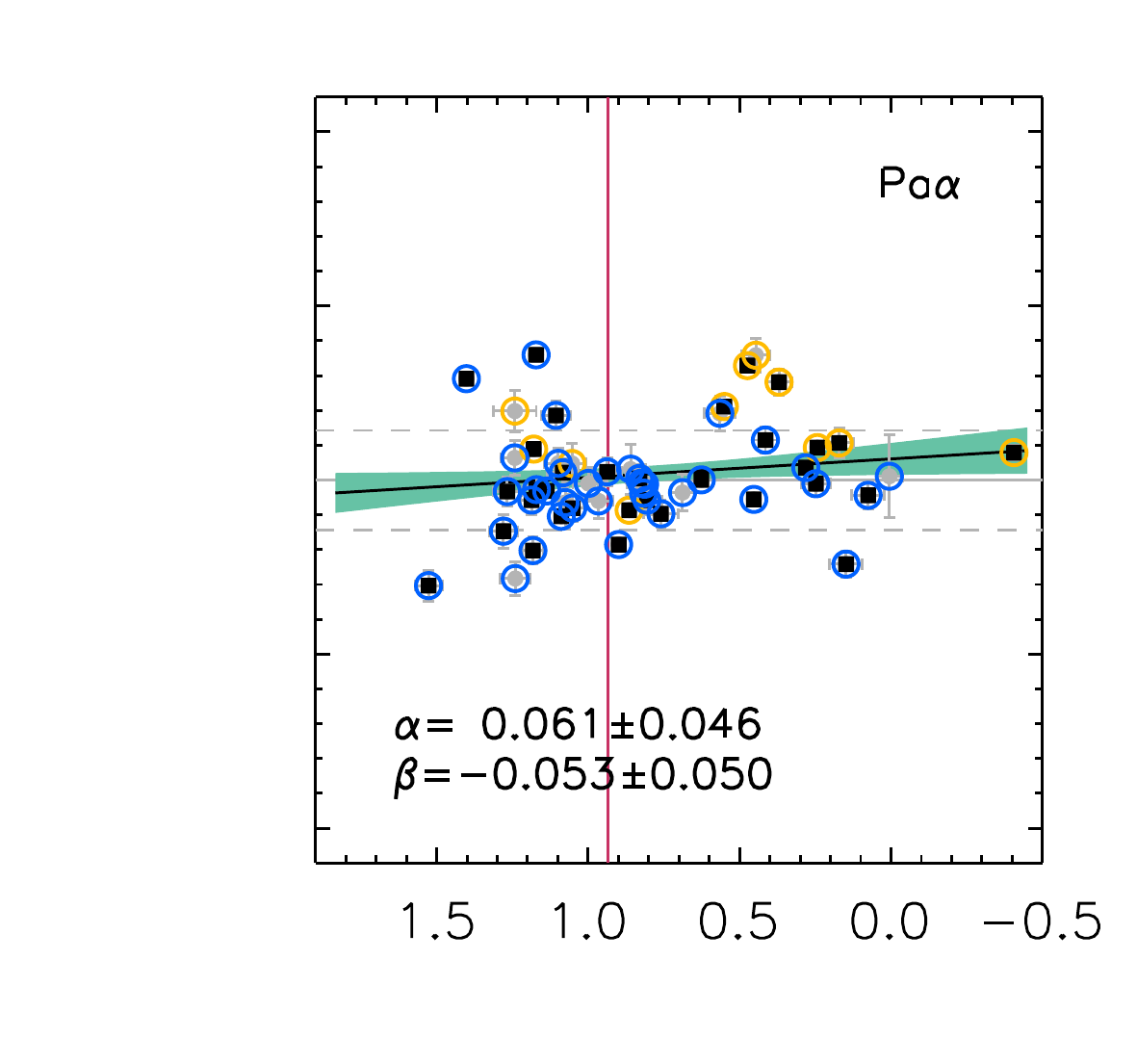}	
		\end{center}
		\vspace{-0.1in}	
		\caption{Ratio of the FWHM velocities measured in the near-infrared to the ones measured from the broad H$\alpha$ 
			as a function of the broad line flux ratios H$\alpha$ to near-infrared.
			The $x$-axis has been plotted showing the direction of increasing extinction (towards the right).
			The panels from left to right compare He~\textsc{i}, Pa$\beta$, Pa$\alpha$, respectively.
			Symbols are plotted with the same color scheme as in Fig. \ref{fig:fwha-fw$NIR$}.
			For each panel, the gray solid line shows the locus where the velocities are equal, and the dashed lines
			mark the $\pm$1$\sigma$ intrinsic scatter. 
			The solid black lines are the best-fit Bayesian relations derived using \texttt{linmix\_err} together with the
			68\% c.l. region shown in green. The best-fit slope $\beta$ and intercept $\alpha$ with their uncertainties are reported as well. The normalizations $y_0$ and $x_0$ in Eq. \ref{eq:offset} are both set to 1. The red line in the center and right panels is the 
			value of the flux ratios expected assuming case B recombination (see text for more details). }
		\label{fig:ratio_hato$NIR$}
	\end{figure*}

	\subsubsection{BLR velocity estimates from H$\alpha$ and near-infrared lines}\label{ss:blr}
	Figure \ref{fig:fwha-fw$NIR$} reports the measured broad H$\alpha$ from BASS DR2 \citep{Mejia_Broadlines}
	compared to the measured broad near-infrared lines from BASS NIR DR1, DR2 and FIRE targets, separated into Sy 1-1.5 (blue circles), Sy 1.8 (cyan circles), and Sy 1.9 (yellow circles) where both optical and NIR BLR components are detected. 
	We recall that we consider the collection of reliably detected broad-line measurements as the $NIR$ line sample (in \textit{italic}, see footnote in Sect. \ref{ss:mvir}), being the weighted average FWHM in case there is more than one near-infrared broad-line detection.  
	The sample of reliable near-infrared, as defined in Sect. \ref{ss:bldet}, and H$\alpha$ broad lines\footnote{From the BASS H$\alpha$ DR2 database we excluded BAT 557 whose observation does not have nightly flux calibrations, thus resulting in higher uncertainties on the flux measurement \citep[e.g.,][]{koss17}.}  
	contains 
	60, 47, 50 and 86
	AGN with He~\textsc{i}, Pa$\beta$, Pa$\alpha$ and $NIR$ broad lines, respectively. 
	The sample with robust optical measurements, i.e., with $<10\%$ uncertainty in the measured FWHM H$\alpha$, contains 
	42, 31, 33 and 60
	AGN with He~\textsc{i}, Pa$\beta$, Pa$\alpha$ and $NIR$ broad lines, respectively (see black filled squares in Fig. \ref{fig:fwha-fw$NIR$}).
	The measurements show some dispersion around the one-to-one locus, with the Paschen-line FWHM being on average higher than the H$\alpha$ one, 
	which might be driven by the less extincted view in the NIR with respect to the optical band, though the averages velocities measured from H$\alpha$ and all the NIR lines considered are
	consistent ($<1 \sigma$) within the uncertainties of the mean (see Tab. \ref{tab:ha$NIR$-stat}). 
	\begin{figure*}
		\begin{center}	
			\hspace{-0.35in}\includegraphics[width=0.37\textwidth]{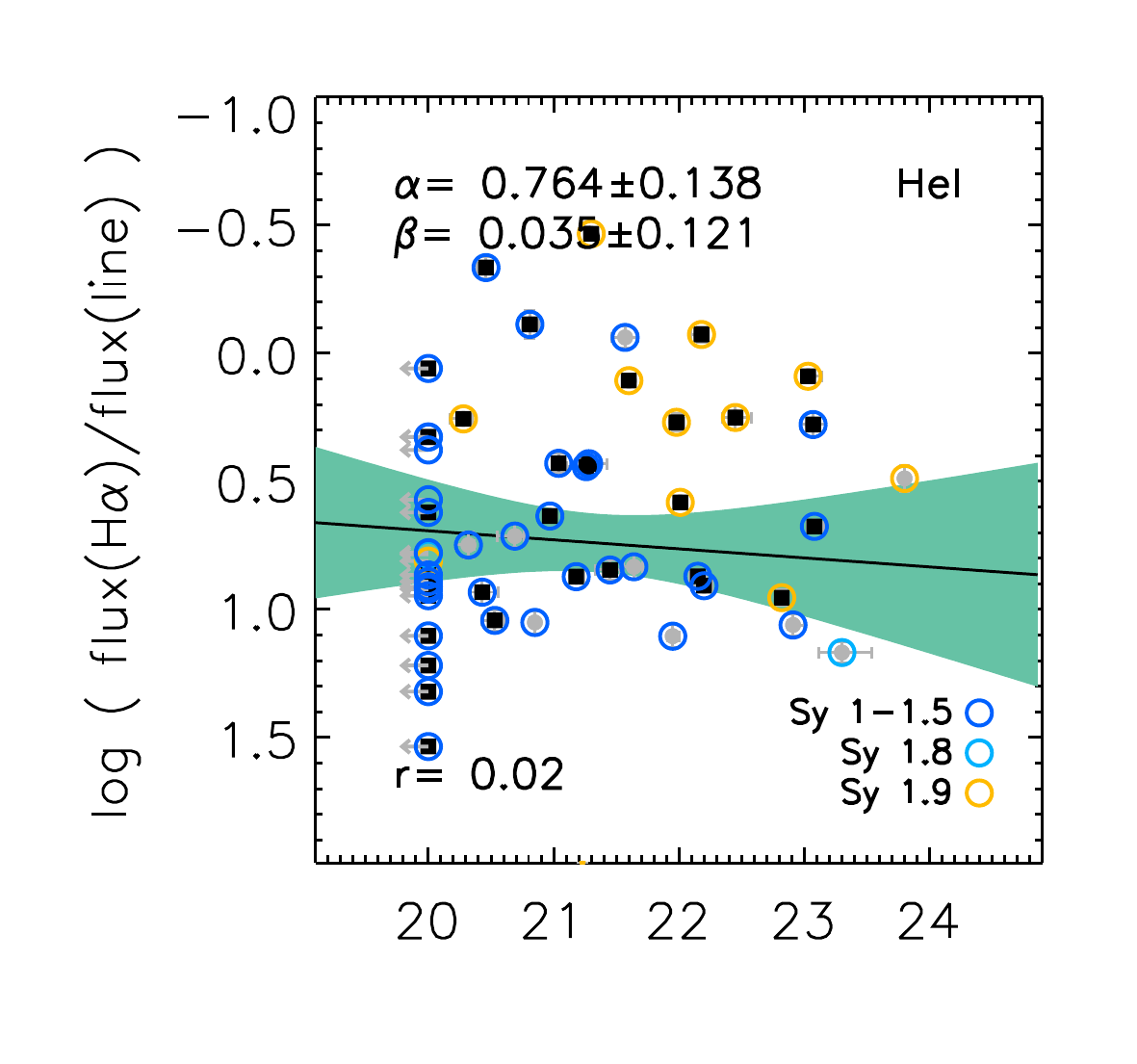}
			\hspace{-.96in}\includegraphics[width=0.37\textwidth]{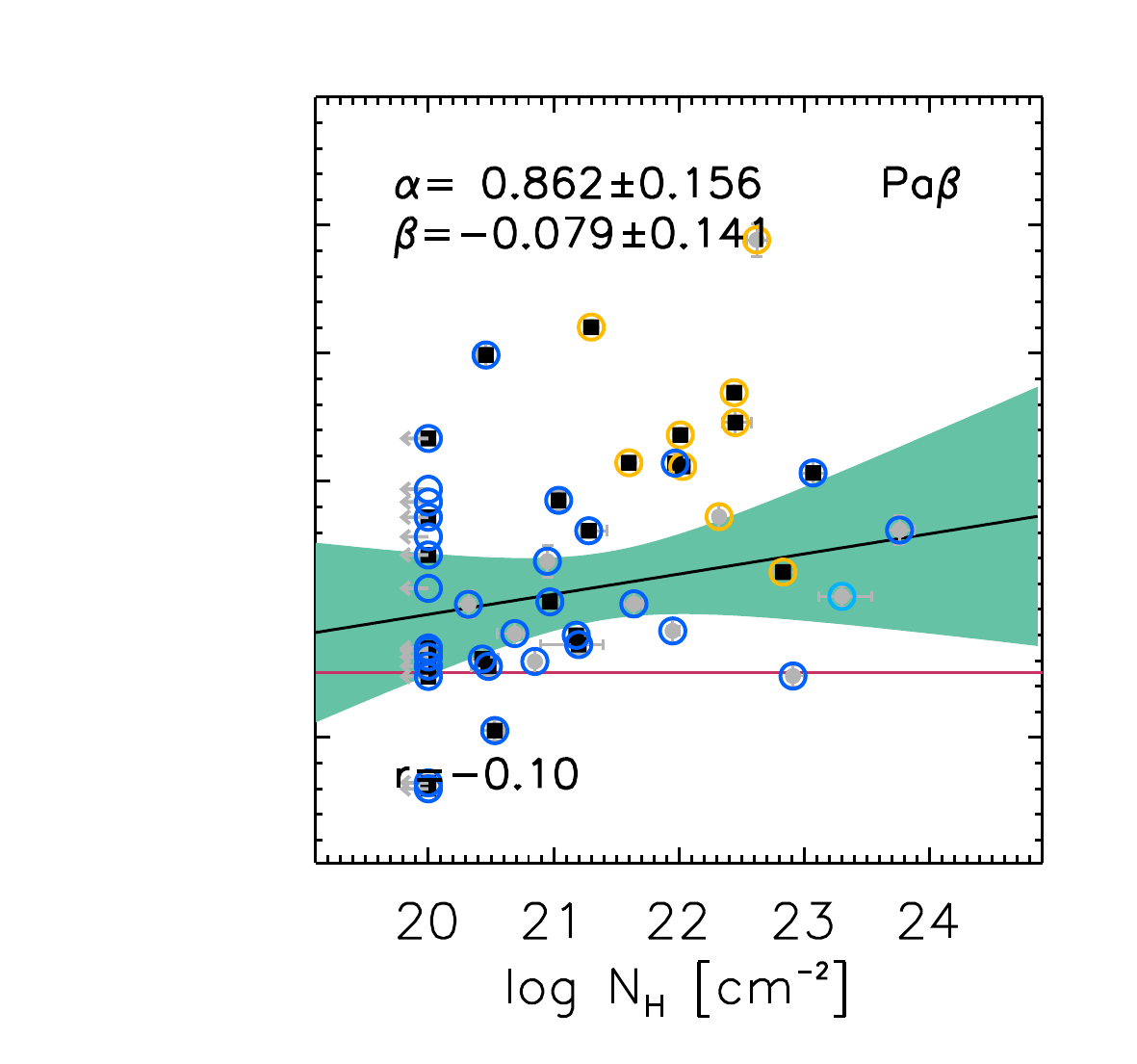}
			\hspace{-.96in}\includegraphics[width=0.37\textwidth]{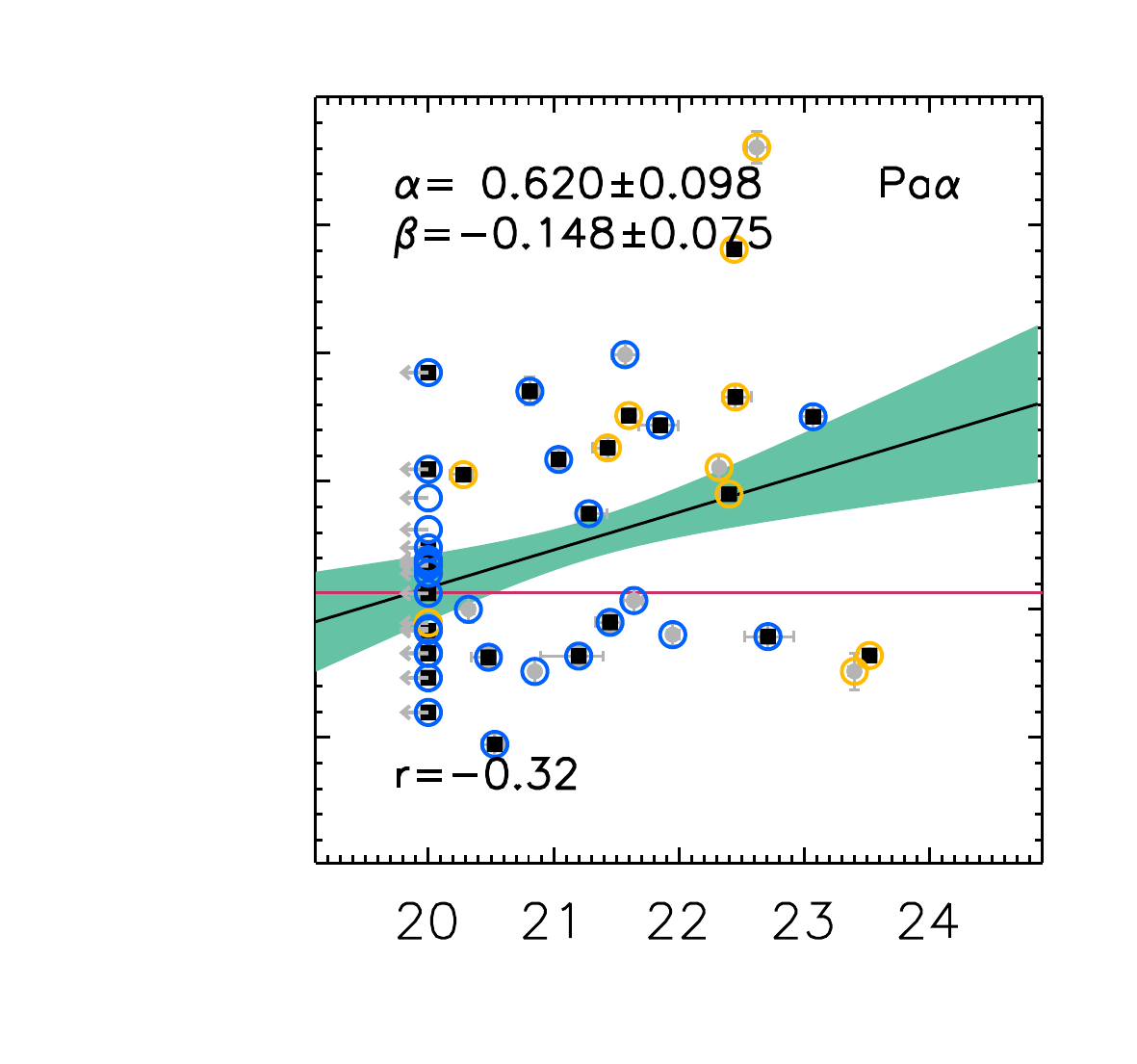}
		\end{center}	
		\vspace{-0.1in}
		\caption{Ratio of the broad line fluxes measured from H$\alpha$ and NIR lines He~\textsc{i} (\textit{left}), Pa$\beta$ (\textit{center}) and Pa$\alpha$ (\textit{right}) as a function of the line-of-sight X-ray column density \citep{cricci17}. 
			The $y$-axis has been plotted showing the direction of increasing extinction (towards the top).
			Symbols are plotted with the same color scheme as in Fig. \ref{fig:fwha-fw$NIR$}. 
			The solid black lines are the best-fit Bayesian relations derived using \texttt{linmix\_err} together with the
			68\% c.l. region shown in green. The best-fit slope $\beta$ and intercept $\alpha$ with their uncertainties are reported as well. The normalizations $y_0$ and $x_0$ in Eq. \ref{eq:offset} are set to 1 and $10^{22}$~cm$^{-2}$, respectively.
			The red line in the center and right panels is the 
			value of the flux ratios expected assuming case B recombination (see text for more details). }
		\label{fig:hato$NIR$_nh}
	\end{figure*}
	
	We found that for the whole sample (see `all' in Tab. \ref{tab:ha$NIR$-stat}), 
	the Pearson coefficients are high ($>$0.66) and statistically significant, as also confirmed by a Student's T-test.	
	This means that the H$\alpha$ and NIR measurements of the BLR velocities are statistically 
	describing the same velocity field. 
	The intrinsic spreads with respect to the H$\alpha$ are 
	0.14, 0.10, 0.15 and 0.13~dex
	for He~\textsc{i}, Pa$\beta$, Pa$\alpha$ and $NIR$, respectively.
	Note that these correlations between optical and NIR broad lines 
	do not depend on our fitting approach, since we are comparing broad lines that come from different spectral regions that are fit independently.

	In order to verify whether some particular classes of Sy are strongly influencing the result, we split the sample into Sy 1-1.5 and Sy 1.8-1.9
	and compare the more
	secure optical and NIR detections (i.e., the black symbols in Fig. \ref{fig:fwha-fw$NIR$} and the `more robust' cases in Tab. \ref{tab:ha$NIR$-stat}). The results 
	are confirmed: the 
	optical and NIR BLR measurements give consistent estimate of the BLR velocity fields.
	Additionally, the FWHM measured in Sy 1-1.5 vs Sy 1.8-1.9
	come statistically 
	from the same parent population in both NIR and optical. 
	Similar statistical conclusions are derived even adopting the Kolmogorov-Smirnov test and a non-parametric test like the Kendall $\tau$ and bootstrapping to consider measured errors on the FWHM \citep[e.g., {\tt pymccorrelation} python package,][]{privon20}.
	Therefore, we can conclude that once a broad H$\alpha$ line is securely detected, the H$\alpha$ width is in agreement within the intrinsic scatter, with the broad-line width as measured in the NIR.
	This scatter might be partially explained by AGN variability since the spectroscopic measurements of the optical and NIR are often obtained from non-simultaneous observations.

	We examine in  Fig. \ref{fig:ratio_nh} the difference in the measured BLR velocities 
	as a function of the line-of-sight column density $N_H$ measured in the X-rays \citep{cricci17}.
	In order to verify the dependencies between the $y$ and $x$ variables, we adopt a standard forward regression using the {\tt linmix\_err} routine of \citet{kelly07}, that employs a fully Bayesian approach, and we fit linear log-log relations 
	\begin{equation}
		\log ( y/y_0 ) = \alpha + \beta \, \log ( x/x_0 ) \, .
		\label{eq:offset}
	\end{equation} 
	The {\tt linmix\_err} routine can include censored data (i.e., upper limits) on the $y$-variable. In our analysis, all the times upper limits are on the $x$-variable, being the $N_H$. The $N_H$ upper limits are objects with $N_H<10^{20}$~cm$^{-2}$, so lacking evidence of absorbing columns in the broad-band X-ray spectral analysis \citep[see BASS X-ray DR1][]{cricci17}. These objects should, in principle, have at least some obscuration along our line of sight coming from their host galaxy, but it is impossible to detect because of the intervening obscuration in the Milky Way. 
	We adopt as minimum $N_H=10^{19}$~cm$^{-2}$ \citep{ryan-weber03,guver09,treister09}, so treating the $N_H$ upper limits as measurements at $N_H=10^{20}$~cm$^{-2}$ with an uncertainty of $\log N_H$=1~dex.
	
	In all the subsamples inspected, 
	the Bayesian best-fit linear regressions (solid black lines in Fig. \ref{fig:ratio_nh} with the 1$\sigma$ c.l. regions in green) are rather flat,
	indicating no significant effect of $N_H$ on the difference in the line widths due to $N_H$.
	We note that the best-fit slope $\beta$ found for He~\textsc{i} has an opposite sign than what is found for the Pa$\beta$ and Pa$\alpha$ samples, though all slopes are statistically consistent
	with zero (p-value $>$0.1). The reason for this might be the ionization structure in the BLR, since the ionization potential of He~\textsc{i} is much higher ($\approx$6 times) than the one of Pa$\alpha$, Pa$\beta$ and H$\alpha$.
	We can conclude that there is a lack of correlation between the NIR-to-optical 
	BLR velocity measurement ratio and the column density $N_H$, at least until $N_H \simeq 10^{23}$~cm$^{-2}$. 
	At higher $N_H$, the sample is too small to derive meaningful conclusions. 
	This result nicely complement what can be seen using the larger statistical sample of optical BASS DR2 \citep{Mejia_Broadlines}
	having both H$\alpha$ broad line and $N_H$ measurements \citep{cricci17}. 
	In particular, the FWHM(H$\alpha$) - $N_H$ plane shows a decrease in the 
	optical H$\alpha$ width in Sy 1.9 at $N_H>10^{23}$~cm$^{-2}$, while below this column density the 
	FWHM - $N_H$ distribution is rather flat (see, e.g., their Fig. 9).

	Observationally, the line-of-sight X-ray column density $N_H$ has shown a correlation with occultation events in the X-rays, due to gas clumps located in a dust-free region or at the inner edge of the dusty torus \citep[see, e.g.,][]{risaliti07,risaliti11x,maiolino10,markowitz14,cricci16,zaino20}. 
	More prevalent eclipsing events are linked to higher covering factors, placing at least some of the X-ray obscuring material in the dust-free BLR \citep[see also][]{schnorr16}. In fact, \citet{gandhi15} have argued that narrow iron K$\alpha$ 
	line-emitting material could reside in between the BLR and the putative dusty torus. In addition, some works
	have suggested that the accretion disk could partially contribute to the observed column density in Compton-thick AGN, as they have large inclination angles \citep[e.g.,][]{masini16,ramosricci17}.
	Those observations imply the presence of gas inside the sublimation radius. 
	However, the X-ray column density $N_H$ is a line-of-sight local measurement, that is very
	unlikely to strongly obscure the full BLR.
	Therefore, we calculated the optical-to-NIR broad line flux ratios as a way to measure the extinction experienced by the broad H$\alpha$ with respect to the broad NIR emission lines. 
	Figure \ref{fig:ratio_hato$NIR$} 
	shows only a mild trend, statistically consistent with being flat given the dispersion (p-value $>$0.05), when comparing
	the BLR velocity ratio with the 
	broad H$\alpha$ to near-infrared broad line fluxes.
	
	From Figs. \ref{fig:ratio_nh} and \ref{fig:ratio_hato$NIR$} we can then conclude that
	neither the X-ray obscuration nor the BLR extinction significantly affect the measurement of the width of the broad H$\alpha$ line and of the NIR broad lines, at least among the total local hard X-ray selected
	AGN sample up to columns $N_H\simeq 10^{23}$~cm$^{-2}$, once the broad H$\alpha$ line detection is secure.

	\begin{figure*}
		\begin{center}	
			\vspace{-2.35in}	
			\hspace*{-1.7in}\includegraphics[width=2.1\textwidth]{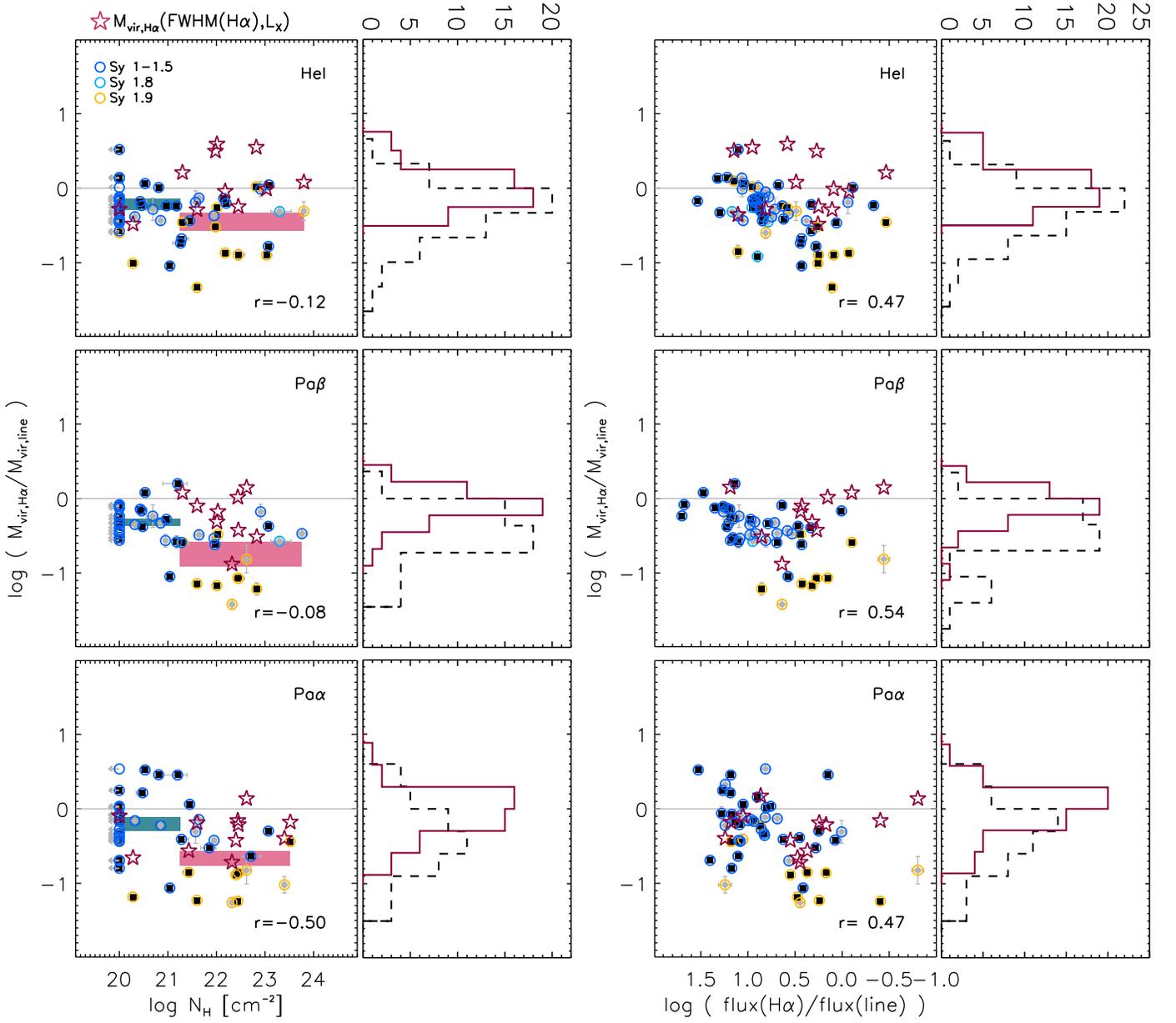}
			
		\end{center}	
		
		\vspace{-10.15in}
		\caption{Difference between the virial BH mass estimates from the optical and NIR as a function of obscuration (\textit{left}) or broad line fluxes (\textit{right}), proxy for the BLR extinction. 
			As previously done in Fig. \ref{fig:ratio_hato$NIR$} and \ref{fig:hato$NIR$_nh}, the $x$-axes in the right panels show the direction of increasing BLR extinction (towards the right).
			Each row reports a NIR line, from top to bottom: He~\textsc{i}, Pa$\beta$ and Pa$\alpha$.
			Symbols are plotted with the same color scheme as in Fig. \ref{fig:fwha-fw$NIR$}. 
			The gray solid lines mark the identity relation.
			The shaded rectangles in the left panels show the average BH mass difference (together with the error on the mean) computed in two bins of $N_H$, low $N_H$ in green and high $N_H$ in magenta.
			Red stars show the difference between the BH mass measurements 
			for Sy 1.9
			when the $M_{vir,H\alpha}$ is 
			estimated using the same prescription as in the NIR, that is to use the hard X-ray luminosity 
			instead of the H$\alpha$ broad line luminosity.
			The vertical panels show the histograms of the BH mass difference in the case of completely H$\alpha$-based (black dashed histogram) or when the hard X-ray luminosity is adopted instead of the broad H$\alpha$ line luminosity (solid red histogram). In the latter case, the distribution of the mass difference is more consistent with scatter around zero. }
		\label{fig:diffMha$NIR$}
	\end{figure*}
	
	\begin{figure}
		\begin{center}	
			\hspace{-0.76in}\includegraphics[width=0.35\textwidth]{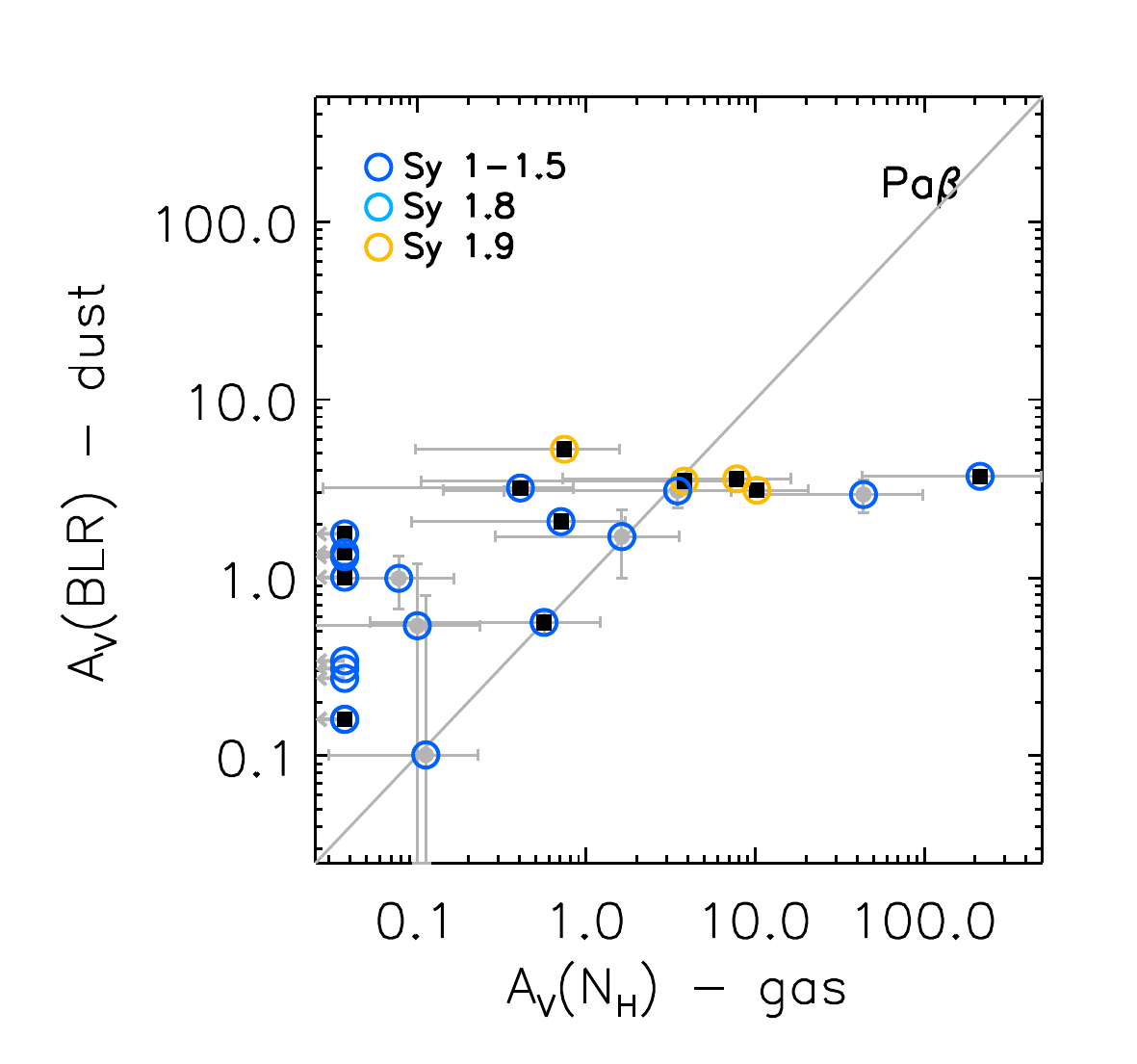}
			\hspace{-.94in}\includegraphics[width=0.35\textwidth]{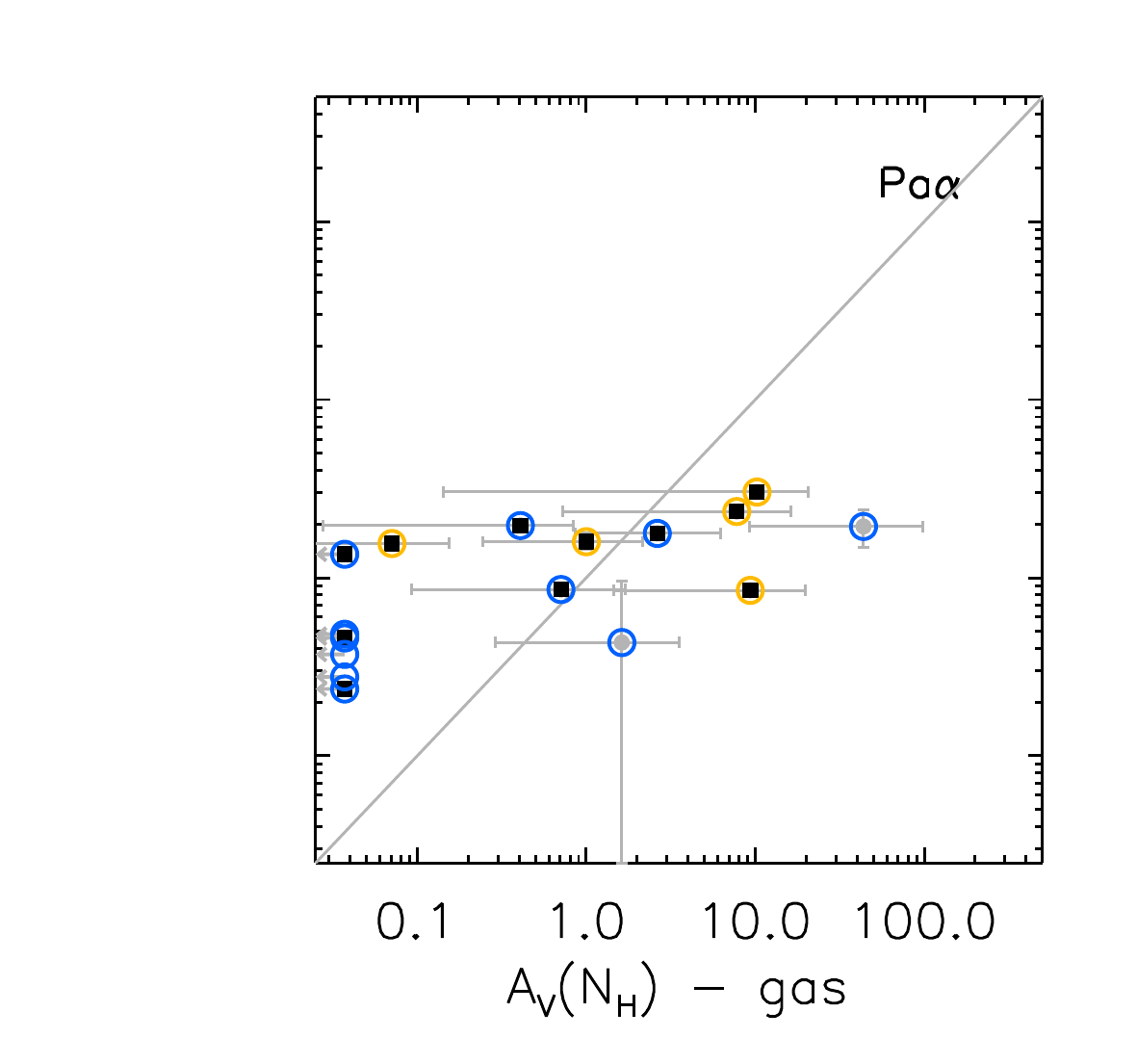}	
			
		\end{center}
		
		\caption{Comparison between $A_V$ expected from gas absorption and $A_V$ towards the BLR due to dust extinction, computed from the broad-line flux ratio of Pa$\beta$ (\textit{left}) and Pa$\alpha$ (\textit{right}) to H$\alpha$. Symbols are plotted with the same color scheme as in Fig. \ref{fig:fwha-fw$NIR$}. The 1:1 is shown in solid gray.  }\label{fig:av}
	\end{figure}
	
	\subsubsection{H$\alpha$ line intensity suppression with $N_H$ in Sy 1.9}\label{sss:habias}
	Rather than affecting the width of the H$\alpha$ broad line, increasing obscuration 
	might simply diminish the entire line intensity, and therefore the flux, of the broad H$\alpha$. 
	Such line suppression would preferentially affect the rest-frame optical lines more than
	the NIR broad lines.
	Figure \ref{fig:hato$NIR$_nh} investigates whether
	there is an
	increasing trend between the optical/NIR BLR extinction and X-ray column obscuration. 
	This correlation is only marginally significant for the Pa$\alpha$ sample (p-value  $\lesssim0.05$), while it is not significant for the He~\textsc{i} and Pa$\beta$ (p-value $>$0.10 for both).
	We find similar results employing the bootstrapping and point perturbation method with the {\tt pymccorrelation} package.
	This observed trend, though  
	only weakly seen for the Pa$\alpha$, is consistent with results from the
	BASS optical DR2 
	\citep[][see their Fig. 8]{Mejia_Broadlines}, 
	where the broad H$\alpha$ to 14-150~keV X-ray luminosity ratio shows a sharp decrease of $\lesssim$1~dex at $N_H \gtrsim 10^{22}$~cm$^{-2}$ for Sy 1.9 AGN, while this ratio remains rather constant 
	up to $N_H \approx 10^{23}$~cm$^{-2}$ for Sy 1-1.5 types. 
	
	\begin{figure*}
		
		\hspace{-0.75in}\includegraphics[width=0.37\textwidth]{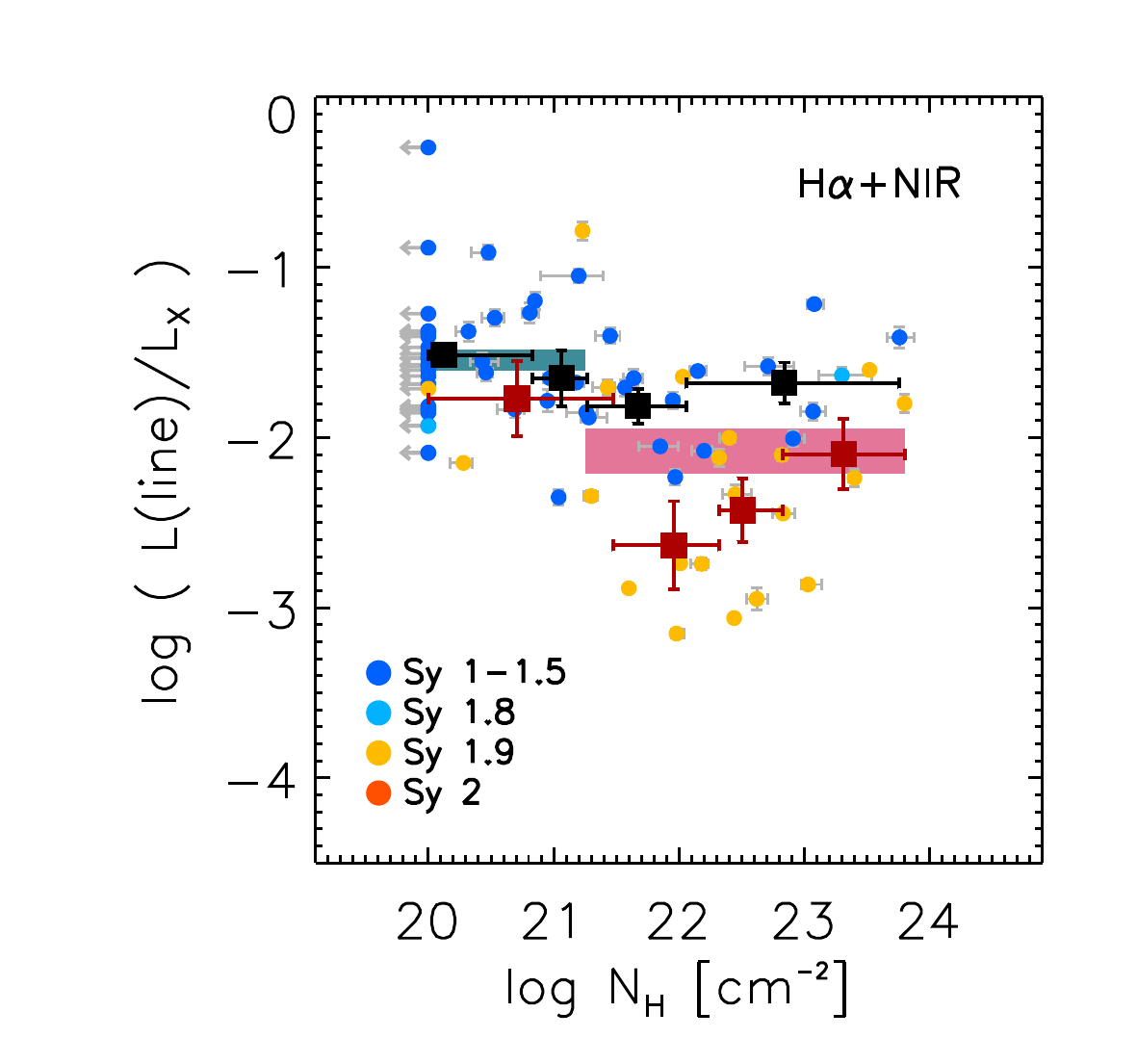}
		\hspace{-0.96in}\includegraphics[width=0.37\textwidth]{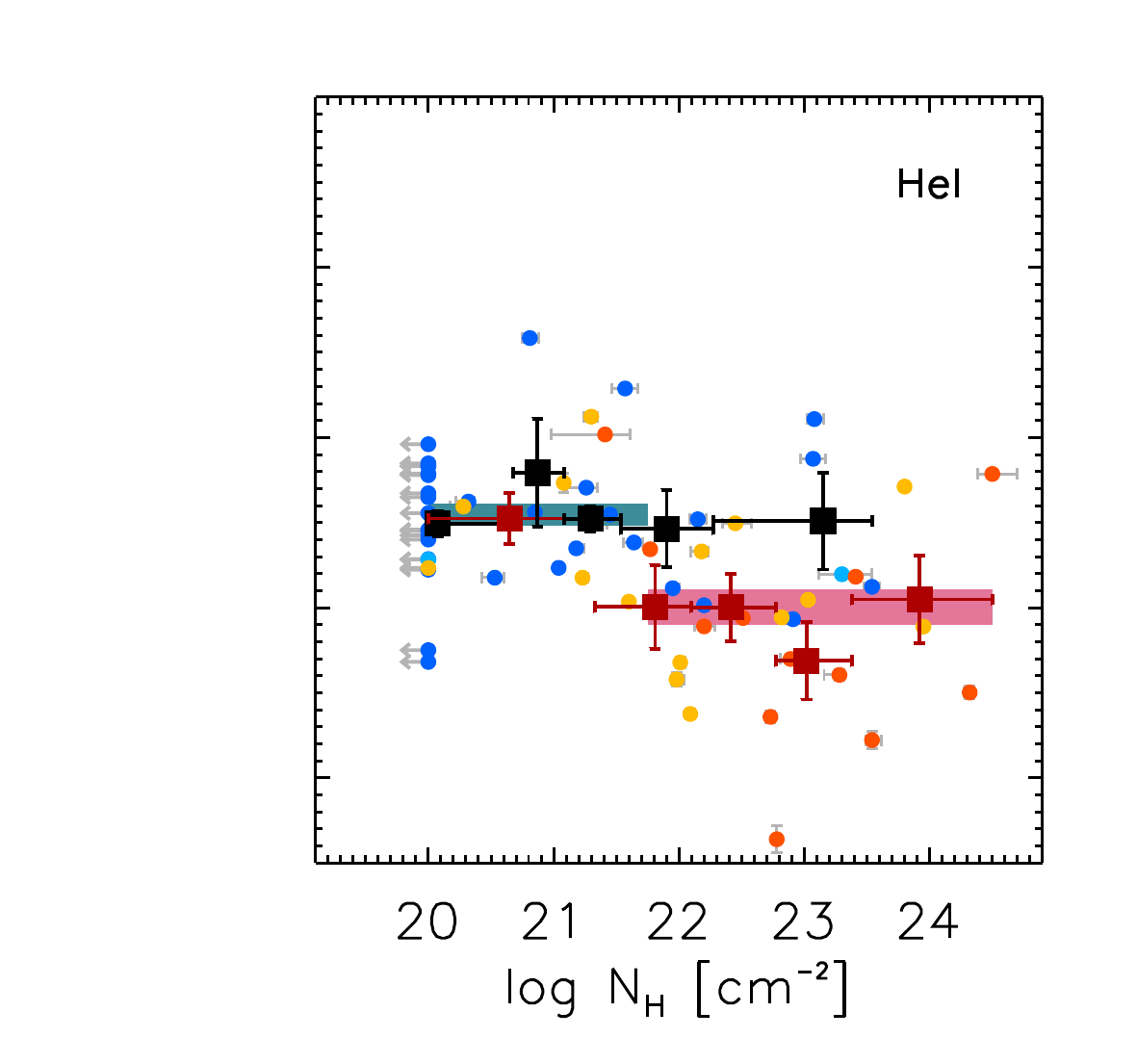}
		\hspace{-.96in}\includegraphics[width=0.37\textwidth]{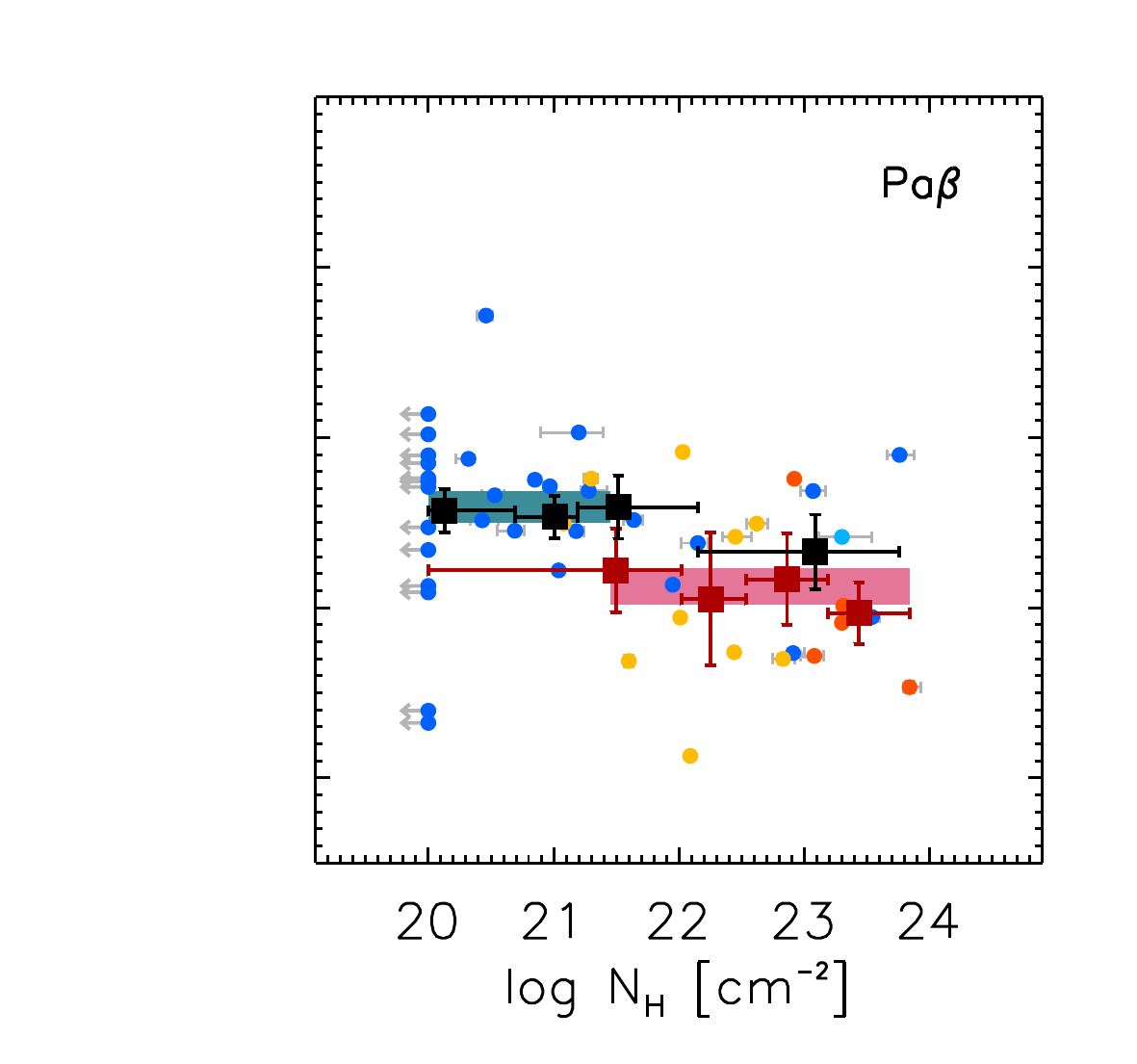}
		\hspace{-.96in}\includegraphics[width=0.37\textwidth]{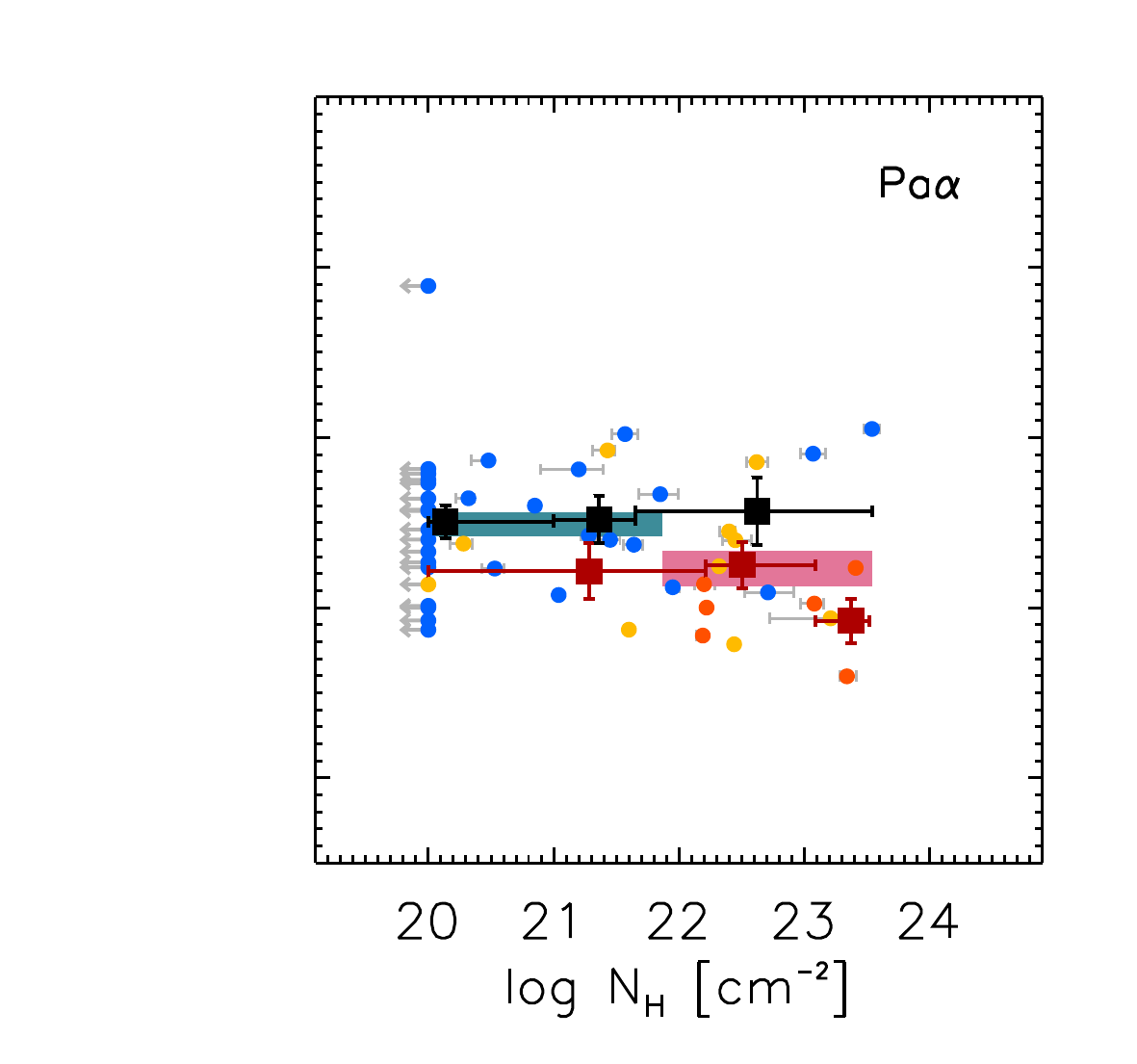}
		\caption{Ratio of the broad-line luminosities to the BAT 14-195~keV hard X-ray luminosities, from left to right: H$\alpha$, He~\textsc{i},  Pa$\beta$ and Pa$\alpha$.
			The H$\alpha$ measurements are BASS targets with 
			robustly detected NIR and H$\alpha$ broad lines (the H$\alpha$+NIR is thus only a subsample of the BASS DR2, see e.g., \citealt{Mejia_Broadlines}). 	 
			Symbols are plotted with the same color scheme as in Fig. \ref{fig:fwha-fw$NIR$}, with the addition of Sy 2 as red circles.	
			The dashed areas show the average luminosity ratio (together with the error on the mean) computed in two bins of $N_H$, low $N_H$ in green and high $N_H$ in magenta (see text for more details).
			The filled squares are the averages in quantiles of $\log N_H$ for Sy 1-1.5 (black) and Sy 1.8-1.9-2 (dark red), with bars reporting the error on the mean of the luminosity ratio and the bin size in $\log N_H$.}
		\label{fig:LirLx_nh}
	\end{figure*}

	\begin{figure*}	
		\begin{center}	
			\hspace{-0.55in}\includegraphics[width=0.37\textwidth]{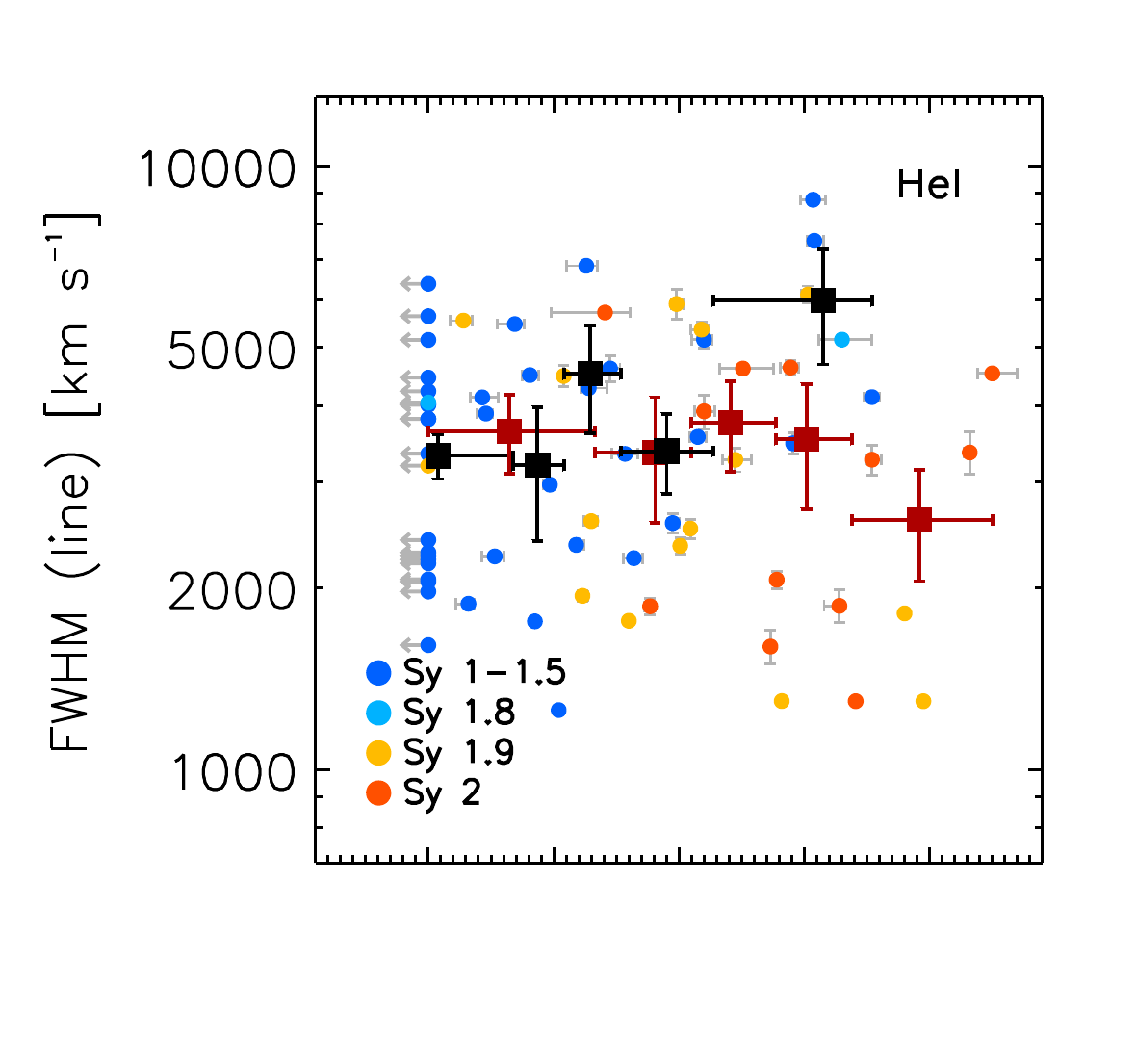}
			\hspace{-.95in}\includegraphics[width=0.37\textwidth]{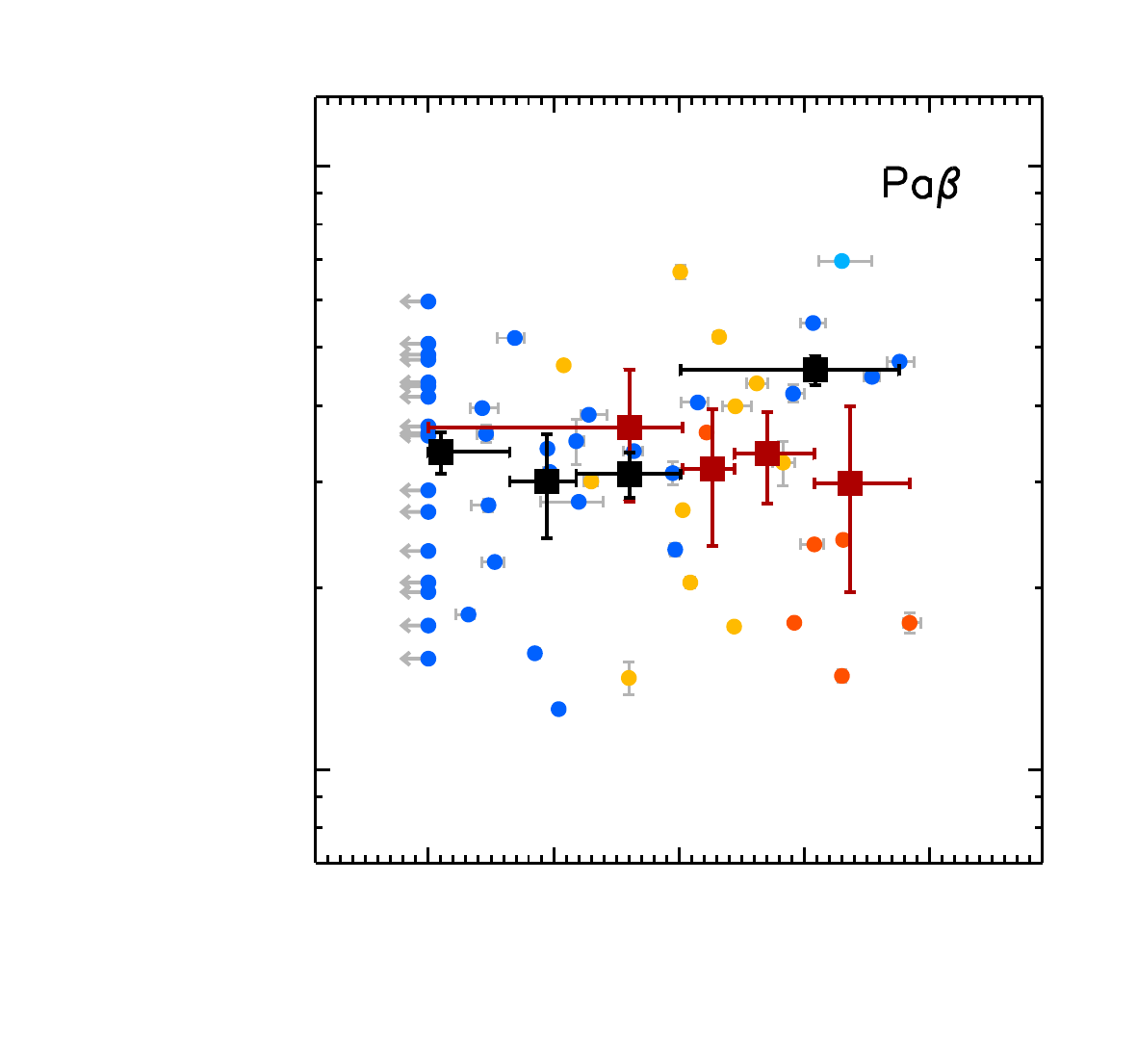}
			\hspace{-.95in}\includegraphics[width=0.37\textwidth]{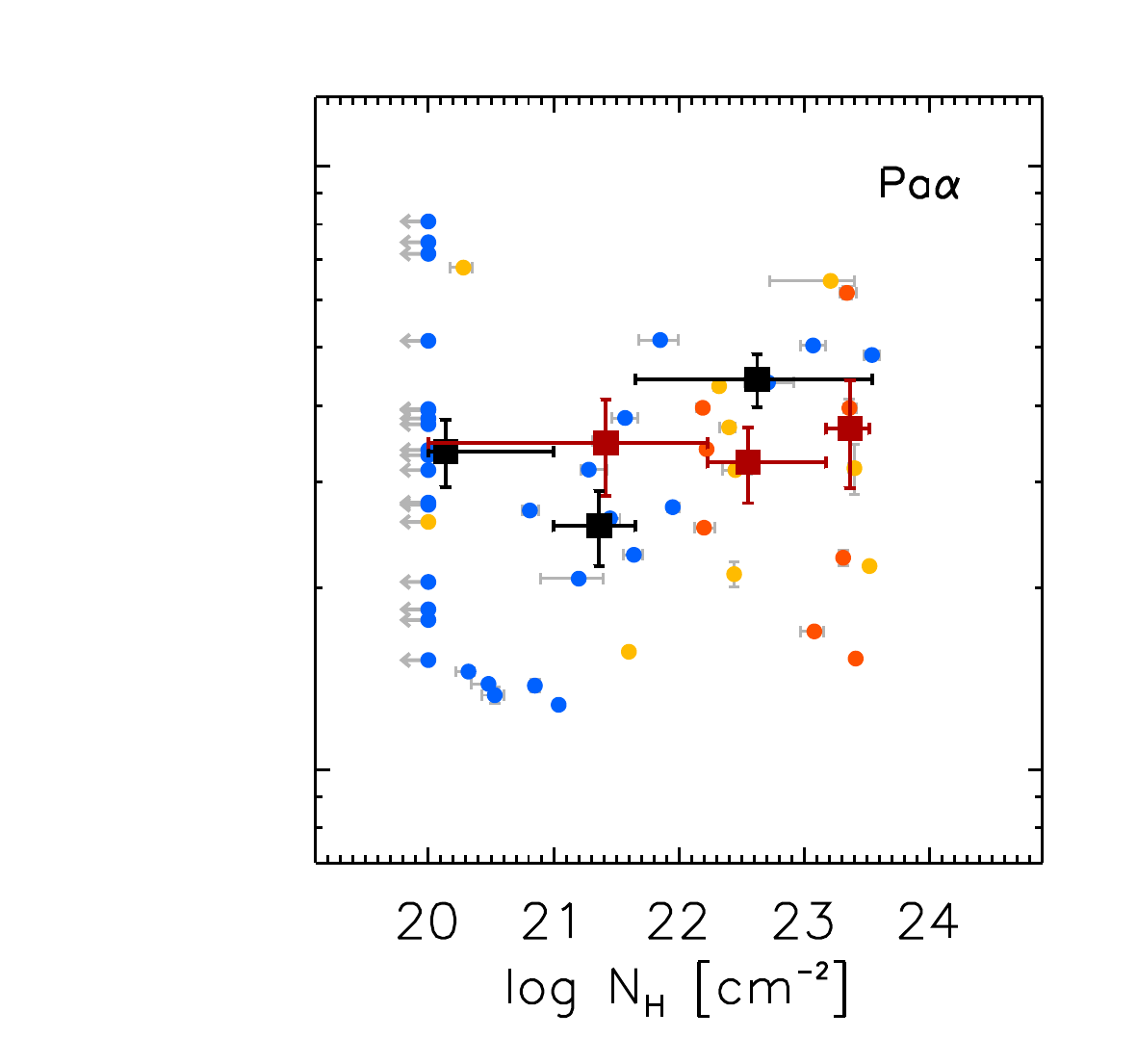}
			
			\vspace{-0.65in}\hspace{-.55in}\includegraphics[width=0.37\textwidth]{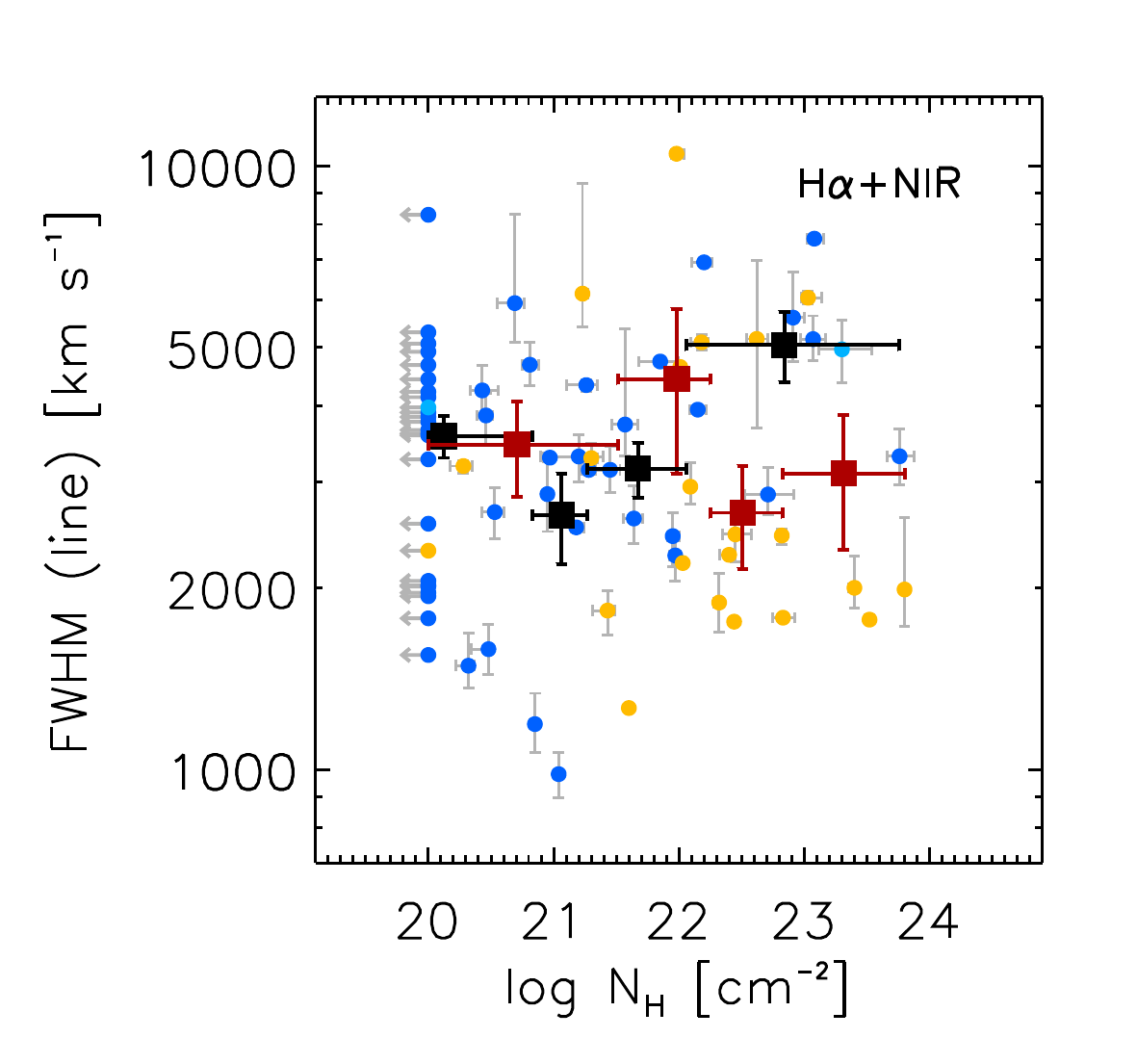}
			\hspace{-.95in}\includegraphics[width=0.37\textwidth]{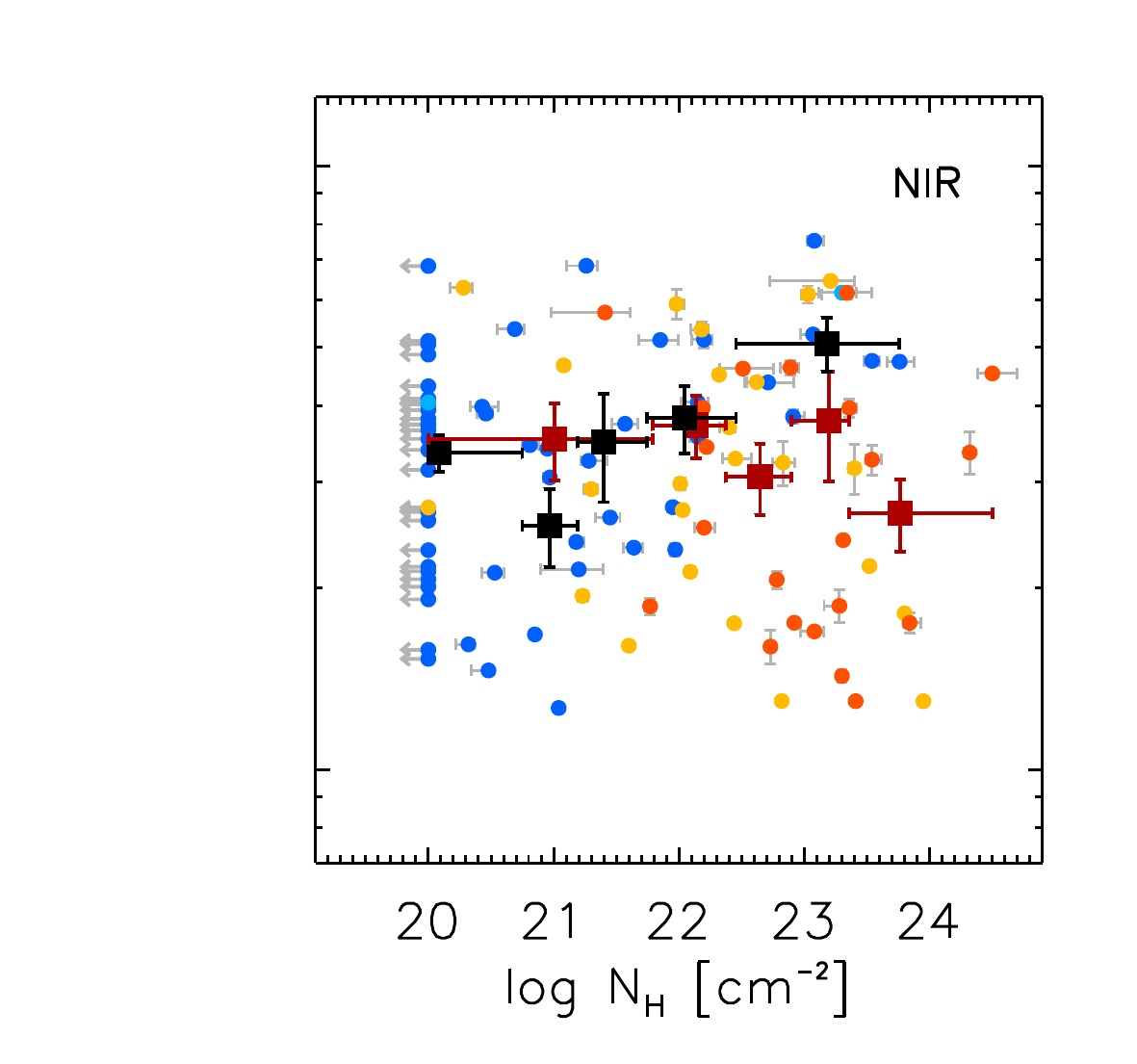}	
			\phantom{\hspace{-.95in}\includegraphics[width=0.37\textwidth]{NIR_Nh.pdf}}
		\end{center} 	
		\caption{The FWHM-$N_H$ plane for BASS AGN having reliable broad detected NIR lines as reported in the labels. 
			Symbols plotted as in Fig. \ref{fig:LirLx_nh}.}
		\label{fig:fw$NIR$_nh}
	\end{figure*} 
	
	A trend of increasing H$\alpha$ line suppression as a function of $N_H$ 
	implies that $M_{BH}$ measurements based on 
	the H$\alpha$ broad line luminosity is biased in presence of obscuration.
	Figure \ref{fig:diffMha$NIR$} shows the ratio between 
	the virial (i.e., $f_0=1$) H$\alpha$-based BH mass, [$M_{vir,H\alpha} = M_{vir}(FWHM(H\alpha), L_{H\alpha} )$], 
	and the virial NIR+$L_X$-based BH mass for each NIR line, [$M_{vir,line} = M_{vir}(FWHM(line), L_X)$], 
	as a function of the X-ray obscuration (left) and broad lines flux ratios, used as proxy for the
	BLR extinction (right). 
	The difference between the BH mass estimates 
	increases with increasing obscuration and extinction, more prominently for Sy 1.9 AGN,
	indicating that the H$\alpha$-based BH mass is biased low in presence of obscuration.
	The trend expected if the NIR-based BH mass was underestimating the true virial $M_{BH}$
	at a given H$\alpha$-based BH mass would be positive with increasing $N_{H}$
	and increasing BLR extinction, while the opposite trend is observed. 
	
	By dividing the sample in two bins of obscuring column and running a Student’s T-statistic test on the values in the two bins, 
	we find a statistically significant decrement of 
	$\Delta=0.23\pm0.14$ ($0.44\pm0.17$, $0.46\pm0.13$)~dex with p-value of $4\times10^{-2}$ ($1.7\times10^{-4}$, $1.3\times10^{-3}$) 
	in the BH mass difference between H$\alpha$ and He~\textsc{i} (Pa$\beta$, Pa$\alpha$) in the two groups at low and high $N_H$.
	There is a significant bias when the H$\alpha$ broad line luminosity is used in SE BH mass estimates in presence of obscuration $N_H>10^{21}$~cm$^{-2}$, as can bee seen in the two shaded bins in the left panels of Fig. \ref{fig:diffMha$NIR$} that report the average BH mass difference in the $N_H$ bin together with the uncertainty on the mean. 
	Indeed the presence of a bias for the H$\alpha$-based BH masses can be seen in the dashed black histogram of the BH mass differences, that are skewed to high (negative) values.

	When the same prescription is adopted to measure the optical BH masses, i.e., using the 
	hard X-ray luminosity instead of the broad H$\alpha$ luminosity as a proxy of the BLR radius
	(see the red stars for Sy 1.9),
	the difference 
	between the BH masses derived in the optical and NIR is consistent 
	with scatter around zero offset (see red solid histograms in Fig. \ref{fig:diffMha$NIR$}), with no obvious trend versus
	obscuration or BLR extinction. 
	Therefore, we can conclude that a mixed BH mass estimate based on the hard X-ray luminosity, as put forward by \citet[][see also \citealt{bongiorno14,LF15}]{ricci17a}, can 
	overcome the biases due to extinction and obscuration of the optical H$\alpha$ broad line in Sy 1.9 and in galaxy-dominated AGN, 
	where the UV/optical continuum emission can be diluted by the host starlight. 
	Whereas, as long as a broad line is reliably detected and the hard X-ray luminosity is higher than what is expected from the emission coming from X-ray binaries and hot extended gas in the host galaxy, the mixed BH mass estimator can be used to get an unbiased BH mass measurement. 
	We note that adopting $L_X$ instead of the UV/optical luminosity may introduce a weak dependence on $M_{BH}$ and/or $L_{Edd}$ \citep[see][albeit there is significant scatter in the correlations]{liu21,lusso10}. However, the extremely tight ($\sim$0.2 dex) non-linear $L_X$-$L_{UV}$ relation has been extensively used to compute Hubble diagrams for quasars, after accounting for flux-limit related biases and testing for additional systematics \citep[][and references therein]{lusso20}. Further investigation on possible $L_{Edd}$ dependencies will be addressed in a forthcoming paper (F. Ricci et al. in prep).
	
	\subsection{Dust extinction toward the BLR and gas absorption in the X-rays}\label{ss:av}
	The central and right panels of Figs. \ref{fig:ratio_hato$NIR$} and \ref{fig:hato$NIR$_nh} report the 
	expected value of the H$\alpha$/Pa$\beta$ and H$\alpha$/Pa$\alpha$ flux
	ratio for case B recombination at $T=10^4$~K and $n = 10^6$~cm$^{-3}$ \citep{osterbrock06}. 
	We see a wide range of these ratios, with the Paschen lines
	ruling out case B recombination in most of the X-ray obscured objects,
	\citep[see also][]{soifer04,glikman06,riffel06,LF15}.
	
	If anyhow we assume that the Paschen lines respect the case B recombination, we can use 
	those broad-line flux ratios to quantify the amount of extinction toward the BLR $A_V(BLR)$. 
	Then we could check whether the decline by dust extinction observed in Fig. \ref{fig:hato$NIR$_nh} is consistent with the gas absorption, by assuming the Galactic dust-to-gas ratio. 
	
	We compute the dust extinction toward the BLR $A_V(BLR)$ following \citet[][see their Eq. 3]{dominguez13}: we assumed the reddening curve $k(\lambda)$ from \citet{calzetti00}, and that the relationship between
	the nebular emission line color excess and the Paschen-to-H$\alpha$
	decrement is given by
	\begin{equation}
		\begin{split}
			E(B-V) &= \frac{E(H\alpha - Pa)}{k(\lambda_{H\alpha}) - k(\lambda_{Pa})} =  \\
			& = \frac{2.5}{k(\lambda_{H\alpha}) - k(\lambda_{Pa})} \log \left[ \frac{(Pa/H\alpha)_{obs}}{(Pa/H\alpha)_{case\,B}}\right]
		\end{split}
	\end{equation}
	where $E(H\alpha - Pa)$ is the $E(H\alpha - Pa\beta)$ and $E(H\alpha - Pa\alpha)$ excess, the $k(\lambda_{H\alpha})$ and $k(\lambda_{Pa})$ are the \citet{calzetti00} reddening curve 
	evaluated at the H$\alpha$, Pa$\beta$ and Pa$\alpha$ wavelengths and (Pa/H$\alpha$) 
	are the broad-line flux ratio Pa$\beta$/H$\alpha$ and Pa$\alpha$/H$\alpha$. 
	We also assume that $A_V(BLR) = 3.1 \times E(B-V)$.
	We then calculate the $A_V(N_H)$ from the Galactic dust-to-gas ratio, $N_H/A_V=2.69\times 10^{21}$~cm$^{-2}$ \citep{nowak12}.
	The comparison between the two independent $A_V$ estimates is presented in Fig. \ref{fig:av}.
	Figure \ref{fig:av} shows a separation with Seyfert subclasses:
	most of the Sy 1s lie above the 1:1, thus the $A_V(N_H)$ is somewhat underestimating the BLR extinction, whereas Sy 1.9s have $A_V(BLR) \sim 1-5$ and their $A_V(BLR)$ are mostly below the 1:1, thus the $A_V$ derived from dust extinction is less than what is expected from gas absorption.
	This result is in agreement with what has been found in \citet[][see, their Fig. 3]{shimizu18}, where the $A_V(BLR)$, estimated from the broad H$\alpha$-to-hard X luminosity ratio, in Sy 1.9 is lower than what is expected from the Galactic dust-to-gas ratio, particularly for  $A_V(BLR)>3$.
	We note that we adopt a single Galactic dust-to-gas ratio of $2.69\times 10^{21}$~cm$^{-2}$, while in the literature it spans a range from $1.79\times 10^{21}$~cm$^{-2}$ \citep{predehl95} to the value we assume. 
	Furthermore, there is evidence that in AGN the dust-to-gas ratio is not Galactic \citep{maiolino01}. If we adopt the average dust-to-gas ratio from \citet{maiolino01}, e.g., $1.1\times10^{22}$~cm$^{-2}$, then the derived $A_V(N_H)$ would be a factor $1.1/0.269\simeq4.1$ smaller than what is shown in Fig. \ref{fig:av}.
	
	Figure \ref{fig:av} suggests that the extinction towards the BLR and line-of-sight X-ray $N_H$ are distinct, i.e. they are due to separate obscuration medium. This difference might arise because the BLR extinction is  a ``global'' diffuse measurement while the $N_H$ is a more ``local'' line-of-sight measurement, or if there is gas within the BLR.
	Alternatively, it might suggest that the dust-to-gas ratio is somehow different from the Galactic value for BASS AGN.
	Given the large spread in Galactic dust-to-gas ratio and the uncertainty in what is the $A_V/N_H$ in AGN environments, we cannot distinguish between these two possibilities.
	
	\subsection{Near-infrared line suppression with obscuration in  Sy 1.8-1.9-2 types?}\label{sec:$NIR$nh}
	We might then ask whether the same effect of line luminosity suppression is similarly experienced by 
	Paschen and helium lines in the NIR as observed for the broad H$\alpha$ line.
	Figure \ref{fig:LirLx_nh} shows the ratio of the broad 
	H$\alpha$ (\textit{left}) and 
	NIR emission lines to the BAT 14-195~keV hard X-ray luminosity as
	a function of the X-ray column density for all BASS targets with reliable NIR broad-line detections and with
	available $N_H$. 
	There is a decreasing trend of the NIR to X-ray luminosity ratio with increasing 
	X-ray obscuration, as similarly observed for the H$\alpha$.
	Binning the data in quantiles of $\log N_H$ and splitting the sample into Sy 1-1.5 (black squares) and 
	Sy 1.8-1.9-2 (dark red squares), Sy 1-1.5 AGN exhibit a roughly constant average $L(line)/L_X$ 
	across the $N_H$ range probed by our sample, 
	while the $L(line)/L_X$ 
	of Sy 1.8-.9-2 types seems to decrease as $N_H$ grows. This effect seems particularly 
	more prominent at shorter wavelengths.
    Running a Student’s T-statistic test 
	we find a statistically significant decrement of $\Delta=0.54\pm0.15$ ($0.54\pm0.12$, $0.46\pm0.14$, $0.27\pm0.12$)~dex with p-value of $6.8\times10^{-6}$
	($4.9\times10^{-5}$, $2.1\times10^{-3}$, $3.3\times10^{-2}$)
	in the 
	H$\alpha$ (He~\textsc{i}, Pa$\beta$, Pa$\alpha$) to hard X-ray luminosity ratio between the two groups 
	divided at $\log (N_H/\rm cm^{-2})=21.25$ (21.75, 21.45, 21.85), as can be seen in the two shaded rectangles in Fig. \ref{fig:LirLx_nh} that report the average luminosity ratio in the two $N_H$ bins together with the uncertainty on the mean. 
	The decrement observed in the H$\alpha$ to X-ray luminosity can explain the bias observed in Fig. \ref{fig:diffMha$NIR$}. 
	We note that the decrement in the H$\alpha$ to X-ray luminosity happens at a column density level that is slightly lower than what observed in the other NIR lines, which might be explained considering that dust attenuation diminishes when moving to longer wavelengths.   
	Near-infrared line luminosity, when available, should be preferred to the H$\alpha$ broad line luminosity as a proxy for the BLR radius when estimating $M_{BH}$ using single-epoch techniques. 
	\begin{figure*}
		\begin{center}
			\hspace{-0.65in}\includegraphics[width=0.34\textwidth]{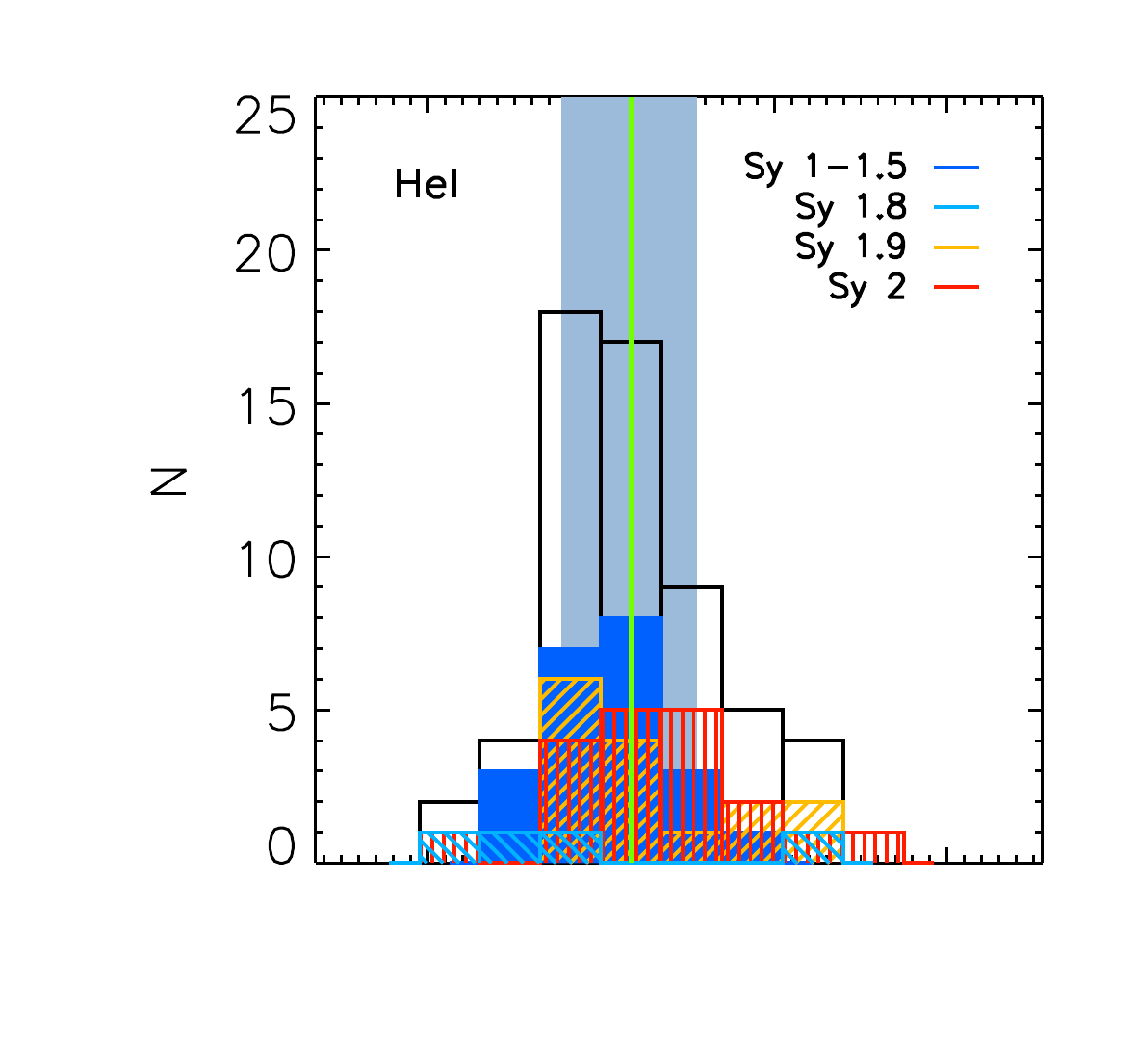}
			\hspace{-.91in}\includegraphics[width=0.34\textwidth]{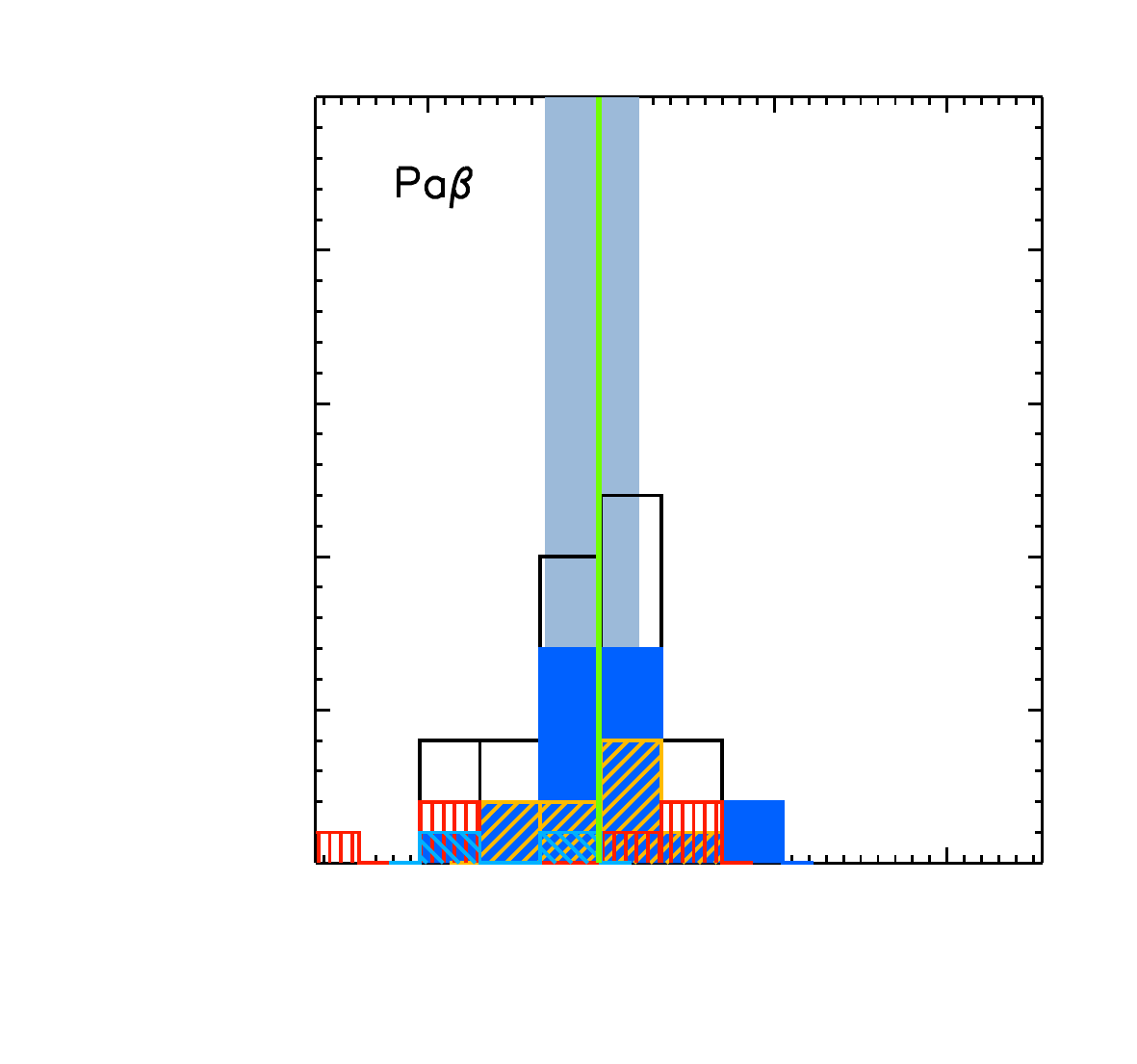}
			
			\vspace{-0.64in}
			\hspace{-.65in}\includegraphics[width=0.34\textwidth]{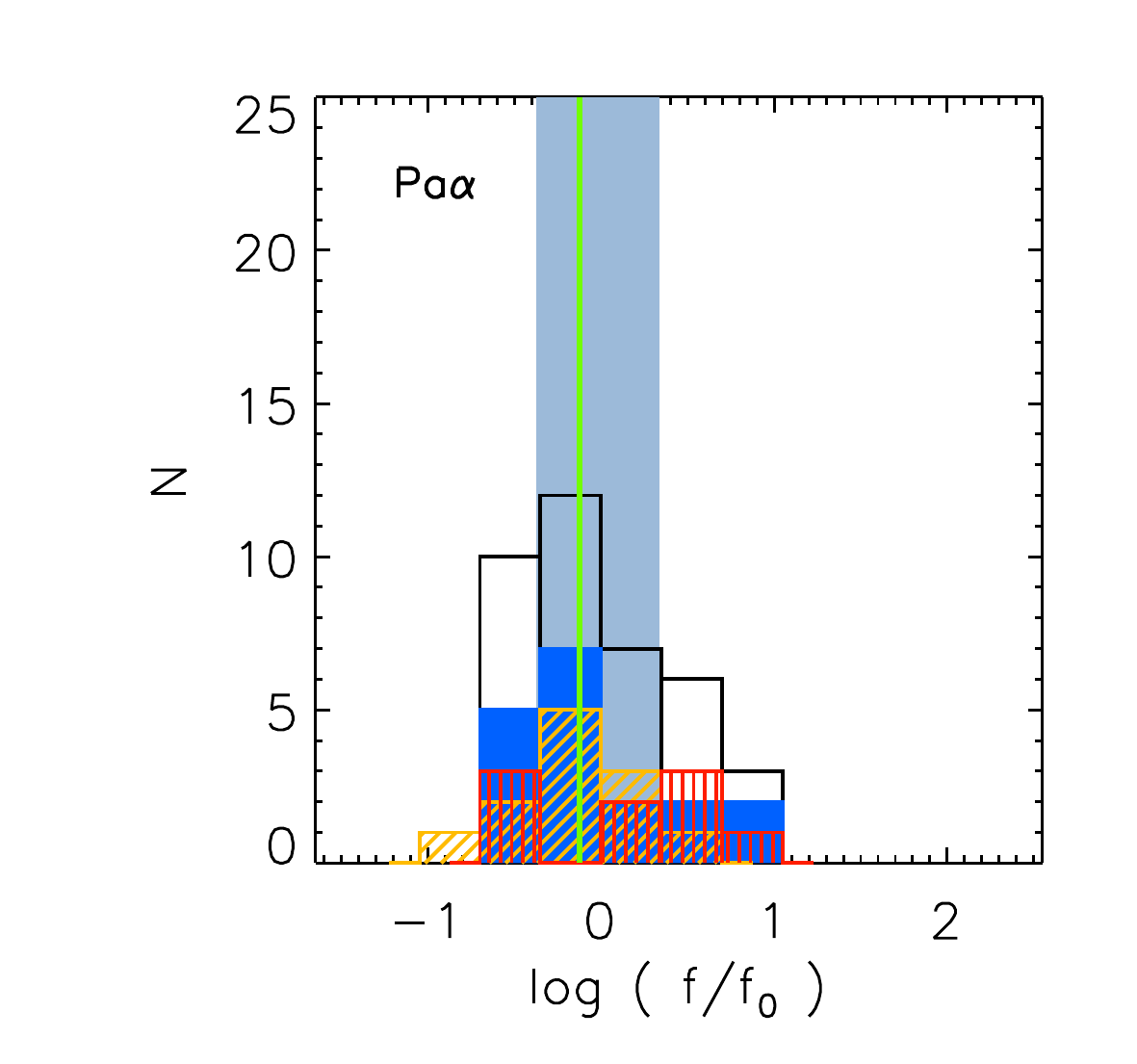}
			\hspace{-.91in}\includegraphics[width=0.34\textwidth]{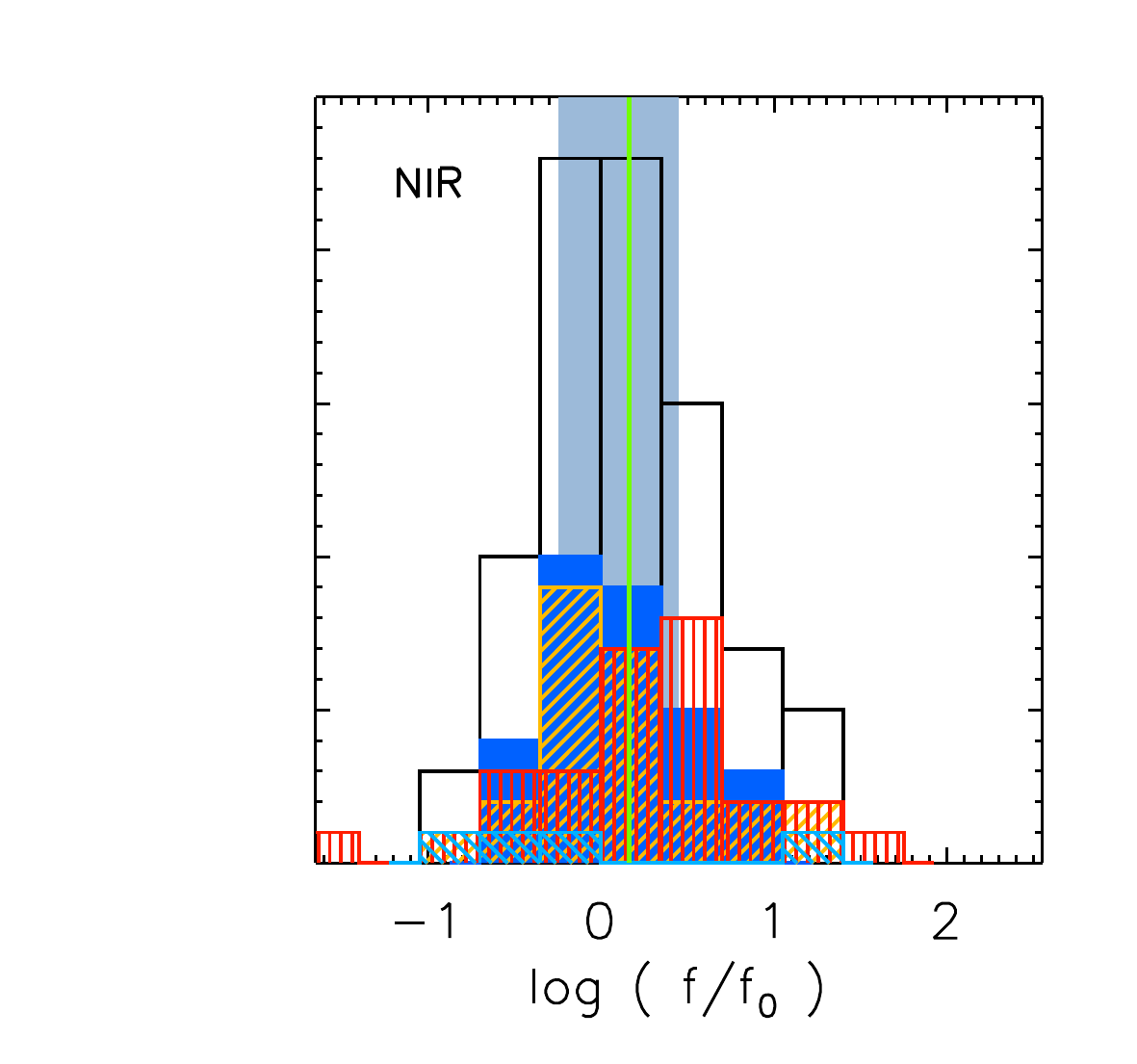}
		\end{center}
		
		\caption{Histograms of the normalized virial factor $f/f_0$, i.e. BH mass deviation $M_{BH,\sigma_\star}/M(line)$, for each line subsample, as labeled 
			in each panel.
			Total sample in black, Sy 1-1.5 in blue, Sy 1.8 in cyan, Sy 1.9 in orange and Sy 2 in red.
			The shaded area defines the $IQR$, that is the region enclosing the 25th to the 75th percentile of the total distribution. The 50th percentile is overplotted with a solid green vertical line. }
		\label{fig:histo}
	\end{figure*}	
	
	Finally, the last question we ask is the following: is NIR really penetrating into the BLR or is the NIR BLR velocity estimate 
	as good as the H$\alpha$ (e.g., see Figs. \ref{fig:fwha-fw$NIR$}, \ref{fig:ratio_nh} and \ref{fig:ratio_hato$NIR$}) just because both are biased in the same way with increasing column density?
	Figure \ref{fig:fw$NIR$_nh} can help address this, showing the FWHM-$N_H$ plane 
	for all the BASS Sy AGN having reliable NIR broad line detection and available $N_H$. 
	The H$\alpha$ BLR velocity estimate is also reported for the subsample of the optical BASS DR2 with $N_H$ and both NIR and H$\alpha$ reliable broad-line detections.
	When binning in quantiles in $\log N_H$, the average FWHM measured in the NIR 
	in Sy 1.8-1.9-2 remains quite constant across the $N_H$ range, with only a slight decrease in the highest $N_H$ bin ($\log (N_H/\rm{cm}^{-2}) \gtrsim 23.5$); the error bars are fully consistent with a constant trend versus obscuration. 
	This behaviour is in agreement with the results presented in \citet{onori17b}, who used a much smaller sample of local hard X-ray BAT selected obscured AGN (N=17).
	Also among Sy 1-1.5 types the average FWHM is rather constant, given the uncertainties, until $\log  (N_H/\rm{cm}^{-2}) \simeq 22$, while in the last $N_H$ bin there is an upturn. We note that a similar behaviour is observed as well in the optical BASS DR2 in the FWHM(H$\beta$) - $N_H$ plane (see, Fig. 9 of \citealt{Mejia_Broadlines}).
	This upturn behaviour in Sy 1-1.5 might be related just to small sample statistics, since type 1 AGN with high $N_H$ are quite rare. However taken at face value, we can speculate that it might be related to a transition in Sy type when the FWHM is smaller than 4000~\kms and $N_H\gtrsim10^{22}$~cm$^{-2}$. At high X-ray columns and high inclinations, the optical photons might in some cases still find their way through the obscuring medium by experiencing multiple scatterings. If this hypothesis were plausible, the fraction of optical polarized light in this particular sample of Sy 1 with high $N_H$ should be higher than what is usually found in normal Sy.

	From Figs. \ref{fig:LirLx_nh}-\ref{fig:fw$NIR$_nh} we conclude that 1) the H$\alpha$ and near-infrared FWHMs are not affected by obscuration, at least up to $\log  (N_H/\rm{cm}^{-2})\approx23.5$, and 2) the near-infrared line luminosities are not as strongly extincted as the 
	H$\alpha$ and can be used in SE BH mass measurements until $\log  (N_H/\rm{cm}^{-2})\approx21.5 - 22$, depending on the specific line chosen.

	\subsection{The virial factor \f }\label{ss:fvir}
	We finally verify if the BH mass estimates obtained from two independent methods, i.e., the NIR+$L_X$ based BH mass $M(line)$ and the $\sigma_\star$-based BH mass $M_{BH,\sigma_{\star}}$, 
	are consistent and, if not, what are the quantities possibly driving the difference.

	\begin{figure*}
		\vspace{-2.7in}
		\includegraphics[width=1.5\textwidth]{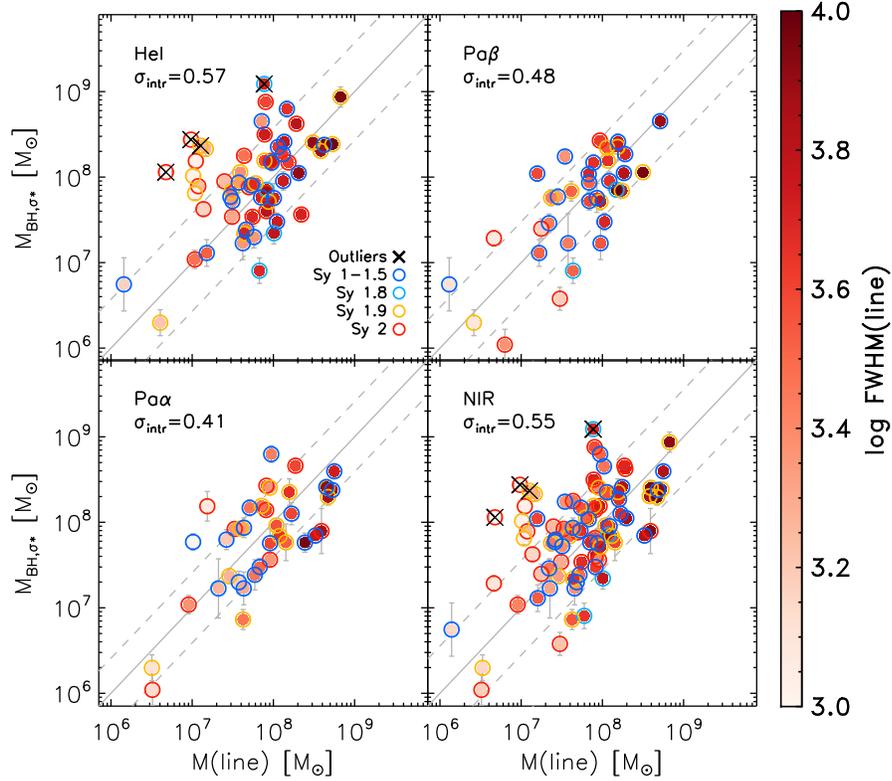}
		
		\vspace{-6.8in}
		\caption{Comparison of $M_{BH,\sigma_\star}$ vs $M(line)$ in BASS Sy AGN color-coded according to the FWHM(line).
			The 1:1 relation is shown in solid gray, with dashed gray lines indicating the intrinsic spread.
			The measurements are evenly distributed around the one-to-one relation, with a clear gradient in FWHM(line): objects with faster BLR velocities are preferentially below the unity locus, while smaller FWHMs are mostly located above the 1:1 relation, in all the lines examined. }
		
		\label{fig:MsMl}
	\end{figure*}

	The resulting sample is composed of 
	60, 37, 39 and 88 local BASS Sy having optical $\sigma_\star$ 
	\citep{Koss_DR2_sigs,Caglar_DR2_Msigma}
	and He~\textsc{i}, Pa$\beta$, Pa$\alpha$ and $NIR$ robustly detected broad lines available, respectively. 
	We were able to gather optical $\sigma_\star$ measurements inside BASS DR2 for $\sim$62-68\% of the sample with near-infrared broad line detections, as there are only 30, 23, 24 and 41 sources without available optical $\sigma_\star$ for the He~\textsc{i}, Pa$\beta$, Pa$\alpha$ and $NIR$ sample, respectively.
	We show in Fig. \ref{fig:histo} the histograms of the normalized virial factor $f/f_0$ which 
	is basically the ratio $M_{BH,\sigma_{\star}}/M(line)$.
	Each panel reports the distribution of $f/f_0$ separately for each line subsample, color coded according to 
	the Sy classification as shown in the legend. 
	The vertical green line marks the 50th percentile, and the shaded area encloses the 25th to 75th percentiles of the whole sample (black histogram), also known as interquartile range $IQR$. 
	The histograms are all roughly centered around 0, meaning that the normalized virial factor is of the order of unity, with a distribution ranging from -1 to 1. Some sources show high $f/f_0$ values, e.g., $1<\log f/f_0<2$ and $-2<\log f/f_0<-1$ , being considered as outliers.
	These sources have the ratio of BH masses deviating from each other more than one order of magnitude. 
	We discuss the possible reasons why these sources are outliers in Appendix \ref{sec:app-outliers}.
	In all the subsequent analysis and plots, these most deviating objects are reported as black crosses, and are 
	omitted when performing linear regression fits. 
	Table \ref{tab:sample} lists the broad line measurements of the samples and ancillary quantities.
	\begin{figure*}
		\vspace{-0.3in}
		\begin{center}
			\hspace*{-0.25in}\includegraphics[width=\textwidth, angle=-90]{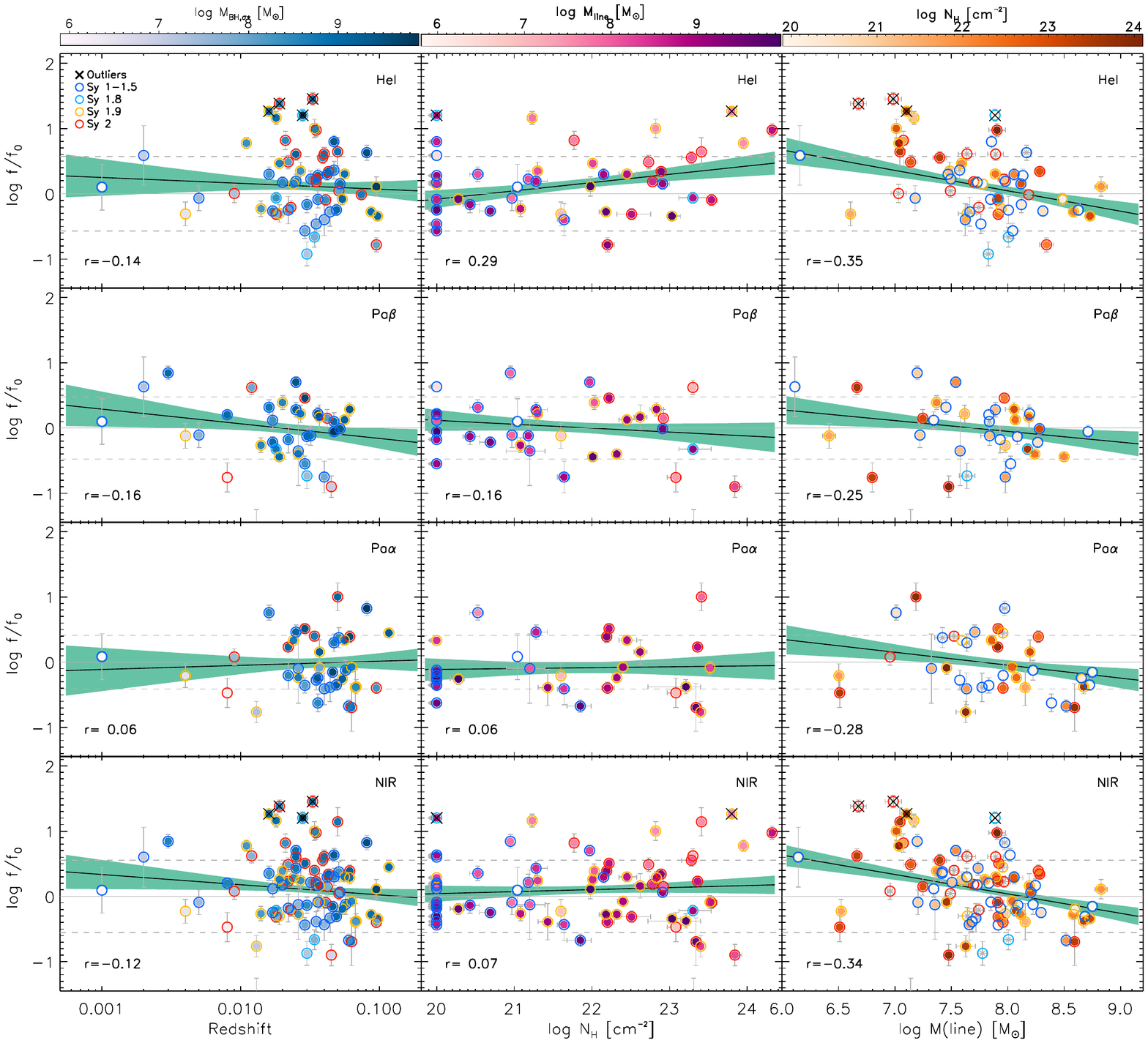}
		\end{center}

		\vspace{-.35in}
		\caption{Comparison between the offset $M_{BH,\sigma_\star}/M(line)$, i.e., the normalized virial factor $f/f_0$, as a function of: (\textit{left}) redshift, color-coded with $M_{BH,\sigma_\star}$; (\textit{center}) $N_H$, color-coded according to $M(line)$; (\textit{right}) $M(line)$, color-coded according to the $N_H$ (when available, if not available the symbols are gray asterisks). 
			The row shows the sample of each line, from the top: He~\textsc{i}; Pa$\beta$; Pa$\alpha$; average $NIR$.
			The color of the outer circle labels different Sy classes, as reported in the legend. Black crosses mark outliers. 
			Black solid lines refer to log-log linear best-fit regressions, with relative 68\% c.l. in green. Pearson correlation coefficients $r$ are reported in the lower left of each panel, see also Tab. \ref{tab:offset-stat}.}
		\label{fig:offset}
	\end{figure*}
	
	In Fig. \ref{fig:MsMl} we compare $M_{BH,\sigma_{\star}}$ and $M(line)$. The panels show separately the BH mass comparison
	for each reliably detected line, e.g., He~\textsc{i} (top-left panel), Pa$\beta$ (top-right panel), Pa$\alpha$ (bottom-left panel), and for the average $NIR$ FWHM (bottom-right panel). 
	The sample is divided according to Sy classification. 
	Each panel also shows the sample color-coded according to the measured broad FWHM of each considered line. 
	The measurements are evenly distributed around the 1:1 relation (solid gray line in Fig. \ref{fig:MsMl}) with some intrinsic scatter of $\sim 0.4-0.57$~dex (marked with dashed lines in the figure). 
	
	As can be seen from the color gradient in Fig. \ref{fig:MsMl}, some of the scatter between $M_{BH,\sigma_{\star}}$ and $M(line)$ estimates might be driven by the observed FWHM; smaller FWHMs (lighter pink) are located above the 1:1 locus while broader lines (darker red) are below it. This difference in the BH mass estimates, i.e., the virial factor \f, 
	is expected in case the FWHM gets broadened with inclination \citep[see, e.g.,][]{collin06,decarli08,shenho14,mejia18}.

	We now investigate whether the scatter between the two different mass estimates depends on additional variables by examining the 
	ratio between the two mass estimates, $M_{BH,\sigma_{\star}}/M(line)$, as a function of either redshift (left panels in Fig. \ref{fig:offset}), X-ray column $N_H$ (center panels in Fig. \ref{fig:offset}) or BH mass estimate based on the virial assumption $M(line)$ (right panels in Fig. \ref{fig:offset}). 
	In all panels, the solid gray line shows the 1:1 zeropoint and the dashed gray lines are indicating the $\pm1\sigma$ scatter computed earlier. Color bars in each panel shows the values of 
	$M_{BH,\sigma_{\star}}$ (left), $M(line)$ (center) and 
	column density $N_H$ (right) when available. 
	The rows in Fig. \ref{fig:offset} show $f/f_0$ for each line sample, respectively, from top to bottom: 
	He~\textsc{i}, Pa$\beta$, Pa$\alpha$ and average $NIR$.
	In order to quantify the presence of possible correlations between $f/f_0$ and other physical quantities, we perform a forward 
	regression Bayesian fit as outlined in Eq. \ref{eq:offset}, 
	where $x$ is either redshift, column density or the NIR+$L_X$ virial-based BH mass estimate $M(line)$ and $y/y_0$ is $f/f_0$.
	Table \ref{tab:offset-stat} reports the best-fit parameters obtained using the IDL routine {\tt linmix\_err}, 
	and solid black lines in Fig. \ref{fig:offset} show the log-log linear best fit regression with 1$\sigma$ c.l. In each panel the Pearson correlation coefficient is also reported.
	
	\begin{deluxetable*}{lcccccc}  %%%%%%%%%%%%%%%%%%%%%%%%%%%%%%%%%%%%%%%%%%%%%%%%%%%%%%%%%%%%%%%%%%%%%%%
		\tablecaption{Best fit log-log linear regression as outlined in Eq. \ref{eq:offset}.\label{tab:offset-stat}}
		\tablehead{
			\colhead{Sample} &
			\multicolumn{5}{c}{$\log z$} &
			\colhead{} \\
			\colhead{}	& 
			\colhead{$\alpha$} & 
			\colhead{$\beta$} & 
			\colhead{r} & 
			\colhead{P(r)}&  
			\colhead{p-value ($\beta\ne$ 0) }&
			\colhead{}\\
			\colhead{(1)}&\colhead{(2)}&\colhead{(3)}&\colhead{(4)}&\colhead{(5)}&\colhead{(6)}&\colhead{}
		}
		\startdata
		He~\textsc{i} &-0.02$\pm$0.29&	-0.09$\pm$0.19&	-0.14&	3.1E-01	&6.3E-01 & \\
		Pa$\beta$&-0.38$\pm$0.34&	-0.22$\pm$0.20&	-0.16	&	3.5E-01	&2.6E-01& \\
		Pa$\alpha$&0.08$\pm$0.32&	0.06$\pm$0.21&	0.065	&	6.9E-01	&7.6E-01& \\
		$NIR$ &-0.13$\pm$0.23&	-0.16$\pm$0.15&	-0.12		&	2.6E-01	&2.9E-01&\\
		\hline		
		\hline
		Sample & \multicolumn{5}{c}{$\log \dfrac{N_H}{10^{22}~{\rm cm^{-2}}}$} & \\
		& $\alpha$ & $\beta$ & r & P(r)& p-value ($\beta\ne$ 0) & \\
		(1) & (2) & (3) & (4) & (5)& (6) &\\
		\hline
		He~\textsc{i} &0.171$\pm$0.071&	0.127$\pm$0.065&	0.29&	5.6E-02		&5.5E-02 & \\
		Pa$\beta$& -0.003$\pm$0.081&	-0.060$\pm$0.081&	-0.16&	3.6E-01		&4.7E-01& \\
		Pa$\alpha$& -0.084$\pm$0.084&	-0.014$\pm$0.079&	-0.062&	7.3E-01		&8.6E-01& \\
		$NIR$ & 0.104$\pm$0.058&	0.031$\pm$0.054&	0.066&	5.9E-01				&5.7E-01&\\
		\hline		
		\hline
		Sample & \multicolumn{5}{c}{$\log \dfrac{M(line)}{10^{7.5}~M_\odot}$} &\\
		& $\alpha$ & $\beta$ & r & P(r)& p-value ($\beta\ne$ 0)& p-value ($\beta\ne$ -1) \\
		(1) & (2) & (3) & (4) & (5)& (6) & (7)\\
		\hline
		He~\textsc{i} & 0.202$\pm$0.066&	-0.31$\pm$0.11&	-0.35&	7.3E-03			&6.3E-03&	6.8E-08\\
		Pa$\beta$& 0.031$\pm$0.078&	-0.16$\pm$0.12&	-0.25&	1.4E-01					&1.8E-01&	5.6E-08\\
		Pa$\alpha$& 0.053$\pm$0.083&	-0.20$\pm$0.12&	-0.28&	8.9E-02				&9.7E-02&	4.0E-08\\
		$NIR$ &0.187$\pm$0.053&	-0.297$\pm$0.088&	-0.34&	1.5E-03					&1.2E-03&	9.7E-12\\
		\enddata
		\tablenotetext{}{Columns are: (1) sample of each line; (2-3) the zero point and slopes of the best-fit relations; (4-5) the Pearson correlation coefficients with the related probability; 
			(6) the probability of the slople being different from zero; (7) same test with the null hypothesis of $\beta=-1$, carried out only for the $f/f_0$ vs $M(line)$ correlation.}
	\end{deluxetable*}
	
	At fixed aperture, with increasing redshift a bigger part of the host (bulge) is sampled in 1D spectra, 
	possibly producing an increase on the measured stellar velocity dispersion and therefore an enhancement of $M_{BH, \sigma_\star}$, particularly relevant in late-type systems \citep{falcon-barroso17}. The BAT AGN are typically found in massive spirals with strong bulges (with high concentration index) in between spirals and ellipticals \citep{koss11}. As such, the aperture corrections go in different directions and it is not straightforward to determine what is the appropriate aperture correction to apply \citep[this issue is further explored by][]{Caglar_DR2_Msigma}.  
	Moreover, line-of-sight velocity dispersion could be broadened due to the disc rotation 
	\citep{bennert11,har12,kang13,caglar20}, and this effect should be higher outside the spheroid.
	In the left panels of Fig. \ref{fig:offset} we can see that 
	there is no strong gradient of $M_{BH, \sigma_\star}$ along the $x$-axis, 
	while there is some gradient along the $y$-axis, meaning that 
	at each redshift there is a range of measured $M_{BH, \sigma_\star}$. 
	This implies that $M_{BH, \sigma_\star}$ has a negligible 
	dependence on redshift. 
	As a matter of fact, the log-log relation (solid black line) has always a flat slope in all cases consistent with 
	zero (see Tab. \ref{tab:offset-stat}) 
	and
	there is no clear correlation between $f/f_0$ and redshift. Therefore we can conclude that aperture effects are most probably not important for this sample. As a matter of fact, the optical slit width is typically $\approx1\farcs5$ in BASS DR2, 
	and at $z=0.1$ the sampled region would be of $1.85$~kpc/arcsec $\times 1\farcs5 \simeq 2.8$~kpc. The average effective radius in SDSS late-type galaxies (Sa, Sb, Sc) is $\sim$2.7~kpc \citep{oohama09},
	therefore the spectral extraction would still be roughly within the bulge or spheroid, even at the highest redshift probed in this work.

	If obscuration was biasing the NIR+$L_X$ virial-based BH masses, the sample should exhibit a gradient with $M(line)$ as a function of the X-ray column, and also we should see a positive correlation between the $f/f_0$ and the column density.
	Our sample shows at most the opposite behaviour, a mild anti-correlation between $f/f_0$ and $N_H$, 
	and no clear gradient in $M(line)$ at increasing $N_H$ (center panels in Fig. \ref{fig:offset}). The derived best-fit relations cannot be statistically distinguished from a relation with a zero slope, thus again there is no statistical evidence for a dependence with X-ray obscuration.
	This is indeed not surprising since we show in Sect. \ref{sec:$NIR$nh} that the FWHM measured 
	from NIR lines is not affected by obscuration until high X-ray column densities $\log  (N_H/\rm{cm}^{-2}) \gtrsim 23.5$.

	Finally, we explore whether there is an anti-correlation between $f/f_0$ and $M(line)$, which could be inherited by the adopted definition of \f, such that $f\propto M(line)^{-1}$ (we recall that $M(line)=f_0\times M_{vir,line}$, see Sect. \ref{ss:mvir}).
	The right panels in Fig. \ref{fig:offset} show that indeed there is an anti-correlation, 
	but the best-fit slopes are always flatter than $-1$. 
	The correlations between \f\ and $M(line)$ are statistically different from the $-1$ slope expected simply by definition (see Tab. \ref{tab:offset-stat}, p-value $<7\times10^{-8}$), 
	while the statistical dependence is significantly different from a zero slope only in the He~\textsc{i} and $NIR$ cases.
	Similar conclusions are reached when boostrapping and point perturbation analysis are adopted.
	We note that this $f-M(line)$ mild anticorrelation might as well be driven by a more fundamental relation with the FWHM, since $M(line)\propto FWHM(line)^2$, and already from Fig. \ref{fig:MsMl} it is evident that there is a dependency of the scatter between the two $M_{BH}$ estimates and FWHM. A thorough investigation of the $f-FWHM$ dependence will be carried out in a separate publication (F. Ricci et al. in prep.).
	
	\section{Discussion and Conclusions}\label{sec:disc}
	In this work we present 0.8--2.5~$\mu$m NIR spectroscopic 
	observations of 65 local BASS selected AGN obtained at Magellan/FIRE, splitted into 13 Sy 1-1.5 and 52 Sy 1.8-1.9-2 types.
	We fit four NIR spectral regions (0.9-0.96~$\mu$m, 1.04-1.15~$\mu$m, 1.15-1.30~$\mu$m, and 1.80-2.00~$\mu$m) to study the most characteristic NIR BLR properties, i.e., the BLR velocities and radii, estimated from the most prominent NIR hydrogen (Pa$\alpha$ and Pa$\beta$) and helium (He~\textsc{i}10830\AA) transitions.
	We combine our NIR FIRE sample with the whole BASS NIR database \citep[DR1;][and DR2; \citealt{denBrok_DR2_NIR}]{lamperti17}, finding NIR broad emission lines in 64/235 Seyferts 1.8-1.9-2. The results of this analysis confirm the possibility of using the NIR band to deeper probe the BLR conditions also in optical narrow-line AGN.
	The line showing the highest success rate of broad-line detection in reddened Seyferts is He~\textsc{i} (43/235), followed by Pa$\alpha$ (24/235) and Pa$\beta$ (20/235), suggesting a possible correlation between the NIR line intensity and the higher ability of isolating faint BLR components reliably in Sy 1.8-1.9-2 types.

	We then complement the NIR BLR view with the one obtained from the optical, namely the H$\alpha$, from BASS DR2 \citep{Mejia_Broadlines}.
	In this way we constructed the largest NIR sample of hard X-ray selected local Sy having robust ancillary data available from BASS.
	In terms of $ \mathbf{L_X}$ and $z$, our sample is representative of the BASS AGN sample, even though the fraction of Seyfert types is not \citep[in the BASS DR1 there is an equal portion of Sy 1 and Sy 2; see, e.g., ][]{koss17}, since our program was aimed at detecting the BLR in the most elusive Seyferts 1.8-1.9-2 (235/314 objects, $\approx$75\%). 
	
	With this obscuration unbiased sample in hand, we investigate the presence of possible systematics in the BLR characterisation taking advantage of the X-ray, H$\alpha$ and NIR spectral information in Sy 1 up to Sy 1.9 (and Sy 2, without the H$\alpha$ but with NIR broad line detection).
	We verify that the FWHM measured from H$\alpha$ and NIR lines are consistent in Sy 1 up to Sy 1.9. 
	%(Fig. \ref{fig:fwha-fw$NIR$}). 
	The same results are found even when splitting the sample according to the Sy subclassification. The H$\alpha$ and NIR FWHM broad line measurements statistically describe the same 
	velocity field in the BLR. Thus, once a broad H$\alpha$ line is reliably detected, its FWHM is in agreement with the NIR FWHM within some scatter ($\approx0.10-0.15$~dex, depending on the specific line).  
	We then demonstrate that the H$\alpha$ and NIR BLR velocity estimates are not 
	significantly affected by neither obscuration nor BLR extinction, as measured by the flux decrement of the broad H$\alpha$ flux to the NIR broad line flux, 
	at least until 
	$\log  (N_H/\rm{cm}^{-2}) \approx 23$, as also consistently found in the optical BASS DR2, where the FWHM(H$\alpha$)-$N_H$ distribution remains constant up to $\log  (N_H/\rm{cm}^{-2}) \approx 23$ \citep{Mejia_Broadlines}. 
	
	Rather than the H$\alpha$ FWHM, it is the entire H$\alpha$ broad line intensity that gets suppressed with increasing obscuration, implying that the optical H$\alpha$ photons are obscured by a uniform screen placed outside the BLR. 
	The ratio of broad H$\alpha$-to-NIR broad lines flux shows a mild decreasing trend with $N_H$ for the Pa$\alpha$ sample, while the statistical evidence is weaker for He~\textsc{i} and Pa$\beta$. 
	We quantify the amount of dust extinction towards the BLR and the level of gas absorption measured in the X-rays, finding a clear separation in Seyferts subclasses. %(Fig. \ref{fig:av}). 
	This has been shown already by several other works \citep{burtscher16,schnorr16,shimizu18}. The comparison of two different $A_V$ estimates suggests that either the material obscuring the BLR is distinct from the one producing the absorption in the X-rays, or that the dust-to-gas ratio in local hard X-ray selected AGN environments is not Galactic \citep[see, e.g., ][]{maiolino01}. There are some limitations in the approach adopted to estimate the $A_V(BLR)$, the most notable being the assumption of case B recombination, which might not hold for Paschen lines \citep{soifer04,glikman06,riffel06,LF15}. 
	Additionally, we stress that the broad-line ratios Paschen-to-H$\alpha$, not only depend on dust extinction but also on collisional effects 
	(i.e., by the ionization parameter $U$ and particle density $n$)
	taking place at the high densities of the BLR, $10^{10-11}$~cm$^{-3}$ \citep[see, e.g., ][]{schnorr16}. 
	Therefore the observed dispersion in broad line flux measurements is 
	likely due to collisional effects in addition to extinction.

	We then explore if the line intensity suppression with increasing X-ray column is similarly experienced by NIR line as found in the H$\alpha$. 
	We find a decrement in the broad lines to hard X-ray luminosity ratio of $0.54\pm0.15$~dex for H$\alpha$, decreasing at longer wavelengths to $0.27\pm0.12$~dex in case of Pa$\alpha$, occurring at a level of obscuring column density $\log (N_H / \rm cm^{-2}) =$21.25 for H$\alpha$, moving to higher $N_H$ levels going to longer wavelengths, up to $\log (N_H / \rm cm^{-2}) =21.85$ for Pa$\alpha$.
	The H$\alpha$ broad line intensity suppression with increasing $N_H$ induce a bias in SE H$\alpha$-based BH masses of $\approx$0.2--0.45~dex, depending on the specific line chosen to compare the H$\alpha$ with.
	The NIR line luminosity should 
	be preferred to the H$\alpha$ line luminosities, when available, to 
	estimate $M_{BH}$ using single-epoch relations \citep{kim10,LF15}.
	Notwithstanding, in presence of substantial obscuration  $\log  (N_H/\rm{cm}^{-2}) \gtrsim 22$
	the NIR line luminosity can be underestimated, particularly in Sy 1.9-2. 
	Thus to overcome these shortcomings, it is preferable to use a more unbiased luminosity as proxy to estimate the 
	BLR radius, and a mixed NIR+$L_X$ approach as put forward by \citet[][see also, \citealt{bongiorno14,LF15}]{ricci17a}
	is a more unbiased estimate of the virial BH mass.
	The NIR FWHMs seem as well to be much less susceptible to obscuration, 
	at least up to columns $\log  (N_H/\rm{cm}^{-2}) \approx 23.5$. %(Fig. \ref{fig:fw$NIR$_nh}).

	We finally evaluate the consistency between $\sigma_\star$-based 
	\citep[from the optical BASS DR2,][]{Koss_DR2_sigs,Caglar_DR2_Msigma} and mixed NIR+$L_X$ virial based BH mass estimates, %(Fig. \ref{fig:MsMl}), 
	finding that the
	two quantities agrees within $\sim$0.4--0.57~dex scatter. 
	The scatter expected in this case is at least the combination of two factors: i) there is an inherent scatter in the $M_{BH}-\sigma_\star$ relation of the order of 0.3~dex, and ii) the intrinsic scatter of virial-based $M_{BH}$ estimates is $\simeq$0.4-0.5~dex. Thus, combining the two log-normal contributions, the expected scatter is of the order of 0.5-0.58~dex, which is consistent with what we find. 
	We note however that there are some works suggesting that the $M_{BH}-\sigma_\star$ intrinsic scatter might be higher if the host morphology is not properly taken into account \citep[e.g., ][]{falcon-barroso17}, thus we caution the reader that the $M_{BH}$ based on $\sigma_\star$ for single BASS targets might be characterised by a systematic uncertainty $>$0.3~dex \citep[e.g., ][]{gultekin09}, as there is an inherent
	difficulty in applying morphology-based corrections to the  BAT AGN sample that is composed primarily by massive spirals and lenticulars with strong bulges, with a high fraction of disturbed mergers \citep{koss10,koss11}, adding further complexity to aperture corrections for stellar kinematics.
	
	The ratio of two independent BH mass estimates is here used to derive the virial factor \f, 
	for the first time for a less biased sample, since we also consider the Sy 1.8-1.9-2 showing BLR components in the NIR. 
	The BH-$\sigma_\star$ scaling relation adopted to derive $M_{BH,\sigma_\star}$ is 
	the one proposed by \citet{kormendyho13} calibrated for elliptical and classical bulges. Still, by normalizing the
	virial factor \f\ with the ensemble virial factor $f_0$ we essentially minimize the effects due to the choice of the BH-$\sigma_\star$ scaling relation. As long as the BH-$\sigma_\star$ relation for pseudo-bulges and/or late-type galaxies differ from the one of elliptical/classical-bulges by only a normalization term, as it occurs for the $M_{BH}-\sigma_\star$ relation of pseudo-bulges proposed by \citet{hk14}, the different normalization is absorbed in $f_0$. If instead the $M_{BH}-\sigma_\star$ relation of pseudo-bulges or late-type galaxies has a different slope than the one observed for classical-bulges, then this effect is not absorbed in the $f_0$ normalization. 
	
	Examining the distributions of normalized virial factors $f/f_0$ divided according to the Seyfert subclassification, our data do not show evidence of different $M_{BH}$ between type 1 and type 2 AGN, as instead claimed by \citet{onori17b} and \citet{ricci17b}.
	These works used a small sample of BAT-selected local Seyferts 1.8-1.9-2 with NIR broad lines 
	and compare their $M_{BH}$  
	to the well-known and best studied BAT-selected type 1 AGN with RM-based BH masses. 
	By matching type 1 and obscured Seyferts in hard X-ray luminosity and
	stellar velocity dispersion they find that Sy 1.8-1.9-2 are undermassive than RM AGN.
	This BH mass difference
	is mainly driven by small sample statistics, since i) their small obscured Seyfert sample do not span the same FWHM range of type 1, being limited to FWHM$\lesssim3600$~\kms, and 
	ii) the type 1 control sample is biased to higher FWHM, having FWHM$\gtrsim1600$~\kms, 
	while in this work, thanks to the additional spectra obtained with FIRE and Xshooter, we much better populate the FWHM-$L_X$ plane for both AGN populations. \\

	Dependencies of the normalized virial factor $f/f_0$ with redshift, obscuration and virial-based BH mass estimate have been explored, only finding a mild dependency with $M(line)$. %(Fig. \ref{fig:offset}).
	This mild anti-correlation might be actually due to a more fundamental relation with the FWHM.
	Indeed, using a sample of about 600 local SDSS optical broad line AGN, 
	\citet{shenho14} show that the virial factor, estimated as  
	$M_{BH,\sigma_{\star}}/M(H\beta)$, is anticorrelated with the FWHM(H$\beta$), 
	and claim that this effect is a by-product of the line broadening due to inclination of clouds moving in a 
	disk-like BLR \citep[see, e.g.,][]{collin06}. 
	Similarly, an earlier attempt by \citet{decarli08} demonstrated that there is an anticorrelation 
	between the virial factor \f, computed as the ratio of the BH mass estimated from the BH-bulge luminosity $M_{BH}-L_{bul}$ relation and the virial H$\beta$-based BH mass, and the broad H$\beta$ FWHM, and this 
	inclination effect might bias single-epoch virial-based BH masses.
	At higher redshift, using accretion-disk models to get a BH mass 
	estimate independent of the virial assumption, \citet{mejia18} 
	have further demonstrated 
	that this \f-FWHM anticorrelation is found not only for the H$\beta$ line but also in H$\alpha$, Mg~\textsc{ii} and C~\textsc{iv}, and can be explained with inclination effects on the measurements of the velocities of clouds moving in a thin disk BLR. 
	We aim to further investigate this possibility with a more detailed analysis on this issue in a separate paper, exploring possible BLR parameters space consistent with our data (F. Ricci et al. in prep.).\\ 
	
	Our findings corroborate that the use of a mixed virial-based NIR BH mass estimator 
	gives a less biased view of the BLR properties, and that the use of a single ensemble
	average virial factor $f_0$ for the whole AGN population, might bias our understanding 
	about the AGN demographics and evolution, since we showed that \f\ is at least related (though weakly) to $M(line)$ in two broad lines samples (He~\textsc{i} and $NIR$ line data sets).
	Hopefully in the near future, the astronomical community will take advantage of the James Webb Space Telescope (JWST), that will essentially allow to explore the rest-frame NIR properties of AGN and galaxies also at higher redshift. Indeed, the Pa$\beta$ line at $z\sim2-3$ can be 
	observed in the $1-5~\micron$ wavelength range with NIRSpec on board JWST. Additionally, 
	being a satellite, the limitations due to ground-based observations will be overcome and more efficient 
	NIR observation will be carried out, opening the path to the construction of large statistical samples of AGN with rest-frame NIR coverage at higher redshifts. 
	
	\section{Summary}\label{sec:sum}
	With the largest and least-biased sample of local hard X-ray selected AGN with restframe NIR spectroscopy, we characterise three key BLR properties directly related to virial-based BH mass measurements, i.e., the BLR velocity, estimated from broad line widths FWHM, the average BLR radius, from broad line and X-ray luminosities, and BLR geometry/dynamics, enclosed in the virial factor \f.
	Our findings are the following:
	\begin{enumerate}
		\item the FWHMs measured from H$\alpha$ and NIR broad lines He~\textsc{i}, Pa$\beta$ and Pa$\alpha$ are consistent in Sy 1 up to Sy 1.9, within an intrinsic scatter of 0.10--0.15~dex, thus the optical and NIR statistically describe the same velocity field in the BLR. Moreover, the FWHM measurements do not depend on the level of BLR extinction or X-ray column density, at least up to $\log (N_H/\rm cm^{-2}) \approx 23.5$;
		\item the H$\alpha$ broad line luminosity gets suppressed with increasing obscuration, with a decrement of 
		0.54$\pm$0.15~dex in the $L(H\alpha)/L_X$ ratio occurring at $\log (N_H/\rm cm^{-2})=$~21.25~cm$^{-2}$. This effect produce a bias in SE H$\alpha$-based $M_{BH}$ when $N_H\gtrsim10^{21}$~cm$^{-2}$;
		\item the material obscuring the BLR is distinct from the one responsible of the X-ray absorption, or the dust-to-gas ratio is not Galactic in hard X-ray selected local Seyfert environments;
		\item NIR broad lines are suppressed with increasing obscuration, but the decrement is smaller than the H$\alpha$ one ($0.54\pm0.12$, $0.46\pm0.14$ and $0.27\pm0.12$ for 
		$L({\rm He~\textsc{i}})/L_X$, $L(Pa\beta)/L_X$ and $L(Pa\alpha)/L_X$, respectively) and occurs at slightly higher $N_H$ levels, $\log (N_H/ \rm cm^{-2}) = 21.75$, 21.45 and 21.85 for  He~\textsc{i}, Pa$\beta$ and Pa$\alpha$, respectively. Even NIR broad line luminosities should not be used in SE $M_{BH}$ estimates when $N_H\gtrsim10^{22}$~cm$^{-2}$, and a less biased BLR radius proxy should be used, as the hard X-ray luminosity $L_X$;
		\item using two obscuration-unbiased BH mass estimates, one based on the $\sigma_\star$ and the other based on the mixed near-infared+$L_X$ virial mass, we show that the two BH mass measurements agree with each other with an intrinsic scatter of 0.4--0.57~dex;
		\item we quantify the virial factors \f\ as the ratio of these two independent BH mass measurements and verify that Sy 1 and Sy 1.8-1.9-2 types have the same distribution of virial factors and that our virial factors are not biased with $z$ or $N_H$ but show a mild anti-correlation with $M(line)$. This last finding might be driven by a more fundamental anti-correlation with the observed FWHM expected due to inclination effects.
	\end{enumerate}

	\acknowledgments
	We thank the anonymous referee for the useful comments that improved our manuscript.
	FR thanks 
	the LCO Instrument \& Operations Support Specialists that supported our runs at LCO using Magellan/FIRE, and in particular G. Prieto for sharing 
	the record of local weather data used to run the molecfit correction on our FIRE observations.
	We acknowledge support from FONDECYT Postdoctorado 3180506 (FR) and 3210157 (ARL); from PRIN MIUR 2017 project “Black Hole winds
	and the Baryon Life Cycle of Galaxies: the stone-guest at the galaxy evolution supper”, contract \#2017PH3WAT (FR); 
	FONDECYT Regular 1190818 (ET,FEB) and 1200495 (FEB,ET);
	ANID grants CATA-Basal AFB-170002 (FR,ET,FEB); 
	ANID Anillo ACT172033 (ET); from Millennium Nucleus NCN19\_058 TITANs
		(ET); and Chile's Ministry of Economy, Development, and Tourism's Millennium Science Initiative through grant IC12\_009, awarded to The Millennium Institute of Astrophysics, MAS (FEB); from NASA through ADAP award NNH16CT03C (MK); the ANID+PAI Convocatoria Nacional subvencion a instalacion en la academia convocatoria a\~no 2017 PAI77170080 (CR); 
		from the State Research Agency
		(AEI-MCINN) and from the Spanish MCIU under grant "Feeding and feedback
		in active galaxies" with reference PID2019-106027GB-C42 (PSB);
	from the Israel Science Foundation grant number 1849/19 (BT);
	the National Research Foundation of Korea (NRF-2020R1C1C1005462, KO);
	the Japan Society for the Promotion of Science (JSPS, ID: 17321, KO); the Jet Propulsion Laboratory, California Institute of Technology, under a contract with NASA.\\

	%% telescopes the AAS Journals has created a group of keywords for telesc
	
	%\hspace{15mm}
	%\vspace{15mm}
	\facilities{Magellan-LCO, Swift (BAT) }
	
	\software{FireHose (v2; Gagne et al. 2015), molecfit (Smette et al. 2015), pymccorrelation package (Privon et al. 2020), PySpecKit (v0.1.21; Ginsburg \& Mirocha 2011)}

	\newpage
	\appendix
	\restartappendixnumbering

	\section{FIRE/Magellan broad line fit measurements}
	In this Sect. we list the spectral emission line fit measurements of the He~\textsc{i} and Pa$\gamma$ (Tab. \ref{tab:bHei}), Pa$\beta$ (Tab. \ref{tab:bPab}), and Pa$\alpha$ (Tab. \ref{tab:bPaa}) broad lines for the whole FIRE/Magellan sample. 
	The few cases in which additional components in the Pa$\gamma$ or Pa$\alpha$ region were required are presented in Tab. \ref{tab:addHei}-\ref{tab:addPaa}.
	
	\startlongtable
	\begin{deluxetable*}{ccccccccc}
		\tablecaption{Broad line measurements of the He~\textsc{i} and Pa$\gamma$.}\label{tab:bHei}
		\tablehead{
			\colhead{BAT ID}&	
			\colhead{flag}&	
			\colhead{noise}&		
			\colhead{He~\textsc{i} flux}&	
			\colhead{Pa$\gamma$ flux}&	
			\colhead{FWHM He~\textsc{i}}&	
			\colhead{FWHM Pa$\gamma$}&
			\colhead{$\Delta$v He~\textsc{i}}&
			\colhead{$\Delta$v Pa$\gamma$} \\
			\colhead{} &
			\colhead{} &
			\colhead{} &
			\colhead{} &
			\colhead{} &
			\colhead{[\kms]} &
			\colhead{[\kms]} & 
			\colhead{[\kms]} & 
			\colhead{[\kms]} \\		
			\colhead{(1)}&\colhead{(2)}&\colhead{(3)}&\colhead{(4)}&\colhead{(5)}&\colhead{(6)}&\colhead{(7)}&\colhead{(8)}&\colhead{(9)}
		}
		\startdata
		7		&2	&4.21E-17	&$<$6.00	 			&$<$2.59	 			&		 		&		 	& 	 	 	 &\\
		10		&2	&1.12E-17	&1.968$\pm$0.064		&$<$0.65	 			&7072$\pm$273		&		 	&-312$\pm$42	& 	 \\
		80		&2	&2.19E-16	&$<$10.91	 			&$<$5.14	 			&		 		&		 	& 	 	 	 &\\
		118		&2	&1.42E-17	&$<$0.69	 			&$<$0.75	 			&		 		&		 	& 	 	 	 &\\
		238		&9	&2.45E-17	&$<$0.72	 			&$<$0.79	 			&		 		&		 	& 	 	 	 &\\
		262		&4	&3.88E-17	&$<$4.61	 			&$<$2.48	 			&		 		&		 	& 	 	 	 &\\
		272		&4	&4.65E-17	&0.92$\pm$0.37			&1.238$\pm$0.082		&1302$\pm$521		&1289$\pm$203	&1000$\pm$232	&1000$\pm$232\\
		305		&2	&3.75E-17	&1.342$\pm$0.028		&$<$2.44	 			&1297.76$\pm$0.03	&		 	&-53.4$\pm$6.3	& 	 \\
		329		&9	&4.92E-17	&$<$4.95	 			&$<$15.78	 			&		 		&		 	& 	 	 	 &\\
		372		&3	&2.91E-17	&$<$4.55	 			&7.44$\pm$0.78			&		 		&11552$\pm$306	& 	 			&534$\pm$142\\
		﻿\enddata
		\tablenotetext{}{Columns: (1) BAT ID, (2) quality fit flag, (3) noise measured in the continuum in erg/cm$^2$/s/\AA, (4)-(6)-(8) He~\textsc{i} flux (in units of 10$^{-14}$ erg/cm$^2$/s), FWHM and velocity shift, respectively, 
			(5)-(7)-(9) Pa$\gamma$ flux (in units of 10$^{-14}$ erg/cm$^2$/s), FWHM and velocity shift, respectively. Values in columns (4) - (5) preceded by `$<$' indicate upper limits. The upper limits on the broad line fluxes have been computed using a FWHM=4200~\kms. (This table is available in its entirety in a machine-readable form in the online journal. A part is shown as guidance for the reader regarding its content.)}
	\end{deluxetable*}
	
	\newpage
	\startlongtable
	\begin{deluxetable*}{cccccc}
		\tablecaption{Broad line measurements of the Pa$\beta$.}\label{tab:bPab}
		\tablehead{
			\colhead{BAT ID}&	
			\colhead{flag}&	
			\colhead{noise}&		
			\colhead{Pa$\beta$ flux}&	
			\colhead{FWHM Pa$\beta$}&	
			\colhead{$\Delta$v Pa$\beta$}\\
			\colhead{} &
			\colhead{} &
			\colhead{} &
			\colhead{} & 
			\colhead{[\kms]} & 
			\colhead{[\kms]} \\
			\colhead{(1)}&\colhead{(2)}&\colhead{(3)}&\colhead{(4)}&\colhead{(5)}&\colhead{(6)}
		}
		\startdata
		7		&4	&3.50E-17	&$<$17.87	 			&		 	& 	\\
		10		&9	&9.81E-18	&$<$13.43	 			&		 	& 	\\
		80		&4	&1.20E-16	&$<$2.79	 			&		 	& 	\\
		118		&9	&1.67E-17	&$<$0.51	 			&		 	& 	\\
		238		&9	&1.54E-15	&$<$3.00	 			&		 	& 	\\
		262		&9	&3.56E-17	&$<$1.78	 			&		 	& 	\\
		272		&2	&4.82E-17	&5.549$\pm$0.019		&1750$\pm$10	&-28.9$\pm$3.5\\
		305		&1	&3.52E-17	&$<$3.40	 			&		 	& 	\\
		329		&9	&6.91E-17	&$<$2.29	 			&		 	& 	\\
		372		&3	&4.26E-17	&$<$3.33	 			&		 	& 	\\
		\enddata
		\tablenotetext{}{Columns: (1) BAT ID, (2) quality fit flag, (3) noise measured in the continuum in erg/cm$^2$/s/\AA, (4)-(5)-(6) Pa$\beta$ flux (in units of 10$^{-14}$ erg/cm$^2$/s), FWHM and velocity shift, respectively. Values in columns (4) preceded by `$<$' indicate upper limits. The upper limits on the broad line fluxes have been computed using a FWHM=4200~\kms (5). (This table is available in its entirety in a machine-readable form in the online journal. A part is shown as guidance for the reader regarding its content.) }
	\end{deluxetable*}
	
	\newpage
	\footnotesize
	\startlongtable
	\begin{deluxetable*}{cccccc}
		\tablecaption{Broad line measurements of the Pa$\alpha$.}\label{tab:bPaa}
		\tablehead{
			\colhead{BAT ID}&	
			\colhead{flag}&	
			\colhead{noise}&		
			\colhead{Pa$\alpha$ flux}&	
			\colhead{FWHM Pa$\alpha$}&	
			\colhead{$\Delta$v Pa$\alpha$}\\
			\colhead{} &
			\colhead{} &
			\colhead{} &
			\colhead{} & 
			\colhead{[\kms]} & 
			\colhead{[\kms]} \\		
			\colhead{(1)}&\colhead{(2)}&\colhead{(3)}&\colhead{(4)}&\colhead{(5)}&\colhead{(6)}
		}
		\startdata
		7		&2	&1.52E-17	&$<$10.65	 			&		 	 		& \\
		10		&3	&6.40E-18	&1.69	 			&8590$\pm$61		&-1000	 \\
		80		&4	&1.89E-17	&$<$3.41	 			&		 	 		& \\
		118		&2	&7.97E-18	&$<$1.00	 			&		 	 		& \\
		238		&2	&1.52E-17	&$<$1.91	 			&		 	 		& \\
		262		&2	&1.49E-17	&$<$3.18	 			&		 	 		& \\
		272		&4	&6.93E-17	&21.81				&2419$\pm$10		&7.1$\pm$2.5\\
		305		&2	&1.25E-17	&1.51$\pm$0.013		&1527$\pm$27		&31.2$\pm$7.2\\
		329		&4	&2.78E-17	&$<$5.49	 			&		 	 		& \\
		372		&3	&9.86E-18	&7.667$\pm$0.024	&11907$\pm$32		&-1000	\\
		\enddata
		\tablenotetext{}{Columns: (1) BAT ID, (2) quality fit flag, (3) noise measured in the continuum in erg/cm$^2$/s/\AA, (4)-(5)-(6) Pa$\alpha$ flux (in units of 10$^{-14}$ erg/cm$^2$/s), FWHM and velocity shift, respectively. Values in columns (4) preceded by `$<$' indicate upper limits.  The upper limits on the broad line fluxes have been computed using a FWHM=4200~\kms. (This table is available in its entirety in a machine-readable form in the online journal. A part is shown as guidance for the reader regarding its content.)}
	\end{deluxetable*}	
	\normalsize
	
	\begin{deluxetable*}{ccccc}
		\tablecaption{Additional Gaussian component of the He~\textsc{i} in the  Pa$\gamma$ spectral region.}\label{tab:addHei}
		\tablehead{
			\colhead{BAT ID}&	
			\colhead{flag}&	
			\colhead{He~\textsc{i} flux}&	
			\colhead{FWHM He~\textsc{i}}&	
			\colhead{$\Delta$v He~\textsc{i}}\\
			\colhead{} &
			\colhead{} &
			\colhead{} & 
			\colhead{[\kms]} & 
			\colhead{[\kms]} \\		
			\colhead{(1)}&\colhead{(2)}&\colhead{(3)}&\colhead{(4)}&\colhead{(5)}
		}
		\startdata
		372		&3	&2.86		&	1957			&277\\
		488		&3	&2.880$\pm$0.059&	1627.1$\pm$0.03		&-324.6$\pm$6.2\\
		577		&3	&0.31$\pm$0.22	&	735$\pm$186			&-397$\pm$16\\
		698		&3	&0.13$\pm$0.0020&	856$\pm$13			&-993.3$\pm$4.2\\
		744		&2	&4.425$\pm$0.062&	3463$\pm$44			&238$\pm$10\\
		1064	&2	&7.97$\pm$0.64	&	1212$\pm$270		&-364$\pm$22\\
		1079	&2	&7.432$\pm$0.027&	1008.8$\pm$3.1		&-287.3$\pm$1.4\\
		\enddata
		\tablenotetext{}{Columns: (1) BAT ID, (2) quality fit flag (same as in Tab. \ref{tab:bHei}), (3)-(4)-(5) flux (in units of 10$^{-14}$ erg/cm$^2$/s), FWHM and velocity shift of the additional Gaussian component in the He~\textsc{i}. These components were not used to measure the virial black hole masses.  }
	\end{deluxetable*}

	\begin{rotatetable*}
		\begin{deluxetable*}{ccccccccccc}
			\tablecaption{Additional Gaussian component of in the Pa$\alpha$ spectral region.}\label{tab:addPaa}
			\tablehead{
				\colhead{BAT ID}&	
				\colhead{flag}&	
				\colhead{Pa$\alpha$ flux}&	
				\colhead{H2 a  flux}&	
				\colhead{H2 b  flux}&	
				\colhead{FWHM Pa$\alpha$}&
				\colhead{FWHM H2 a }&
				\colhead{FWHM H2 b }&	
				\colhead{$\Delta$v Pa$\alpha$}& 
				\colhead{$\Delta$v H2 a}& 
				\colhead{$\Delta$v H2 b}\\
				\colhead{} &
				\colhead{} &
				\colhead{} &
				\colhead{} &
				\colhead{} &
				\colhead{[\kms]} &
				\colhead{[\kms]} & 
				\colhead{[\kms]} & 
				\colhead{[\kms]} & 
				\colhead{[\kms]} & 
				\colhead{[\kms]} \\		
				\colhead{(1)}&\colhead{(2)}&\colhead{(3)}&\colhead{(4)}&\colhead{(5)}&\colhead{(6)}&\colhead{(7)}&\colhead{(8)}&\colhead{(9)}&\colhead{(10)}&\colhead{(11)}
			}
			\startdata
			372		&3&1.609$\pm$0.014	&			 	&				&3017.48$\pm$0.05	&			  &			  &593.3$\pm$5.4	&				 &	\\
			1079	&4&0.73$\pm$0.24	&2.599$\pm$0.013&0.642$\pm$0.011&299.45$\pm$0.01	&2974.68&29758&-148.25$\pm$8.9	&-453.62$\pm$6.86&-999\\
			\enddata
			\tablenotetext{}{Columns: (1) BAT ID, (2) quality fit flag (same as in Tab. \ref{tab:bPaa}), (3)-(6)-(9) flux (in units of 10$^{-14}$ erg/cm$^2$/s), FWHM and velocity shift of the additional gaussian component in the Pa$\alpha$, respectively, 
				(4)-(7)-(10) flux (in units of 10$^{-14}$ erg/cm$^2$/s), FWHM and velocity shift, of the H2 a, i.e. H2 $\lambda$18345\AA, and 
				(5)-(8)-(11) flux (in units of 10$^{-14}$ erg/cm$^2$/s), FWHM and velocity shift of the H2 b, i.e. H2 $\lambda$18920\AA. These additional Pa$\alpha$ components were not used to measure the virial black hole masses. }
		\end{deluxetable*}	
	\end{rotatetable*}

	\section{$f/f_0$ outliers}\label{sec:app-outliers}
	There are some targets showing an outlier nature in the $f/f_0$ distributions that have been omitted in our analysis. These sources are five in total, namely: IDs 670, 677, 1131, 1465 and 1470.
	The outliers  
	are mostly (4/5) located above the 1:1 line in Fig. \ref{fig:MsMl}, meaning that either the $M_{BH,\sigma_{\star}}$ is overestimated, or that the NIR BH mass is underestimated. The former case would indicate that the measured optical stellar velocity dispersion is unreliable for those sources, due to AGN contamination of the stellar absorption features, aperture effects or to host galaxy rotation contamination, while the latter case would happen if there is substantial obscuration along the line of sight, such that the highest velocity clouds of the BLR are not observable, being embedded in some obscuring medium. 
	These four outliers are IDs 670, 1131, 1465 and 1470.

	For BAT IDs 670, 1131,  
	1465  and 1470 the NIR+$L_X$ based BH mass is somehow underestimated. 
	Those are part of the NIR DR2 \citep{denBrok_DR2_NIR}: those NIR spectra show only broad He~\textsc{i} line, as Pa$\alpha$ is located in a heavily affected telluric region and Pa$\beta$ is only detected, marginally, as a narrow line. All the other higher-order Paschen lines are missing as well. For ID 670, the broad He~\textsc{i} transition might not be (fully) tracing the BLR, since a blueshifted outflow in the [O~\textsc{iii}] has been detected, with $v_{max}=1072^{+80}_{-990}$~\kms\, \citep{rojas20}. 
	
	For IDs 1131, 1465 and 1470 (Sy 1.8, 2 and 2, respectively), such information is not available since they were not part of the sample studied in \citet{rojas20}. However, the optical BASS DR2 spectrum of BAT 1131 appears to have some ionized outflows in the [O~\textsc{iii}], as the residuals from a single Gaussian fit are rather prominent, blueshifted and asymmetric. Therefore the broad component detected in the He~\textsc{i} might not be describing the BLR. Additionally, this source, a Sy 1.8, is particularly AGN-dominated in the optical spectrum, making the stellar velocity dispersion measurement a bit tricky.
	
	Finally, we note that for ID 1470 the Galactic extinction is fairly important, $E(B-V)=0.48$ (corresponding to an $A_J=0.354$~mag according to the IRSA dust database\footnote{https://irsa.ipac.caltech.edu/applications/DUST/}), while for 1465 is quite relevant, $E(B-V)=3.03$ ($A_J=2.150$~mag according to the IRSA database), probably affecting the NIR spectrum measurement. 
	
	One source is instead located below the 1:1 relation, at $\log f/f_0< -1$, namely BAT 677. In this case, the NIR+$L_X$ based $M(line)$ (e.g., $M(NIR)=10^{7.140\pm0.023}$~M$_\odot$) is about two orders of magnitude higher than the $\sigma_\star$-based BH mass estimate (e.g., $M_{BH,\sigma_\star}=10^{5.47\pm0.29}$~M$_\odot$). BAT 677 is part of the FIRE sample, and has a good Pa$\beta$ fit (FWHM=2401$\pm$8~\kms, fit quality 2). The $M_{BH,\sigma_\star}$ is computed from a rather small velocity dispersion measurement $\sigma_\star =41\pm5$~\kms\ performed on the CaT region. We note that the BH-$\sigma_\star$ relation is not calibrated for $\sigma_\star<65$~\kms, thus extrapolating to such low velocity dispersion might introduce a higher error budget on the $M_{BH,\sigma_\star}$ estimation. We finally note that if instead of $\sigma_\star =41$ we adopt $\sigma_\star =68$~\kms, as measured from the Ca~\textsc{ii} and Mg~\textsc{i} absorptions, the resulting $M_{BH,\sigma_\star}$ is about an order of magnitude higher ($M_{BH,\sigma_\star}=10^{6.44\pm0.20}$~M$_\odot$) and it is more consistent with the virial NIR+$L_X$ based estimate.

	Therefore,  
	for the former four sources the NIR+$L_X$ based $M(line)$ is not reliable, either due to  
	the possible presence of outflows in ionized material disguised as BLR components (670, also possibly 1465 and 1470) or high Galactic extinction in our line of sight (1465, 1470). For BAT 1131 both the $\sigma_\star$ and the NIR measurements are not trustable, for strong AGN contamination in the optical spectrum and for possible outflows in ionized material. Finally for BAT 677 the $M_{BH,\sigma_\star}$ is more uncertain being located in the low-velocity dispersion extrapolation of the BH-$\sigma_\star$ scaling relation.
	
	\startlongtable
	\begin{deluxetable*}{llllllccr}   %%%%%%%%%%%%%%%%%%%%%%%%%%%%%%%%%%%%%%%%%%%%%%%%%%%%%%%%%%%%%%%%%%%%%%%
		\tablecaption{Physical properties of the sample having both NIR reliable broad line detection and optical stellar velocity dispersion measurements available inside BASS.\label{tab:sample}}
		\tablehead{
			\colhead{BAT ID}&
			\colhead{Sy class}& 
			\multicolumn{4}{c}{FWHM [\kms] }  & 
			\colhead{$\log M(NIR)$} & 
			\colhead{$\log f_{NIR}$} &
			\colhead{$\log N_H$}\\
			\colhead{}& 
			\colhead{}&
			\colhead{He~\textsc{i}}& 
			\colhead{Pa$\beta$}&
			\colhead{Pa$\alpha$}&
			\colhead{$NIR$}& 
			\colhead{[M$_\odot$]} &
			\colhead{}&
			\colhead{[cm$^{-2}$]}\\
			\colhead{(1)}&\colhead{(2)}&\colhead{(3)}&\colhead{(4)}&\colhead{(5)}&\colhead{(6)}&\colhead{(7)}&\colhead{(8)}&\colhead{(9)}
		}
		\startdata 
10	&1.9	&5906$\pm$336	&			&		&5906$\pm$336	&8.827$\pm$0.054	&0.91$\pm$0.15	&21.98\\
53	&2	&3260$\pm$188	&			&		&3260$\pm$188	&7.910$\pm$0.055	&0.705$\pm$0.092	&23.54\\
63	&2	&			&1429$\pm$36	&		&1429$\pm$36	&6.664$\pm$0.032	&1.42$\pm$0.11	&23.30\\
72	&1.9	&2347$\pm$76	&6673$\pm$183	&		&2975$\pm$70	&7.797$\pm$0.031	&1.059$\pm$0.068	&22.01\\
73	&1.2	&2226$\pm$30	&5960$\pm$80	&		&2690$\pm$28	&8.020$\pm$0.025	&1.43$\pm$0.10	&20.00\\
193	&2	&3351$\pm$271	&			&		&3351$\pm$271	&7.904$\pm$0.074	&1.77$\pm$0.10	&24.32\\
202	&1.9	&6122$\pm$203	&			&		&6122$\pm$203	&8.728$\pm$0.037	&0.457$\pm$0.065	&23.03\\
217	&2	&4634$\pm$128	&			&		&4634$\pm$128	&8.283$\pm$0.033	&1.141$\pm$0.063	&22.89\\
218	&2	&			&1750$\pm$68	&		&1750$\pm$68	&7.477$\pm$0.041	&-0.10$\pm$0.16	&23.84\\
246	&1.9	&5355$\pm$155	&			&		&5355$\pm$155	&8.582$\pm$0.034	&0.522$\pm$0.064	&22.18\\
		\enddata
		\tablenotetext{}{Columns are: (1) BAT ID; (2) optical Seyfert classification; (3-6) near-infrared broad line measurements, from either BASS DR1 \citep{lamperti17}, DR2 \citep{denBrok_DR2_NIR} or from this work, i.e., FIRE spectra; (7) logarithm of the mixed NIR+$L_X$-based BH mass, calculated using the FWHM($NIR$); 
			(8) logarithm of the virial factor computed as the ratio of the $\sigma_\star$-based BH mass with velocity dispersions from the BASS DR2 
			\citep[either from][or \citealt{Caglar_DR2_Msigma}]{Koss_DR2_sigs}
			and $M(NIR)$;
			(9) column density derived from X-ray spectral fitting \citep{cricci17}. Upper limits on the X-ray columns are denoted with $<$. (This table is available in its entirety in a machine-readable form in the online journal. A part is shown as guidance for the reader regarding its content.)}
	\end{deluxetable*}
	
	\bibliography{bibfinal.bib}{}
	\bibliographystyle{aasjournal}

\end{document}